\documentclass[longauth]{aaEC}

\usepackage{graphicx}
\usepackage{natbib}
\usepackage{scalerel}

\usepackage[table]{xcolor}

\usepackage{txfonts}
\usepackage[pdfencoding=auto,psdextra]{hyperref}
\hypersetup{
    colorlinks=true,
    linkcolor=blue,
    filecolor=magenta,      
    urlcolor=blue,
    citecolor=blue
}
\makeatletter
\renewcommand*\aa@pageof{, page \thepage{} of \pageref*{LastPage}}
\makeatother

\usepackage[utf8]{inputenc}

\usepackage[switch, modulo]{lineno}
\linenumbers

\usepackage{euclid}

\def\siiv     {\ensuremath{\text{Si\,\textsc{iv}}}\xspace}

\def\civ     {\ensuremath{\text{C\,\textsc{iv}}}\xspace }

\def\mgii     {\ensuremath{\text{Mg\,\textsc{ii}}}\xspace }

\def\heii     {\ensuremath{\text{He\,\textsc{ii}}}\xspace}

\def\ciii     {\ensuremath{\text{C\,\textsc{iii]}}}\xspace}

\def\heii    {\ensuremath{\text{He\,\textsc{ii}}}\xspace}

\def\lya {\ensuremath{\text{Ly}\alpha}\xspace}
\def\nv {\ensuremath{\text{N\,\textsc{v}}}\xspace }
\def\oi {\ensuremath{\text{O\,\textsc{i}}}\xspace }

\usepackage{booktabs}
\usepackage{threeparttable}
\usepackage{multirow}
\usepackage{physics}

\usepackage{enumitem}
\usepackage{rotating}
\usepackage{float}

\usepackage[final]{changes} % clean
\setaddedmarkup{\textcolor{red}{#1}}
\setdeletedmarkup{\textcolor{red}{\sout{#1}}}

\begin{document}
%
% Put the title and authors of your (Standard Project) paper here
%

\title{\Euclid\/: Discovery of 31 new quasars at $6.6 < z < 7.8$\thanks{This paper is published on
       behalf of the Euclid Consortium}}

\newcommand{\orcid}[1]{} %% if already defined in aa.cls: comment, or use renewcommand			   
\author{D.~Yang\orcid{0000-0002-6769-0910}\thanks{\email{dyang@strw.leidenuniv.nl}}\inst{\ref{aff1}}
\and J.~F.~Hennawi\orcid{0000-0002-7054-4332}\inst{\ref{aff1},\ref{aff2}}
\and F.~Guarneri\orcid{0000-0003-4740-9762}\inst{\ref{aff3},\ref{aff4}}
\and J.~Wolf\orcid{0000-0003-0643-7935}\inst{\ref{aff5}}
\and S.~Belladitta\orcid{0000-0003-4747-4484}\inst{\ref{aff5},\ref{aff6}}
\and J.-T.~Schindler\orcid{0000-0002-4544-8242}\inst{\ref{aff3}}
\and A.~C.~N.~Hughes\orcid{0000-0001-9294-3089}\inst{\ref{aff7}}
\and E.~Ba\~nados\orcid{0000-0002-2931-7824}\inst{\ref{aff5}}
\and D.~J.~Mortlock\orcid{0000-0002-0041-3783}\inst{\ref{aff7},\ref{aff8}}
\and J.~Yang\orcid{0000-0001-5287-4242}\inst{\ref{aff9}}
\and F.~Wang\orcid{0000-0002-7633-431X}\inst{\ref{aff9}}
\and X.~Fan\orcid{0000-0003-3310-0131}\inst{\ref{aff10}}
\and K.~Jahnke\orcid{0000-0003-3804-2137}\inst{\ref{aff5}}
\and D.~Stern\orcid{0000-0003-2686-9241}\inst{\ref{aff11}}
\and C.~J.~Willott\orcid{0000-0002-4201-7367}\inst{\ref{aff12}}
\and A.~J.~Barth\orcid{0000-0002-3026-0562}\inst{\ref{aff13}}
\and H.~J.~A.~Rottgering\orcid{0000-0001-8887-2257}\inst{\ref{aff1}}
\and R.~G.~Varadaraj\orcid{0009-0006-9953-6471}\inst{\ref{aff14}}
\and R.~Decarli\orcid{0000-0002-2662-8803}\inst{\ref{aff6}}
\and A.-C.~Eilers\orcid{0000-0003-2895-6218}\inst{\ref{aff15},\ref{aff16}}
\and M.~Ezziati\orcid{0009-0003-6065-1585}\inst{\ref{aff17}}
\and Y.~Fu\orcid{0000-0002-0759-0504}\inst{\ref{aff1},\ref{aff18}}
\and J.~Huang\orcid{0000-0002-5721-0709}\inst{\ref{aff2}}
\and X.~Jin\orcid{0000-0002-5768-738X}\inst{\ref{aff9},\ref{aff19}}
\and Y.~Kang\orcid{0009-0001-7010-5989}\inst{\ref{aff2}}
\and L.~N.~Martinez-Ramirez\orcid{0009-0003-5506-5469}\inst{\ref{aff5},\ref{aff20},\ref{aff21}}
\and Y.~Matsuoka\orcid{0000-0001-5063-0340}\inst{\ref{aff22}}
\and M.~Onoue\orcid{0000-0003-2984-6803}\inst{\ref{aff23},\ref{aff24}}
\and R.~Pello\orcid{0000-0003-0858-6109}\inst{\ref{aff17}}
\and R.~P.~Remigio\orcid{0000-0002-0164-8795}\inst{\ref{aff13}}
\and W.~L.~Tee\orcid{0000-0003-0747-1780}\inst{\ref{aff25}}
\and B.~Venemans\orcid{0000-0001-9024-8322}\inst{\ref{aff1}}
\and G.~Vietri\orcid{0000-0001-9155-8875}\inst{\ref{aff26}}
\and B.~Wang\orcid{0000-0003-4877-1659}\inst{\ref{aff27},\ref{aff1}}
\and L.~J.~Abbo\orcid{0009-0004-4865-3614}\inst{\ref{aff1}}
\and H.~Atek\orcid{0000-0002-7570-0824}\inst{\ref{aff28}}
\and S.~Bisogni\orcid{0000-0003-3746-4565}\inst{\ref{aff26}}
\and S.~E.~I.~Bosman\orcid{0000-0001-8582-7012}\inst{\ref{aff29},\ref{aff5}}
\and R.~A.~A.~Bowler\orcid{0000-0003-3917-1678}\inst{\ref{aff30}}
\and C.~J.~Conselice\orcid{0000-0003-1949-7638}\inst{\ref{aff30}}
\and F.~B.~Davies\orcid{0000-0003-0821-3644}\inst{\ref{aff5}}
\and C.~M.~Gutierrez\orcid{0000-0001-7854-783X}\inst{\ref{aff31},\ref{aff32}}
\and Y.~Harikane\orcid{0000-0002-6047-430X}\inst{\ref{aff33}}
\and K.~Rubinur\orcid{0000-0001-5574-5104}\inst{\ref{aff34}}
\and C.~C.~Lovell\orcid{0000-0001-7964-5933}\inst{\ref{aff35},\ref{aff36}}
\and M.~Magliocchetti\orcid{0000-0001-9158-4838}\inst{\ref{aff37}}
\and J.~Matthee\orcid{0000-0003-2871-127X}\inst{\ref{aff38}}
\and F.~Ricci\orcid{0000-0001-5742-5980}\inst{\ref{aff39},\ref{aff40}}
\and M.~Scialpi\orcid{0009-0006-5100-4986}\inst{\ref{aff41},\ref{aff42},\ref{aff43}}
\and D.~Scott\orcid{0000-0002-6878-9840}\inst{\ref{aff44}}
\and L.~Spinoglio\orcid{0000-0001-8840-1551}\inst{\ref{aff37}}
\and F.~Tarsitano\orcid{0000-0002-5919-0238}\inst{\ref{aff45},\ref{aff46}}
\and Y.~Toba\orcid{0000-0002-3531-7863}\inst{\ref{aff47},\ref{aff48}}
\and F.~Walter\orcid{0000-0003-4793-7880}\inst{\ref{aff5}}
\and J.~R.~Weaver\orcid{0000-0003-1614-196X}\inst{\ref{aff16}}
\and G.~Zamorani\orcid{0000-0002-2318-301X}\inst{\ref{aff6}}
\and B.~Altieri\orcid{0000-0003-3936-0284}\inst{\ref{aff49}}
\and A.~Amara\inst{\ref{aff50}}
\and S.~Andreon\orcid{0000-0002-2041-8784}\inst{\ref{aff51}}
\and H.~Aussel\orcid{0000-0002-1371-5705}\inst{\ref{aff52}}
\and C.~Baccigalupi\orcid{0000-0002-8211-1630}\inst{\ref{aff53},\ref{aff4},\ref{aff54},\ref{aff55}}
\and M.~Baldi\orcid{0000-0003-4145-1943}\inst{\ref{aff56},\ref{aff6},\ref{aff57}}
\and A.~Balestra\orcid{0000-0002-6967-261X}\inst{\ref{aff58}}
\and S.~Bardelli\orcid{0000-0002-8900-0298}\inst{\ref{aff6}}
\and P.~Battaglia\orcid{0000-0002-7337-5909}\inst{\ref{aff6}}
\and A.~Biviano\orcid{0000-0002-0857-0732}\inst{\ref{aff4},\ref{aff53}}
\and E.~Branchini\orcid{0000-0002-0808-6908}\inst{\ref{aff59},\ref{aff60},\ref{aff51}}
\and M.~Brescia\orcid{0000-0001-9506-5680}\inst{\ref{aff61},\ref{aff62}}
\and S.~Camera\orcid{0000-0003-3399-3574}\inst{\ref{aff63},\ref{aff64},\ref{aff65}}
\and G.~Ca\~nas-Herrera\orcid{0000-0003-2796-2149}\inst{\ref{aff66},\ref{aff1}}
\and V.~Capobianco\orcid{0000-0002-3309-7692}\inst{\ref{aff65}}
\and C.~Carbone\orcid{0000-0003-0125-3563}\inst{\ref{aff26}}
\and J.~Carretero\orcid{0000-0002-3130-0204}\inst{\ref{aff67},\ref{aff68}}
\and M.~Castellano\orcid{0000-0001-9875-8263}\inst{\ref{aff40}}
\and G.~Castignani\orcid{0000-0001-6831-0687}\inst{\ref{aff6}}
\and S.~Cavuoti\orcid{0000-0002-3787-4196}\inst{\ref{aff62},\ref{aff69}}
\and K.~C.~Chambers\orcid{0000-0001-6965-7789}\inst{\ref{aff70}}
\and A.~Cimatti\inst{\ref{aff71}}
\and C.~Colodro-Conde\inst{\ref{aff31}}
\and G.~Congedo\orcid{0000-0003-2508-0046}\inst{\ref{aff66}}
\and L.~Conversi\orcid{0000-0002-6710-8476}\inst{\ref{aff72},\ref{aff49}}
\and Y.~Copin\orcid{0000-0002-5317-7518}\inst{\ref{aff73}}
\and F.~Courbin\orcid{0000-0003-0758-6510}\inst{\ref{aff74},\ref{aff75},\ref{aff76}}
\and H.~M.~Courtois\orcid{0000-0003-0509-1776}\inst{\ref{aff77}}
\and M.~Cropper\orcid{0000-0003-4571-9468}\inst{\ref{aff78}}
\and J.-C.~Cuillandre\orcid{0000-0002-3263-8645}\inst{\ref{aff52}}
\and H.~Degaudenzi\orcid{0000-0002-5887-6799}\inst{\ref{aff46}}
\and G.~De~Lucia\orcid{0000-0002-6220-9104}\inst{\ref{aff4}}
\and C.~Dolding\orcid{0009-0003-7199-6108}\inst{\ref{aff78}}
\and H.~Dole\orcid{0000-0002-9767-3839}\inst{\ref{aff79}}
\and M.~Douspis\orcid{0000-0003-4203-3954}\inst{\ref{aff79}}
\and F.~Dubath\orcid{0000-0002-6533-2810}\inst{\ref{aff46}}
\and X.~Dupac\inst{\ref{aff49}}
\and S.~Dusini\orcid{0000-0002-1128-0664}\inst{\ref{aff80}}
\and S.~Escoffier\orcid{0000-0002-2847-7498}\inst{\ref{aff81}}
\and M.~Farina\orcid{0000-0002-3089-7846}\inst{\ref{aff37}}
\and R.~Farinelli\inst{\ref{aff6}}
\and S.~Ferriol\inst{\ref{aff73}}
\and F.~Finelli\orcid{0000-0002-6694-3269}\inst{\ref{aff6},\ref{aff82}}
\and N.~Fourmanoit\orcid{0009-0005-6816-6925}\inst{\ref{aff81}}
\and M.~Frailis\orcid{0000-0002-7400-2135}\inst{\ref{aff4}}
\and E.~Franceschi\orcid{0000-0002-0585-6591}\inst{\ref{aff6}}
\and M.~Fumana\orcid{0000-0001-6787-5950}\inst{\ref{aff26}}
\and S.~Galeotta\orcid{0000-0002-3748-5115}\inst{\ref{aff4}}
\and K.~George\orcid{0000-0002-1734-8455}\inst{\ref{aff83}}
\and B.~Gillis\orcid{0000-0002-4478-1270}\inst{\ref{aff66}}
\and C.~Giocoli\orcid{0000-0002-9590-7961}\inst{\ref{aff6},\ref{aff57}}
\and P.~G\'omez-Alvarez\orcid{0000-0002-8594-5358}\inst{\ref{aff84},\ref{aff49}}
\and J.~Gracia-Carpio\inst{\ref{aff85}}
\and A.~Grazian\orcid{0000-0002-5688-0663}\inst{\ref{aff58}}
\and F.~Grupp\inst{\ref{aff85},\ref{aff86}}
\and L.~Guzzo\orcid{0000-0001-8264-5192}\inst{\ref{aff87},\ref{aff51},\ref{aff88}}
\and S.~Gwyn\orcid{0000-0001-8221-8406}\inst{\ref{aff12}}
\and S.~V.~H.~Haugan\orcid{0000-0001-9648-7260}\inst{\ref{aff34}}
\and H.~Hoekstra\orcid{0000-0002-0641-3231}\inst{\ref{aff1}}
\and W.~Holmes\inst{\ref{aff11}}
\and I.~M.~Hook\orcid{0000-0002-2960-978X}\inst{\ref{aff89}}
\and F.~Hormuth\inst{\ref{aff90}}
\and A.~Hornstrup\orcid{0000-0002-3363-0936}\inst{\ref{aff91},\ref{aff92}}
\and M.~Jhabvala\inst{\ref{aff93}}
\and S.~Kermiche\orcid{0000-0002-0302-5735}\inst{\ref{aff81}}
\and B.~Kubik\orcid{0009-0006-5823-4880}\inst{\ref{aff73}}
\and K.~Kuijken\orcid{0000-0002-3827-0175}\inst{\ref{aff1}}
\and M.~K\"ummel\orcid{0000-0003-2791-2117}\inst{\ref{aff86}}
\and M.~Kunz\orcid{0000-0002-3052-7394}\inst{\ref{aff94}}
\and H.~Kurki-Suonio\orcid{0000-0002-4618-3063}\inst{\ref{aff95},\ref{aff96}}
\and A.~M.~C.~Le~Brun\orcid{0000-0002-0936-4594}\inst{\ref{aff97}}
\and S.~Ligori\orcid{0000-0003-4172-4606}\inst{\ref{aff65}}
\and P.~B.~Lilje\orcid{0000-0003-4324-7794}\inst{\ref{aff34}}
\and V.~Lindholm\orcid{0000-0003-2317-5471}\inst{\ref{aff95},\ref{aff96}}
\and I.~Lloro\orcid{0000-0001-5966-1434}\inst{\ref{aff98}}
\and G.~Mainetti\orcid{0000-0003-2384-2377}\inst{\ref{aff99}}
\and D.~Maino\inst{\ref{aff87},\ref{aff26},\ref{aff88}}
\and E.~Maiorano\orcid{0000-0003-2593-4355}\inst{\ref{aff6}}
\and O.~Mansutti\orcid{0000-0001-5758-4658}\inst{\ref{aff4}}
\and O.~Marggraf\orcid{0000-0001-7242-3852}\inst{\ref{aff100}}
\and M.~Martinelli\orcid{0000-0002-6943-7732}\inst{\ref{aff40},\ref{aff101}}
\and N.~Martinet\orcid{0000-0003-2786-7790}\inst{\ref{aff17}}
\and F.~Marulli\orcid{0000-0002-8850-0303}\inst{\ref{aff102},\ref{aff6},\ref{aff57}}
\and R.~J.~Massey\orcid{0000-0002-6085-3780}\inst{\ref{aff103}}
\and H.~J.~McCracken\orcid{0000-0002-9489-7765}\inst{\ref{aff28}}
\and E.~Medinaceli\orcid{0000-0002-4040-7783}\inst{\ref{aff6}}
\and S.~Mei\orcid{0000-0002-2849-559X}\inst{\ref{aff104},\ref{aff105}}
\and Y.~Mellier\inst{\ref{aff106},\ref{aff28}}
\and M.~Meneghetti\orcid{0000-0003-1225-7084}\inst{\ref{aff6},\ref{aff57}}
\and E.~Merlin\orcid{0000-0001-6870-8900}\inst{\ref{aff40}}
\and G.~Meylan\inst{\ref{aff107}}
\and J.~J.~Mohr\orcid{0000-0002-6875-2087}\inst{\ref{aff83}}
\and A.~Mora\orcid{0000-0002-1922-8529}\inst{\ref{aff108}}
\and M.~Moresco\orcid{0000-0002-7616-7136}\inst{\ref{aff102},\ref{aff6}}
\and L.~Moscardini\orcid{0000-0002-3473-6716}\inst{\ref{aff102},\ref{aff6},\ref{aff57}}
\and E.~Munari\orcid{0000-0002-1751-5946}\inst{\ref{aff4},\ref{aff53}}
\and R.~Nakajima\orcid{0009-0009-1213-7040}\inst{\ref{aff100}}
\and C.~Neissner\orcid{0000-0001-8524-4968}\inst{\ref{aff109},\ref{aff68}}
\and R.~C.~Nichol\orcid{0000-0003-0939-6518}\inst{\ref{aff50}}
\and S.-M.~Niemi\orcid{0009-0005-0247-0086}\inst{\ref{aff110}}
\and C.~Padilla\orcid{0000-0001-7951-0166}\inst{\ref{aff109}}
\and S.~Paltani\orcid{0000-0002-8108-9179}\inst{\ref{aff46}}
\and F.~Pasian\orcid{0000-0002-4869-3227}\inst{\ref{aff4}}
\and K.~Pedersen\inst{\ref{aff111}}
\and W.~J.~Percival\orcid{0000-0002-0644-5727}\inst{\ref{aff112},\ref{aff113},\ref{aff114}}
\and V.~Pettorino\orcid{0000-0002-4203-9320}\inst{\ref{aff110}}
\and S.~Pires\orcid{0000-0002-0249-2104}\inst{\ref{aff52}}
\and G.~Polenta\orcid{0000-0003-4067-9196}\inst{\ref{aff115}}
\and M.~Poncet\inst{\ref{aff116}}
\and L.~A.~Popa\inst{\ref{aff117}}
\and L.~Pozzetti\orcid{0000-0001-7085-0412}\inst{\ref{aff6}}
\and G.~D.~Racca\orcid{0000-0002-9883-8981}\inst{\ref{aff110},\ref{aff1}}
\and F.~Raison\orcid{0000-0002-7819-6918}\inst{\ref{aff85}}
\and R.~Rebolo\orcid{0000-0003-3767-7085}\inst{\ref{aff31},\ref{aff118},\ref{aff32}}
\and A.~Renzi\orcid{0000-0001-9856-1970}\inst{\ref{aff119},\ref{aff80}}
\and J.~Rhodes\orcid{0000-0002-4485-8549}\inst{\ref{aff11}}
\and G.~Riccio\inst{\ref{aff62}}
\and H.-W.~Rix\orcid{0000-0003-4996-9069}\inst{\ref{aff5}}
\and E.~Romelli\orcid{0000-0003-3069-9222}\inst{\ref{aff4}}
\and M.~Roncarelli\orcid{0000-0001-9587-7822}\inst{\ref{aff6}}
\and C.~Rosset\orcid{0000-0003-0286-2192}\inst{\ref{aff104}}
\and B.~Rusholme\orcid{0000-0001-7648-4142}\inst{\ref{aff120}}
\and R.~Saglia\orcid{0000-0003-0378-7032}\inst{\ref{aff86},\ref{aff85}}
\and Z.~Sakr\orcid{0000-0002-4823-3757}\inst{\ref{aff29},\ref{aff121},\ref{aff122}}
\and D.~Sapone\orcid{0000-0001-7089-4503}\inst{\ref{aff123}}
\and M.~Sauvage\orcid{0000-0002-0809-2574}\inst{\ref{aff52}}
\and M.~Schirmer\orcid{0000-0003-2568-9994}\inst{\ref{aff5}}
\and P.~Schneider\orcid{0000-0001-8561-2679}\inst{\ref{aff100}}
\and T.~Schrabback\orcid{0000-0002-6987-7834}\inst{\ref{aff124}}
\and A.~Secroun\orcid{0000-0003-0505-3710}\inst{\ref{aff81}}
\and G.~Seidel\orcid{0000-0003-2907-353X}\inst{\ref{aff5}}
\and S.~Serrano\orcid{0000-0002-0211-2861}\inst{\ref{aff125},\ref{aff126},\ref{aff127}}
\and E.~Sihvola\orcid{0000-0003-1804-7715}\inst{\ref{aff128}}
\and P.~Simon\inst{\ref{aff100}}
\and C.~Sirignano\orcid{0000-0002-0995-7146}\inst{\ref{aff119},\ref{aff80}}
\and G.~Sirri\orcid{0000-0003-2626-2853}\inst{\ref{aff57}}
\and L.~Stanco\orcid{0000-0002-9706-5104}\inst{\ref{aff80}}
\and J.~Steinwagner\orcid{0000-0001-7443-1047}\inst{\ref{aff85}}
\and P.~Tallada-Cresp\'{i}\orcid{0000-0002-1336-8328}\inst{\ref{aff67},\ref{aff68}}
\and I.~Tereno\orcid{0000-0002-4537-6218}\inst{\ref{aff129},\ref{aff130}}
\and N.~Tessore\orcid{0000-0002-9696-7931}\inst{\ref{aff78}}
\and S.~Toft\orcid{0000-0003-3631-7176}\inst{\ref{aff131},\ref{aff132}}
\and R.~Toledo-Moreo\orcid{0000-0002-2997-4859}\inst{\ref{aff133}}
\and F.~Torradeflot\orcid{0000-0003-1160-1517}\inst{\ref{aff68},\ref{aff67}}
\and I.~Tutusaus\orcid{0000-0002-3199-0399}\inst{\ref{aff127},\ref{aff125},\ref{aff121}}
\and L.~Valenziano\orcid{0000-0002-1170-0104}\inst{\ref{aff6},\ref{aff82}}
\and J.~Valiviita\orcid{0000-0001-6225-3693}\inst{\ref{aff95},\ref{aff96}}
\and T.~Vassallo\orcid{0000-0001-6512-6358}\inst{\ref{aff4}}
\and Y.~Wang\orcid{0000-0002-4749-2984}\inst{\ref{aff120}}
\and J.~Weller\orcid{0000-0002-8282-2010}\inst{\ref{aff86},\ref{aff85}}
\and F.~M.~Zerbi\inst{\ref{aff51}}
\and E.~Zucca\orcid{0000-0002-5845-8132}\inst{\ref{aff6}}
\and G.~Fabbian\orcid{0000-0002-3255-4695}\inst{\ref{aff79}}
\and M.~Huertas-Company\orcid{0000-0002-1416-8483}\inst{\ref{aff31},\ref{aff134},\ref{aff135}}
\and J.~Mart\'{i}n-Fleitas\orcid{0000-0002-8594-569X}\inst{\ref{aff136}}
\and P.~Monaco\orcid{0000-0003-2083-7564}\inst{\ref{aff137},\ref{aff4},\ref{aff54},\ref{aff53}}
\and V.~Scottez\orcid{0009-0008-3864-940X}\inst{\ref{aff106},\ref{aff138}}
\and M.~Viel\orcid{0000-0002-2642-5707}\inst{\ref{aff53},\ref{aff4},\ref{aff55},\ref{aff54},\ref{aff139}}}
										   
%%%% please do not edit the affiliation list -- contact ECEB Bureau for changes
\institute{Leiden Observatory, Leiden University, Einsteinweg 55, 2333 CC Leiden, The Netherlands\label{aff1}
\and
Department of Physics, University of California, Santa Barbara, CA 93106, USA\label{aff2}
\and
Hamburger Sternwarte, University of Hamburg, Gojenbergsweg 112, 21029 Hamburg, Germany\label{aff3}
\and
INAF-Osservatorio Astronomico di Trieste, Via G. B. Tiepolo 11, 34143 Trieste, Italy\label{aff4}
\and
Max-Planck-Institut f\"ur Astronomie, K\"onigstuhl 17, 69117 Heidelberg, Germany\label{aff5}
\and
INAF-Osservatorio di Astrofisica e Scienza dello Spazio di Bologna, Via Piero Gobetti 93/3, 40129 Bologna, Italy\label{aff6}
\and
Astrophysics Group, Blackett Laboratory, Imperial College London, London SW7 2AZ, UK\label{aff7}
\and
Department of Mathematics, Imperial College London, London SW7 2AZ, UK\label{aff8}
\and
Department of Astronomy, University of Michigan, 1085 S. University Ave., Ann Arbor, MI 48109, USA\label{aff9}
\and
Steward Observatory, University of Arizona, 933 N. Cherry Ave, Tucson, AZ 85750, USA\label{aff10}
\and
Jet Propulsion Laboratory, California Institute of Technology, 4800 Oak Grove Drive, Pasadena, CA, 91109, USA\label{aff11}
\and
Herzberg Astronomy and Astrophysics Research Centre, 5071 W. Saanich Rd. Victoria, BC, V9E 2E7, Canada\label{aff12}
\and
Department of Physics \& Astronomy, University of California Irvine, Irvine CA 92697, USA\label{aff13}
\and
Department of Physics, Oxford University, Keble Road, Oxford OX1 3RH, UK\label{aff14}
\and
Department of Physics, Massachusetts Institute of Technology, Cambridge, MA 02139, USA\label{aff15}
\and
MIT Kavli Institute for Astrophysics and Space Research, Massachusetts Institute of Technology, Cambridge, MA 02139, USA\label{aff16}
\and
Aix-Marseille Universit\'e, CNRS, CNES, LAM, Marseille, France\label{aff17}
\and
Kapteyn Astronomical Institute, University of Groningen, PO Box 800, 9700 AV Groningen, The Netherlands\label{aff18}
\and
Department of Astronomy/Steward Observatory, University of Arizona, 933 N. Cherry Avenue, Tucson, AZ 85721, USA\label{aff19}
\and
Instituto de Astrof\'isica, Facultad de F\'isica, Pontificia Universidad Cat\'olica de Chile Av. Vicu\~na Mackenna 4860, 7820436 Macul, Santiago, Chile\label{aff20}
\and
Fakult\"at f\"ur Physik und Astronomie, Universit\"at Heidelberg Im Neuenheimer Feld 226, 69115 Heidelberg, Germany\label{aff21}
\and
Research Center for Space and Cosmic Evolution, Ehime University, 2-5 Bunkyo-cho, Matsuyama, Ehime 790-8577, Japan\label{aff22}
\and
Waseda Institute for Advanced Study (WIAS), Waseda University, 1-21-1, Nishi-Waseda, Shinjuku, Tokyo 169-0051, Japan\label{aff23}
\and
Kavli Institute for the Physics and Mathematics of the Universe (WPI), University of Tokyo, Kashiwa, Chiba 277-8583, Japan\label{aff24}
\and
Department of Astronomy and Astrophysics, 525 Davey Lab, The Pennsylvania State University, University Park, PA 16802, USA\label{aff25}
\and
INAF-IASF Milano, Via Alfonso Corti 12, 20133 Milano, Italy\label{aff26}
\and
Department of Astronomy, Tsinghua University, Beijing 100084, China\label{aff27}
\and
Institut d'Astrophysique de Paris, UMR 7095, CNRS, and Sorbonne Universit\'e, 98 bis boulevard Arago, 75014 Paris, France\label{aff28}
\and
Institut f\"ur Theoretische Physik, University of Heidelberg, Philosophenweg 16, 69120 Heidelberg, Germany\label{aff29}
\and
Jodrell Bank Centre for Astrophysics, Department of Physics and Astronomy, University of Manchester, Oxford Road, Manchester M13 9PL, UK\label{aff30}
\and
Instituto de Astrof\'{\i}sica de Canarias, E-38205 La Laguna, Tenerife, Spain\label{aff31}
\and
Universidad de La Laguna, Dpto. Astrof\'\i sica, E-38206 La Laguna, Tenerife, Spain\label{aff32}
\and
Institute for Cosmic Ray Research, The University of Tokyo, 5-1-5 Kashiwanoha, Kashiwa, Chiba 277-8582, Japan\label{aff33}
\and
Institute of Theoretical Astrophysics, University of Oslo, P.O. Box 1029 Blindern, 0315 Oslo, Norway\label{aff34}
\and
Kavli Institute for Cosmology Cambridge, Madingley Road, Cambridge, CB3 0HA, UK\label{aff35}
\and
Institute of Astronomy, University of Cambridge, Madingley Road, Cambridge CB3 0HA, UK\label{aff36}
\and
INAF-Istituto di Astrofisica e Planetologia Spaziali, via del Fosso del Cavaliere, 100, 00100 Roma, Italy\label{aff37}
\and
Institute of Science and Technology Austria (ISTA), Am Campus 1, 3400 Klosterneuburg, Austria\label{aff38}
\and
Department of Mathematics and Physics, Roma Tre University, Via della Vasca Navale 84, 00146 Rome, Italy\label{aff39}
\and
INAF-Osservatorio Astronomico di Roma, Via Frascati 33, 00078 Monteporzio Catone, Italy\label{aff40}
\and
Dipartimento di Fisica e Astronomia, Universit\`{a} di Firenze, via G. Sansone 1, 50019 Sesto Fiorentino, Firenze, Italy\label{aff41}
\and
University of Trento, Via Sommarive 14, I-38123 Trento, Italy\label{aff42}
\and
INAF-Osservatorio Astrofisico di Arcetri, Largo E. Fermi 5, 50125, Firenze, Italy\label{aff43}
\and
Department of Physics and Astronomy, University of British Columbia, Vancouver, BC V6T 1Z1, Canada\label{aff44}
\and
Institute for Particle Physics and Astrophysics, Dept. of Physics, ETH Zurich, Wolfgang-Pauli-Strasse 27, 8093 Zurich, Switzerland\label{aff45}
\and
Department of Astronomy, University of Geneva, ch. d'Ecogia 16, 1290 Versoix, Switzerland\label{aff46}
\and
Department of Physical Sciences, Ritsumeikan University, Kusatsu, Shiga 525-8577, Japan\label{aff47}
\and
Academia Sinica Institute of Astronomy and Astrophysics (ASIAA), 11F of ASMAB, No.~1, Section 4, Roosevelt Road, Taipei 10617, Taiwan\label{aff48}
\and
ESAC/ESA, Camino Bajo del Castillo, s/n., Urb. Villafranca del Castillo, 28692 Villanueva de la Ca\~nada, Madrid, Spain\label{aff49}
\and
School of Mathematics and Physics, University of Surrey, Guildford, Surrey, GU2 7XH, UK\label{aff50}
\and
INAF-Osservatorio Astronomico di Brera, Via Brera 28, 20122 Milano, Italy\label{aff51}
\and
Universit\'e Paris-Saclay, Universit\'e Paris Cit\'e, CEA, CNRS, AIM, 91191, Gif-sur-Yvette, France\label{aff52}
\and
IFPU, Institute for Fundamental Physics of the Universe, via Beirut 2, 34151 Trieste, Italy\label{aff53}
\and
INFN, Sezione di Trieste, Via Valerio 2, 34127 Trieste TS, Italy\label{aff54}
\and
SISSA, International School for Advanced Studies, Via Bonomea 265, 34136 Trieste TS, Italy\label{aff55}
\and
Dipartimento di Fisica e Astronomia, Universit\`a di Bologna, Via Gobetti 93/2, 40129 Bologna, Italy\label{aff56}
\and
INFN-Sezione di Bologna, Viale Berti Pichat 6/2, 40127 Bologna, Italy\label{aff57}
\and
INAF-Osservatorio Astronomico di Padova, Via dell'Osservatorio 5, 35122 Padova, Italy\label{aff58}
\and
Dipartimento di Fisica, Universit\`a di Genova, Via Dodecaneso 33, 16146, Genova, Italy\label{aff59}
\and
INFN-Sezione di Genova, Via Dodecaneso 33, 16146, Genova, Italy\label{aff60}
\and
Department of Physics "E. Pancini", University Federico II, Via Cinthia 6, 80126, Napoli, Italy\label{aff61}
\and
INAF-Osservatorio Astronomico di Capodimonte, Via Moiariello 16, 80131 Napoli, Italy\label{aff62}
\and
Dipartimento di Fisica, Universit\`a degli Studi di Torino, Via P. Giuria 1, 10125 Torino, Italy\label{aff63}
\and
INFN-Sezione di Torino, Via P. Giuria 1, 10125 Torino, Italy\label{aff64}
\and
INAF-Osservatorio Astrofisico di Torino, Via Osservatorio 20, 10025 Pino Torinese (TO), Italy\label{aff65}
\and
Institute for Astronomy, University of Edinburgh, Royal Observatory, Blackford Hill, Edinburgh EH9 3HJ, UK\label{aff66}
\and
Centro de Investigaciones Energ\'eticas, Medioambientales y Tecnol\'ogicas (CIEMAT), Avenida Complutense 40, 28040 Madrid, Spain\label{aff67}
\and
Port d'Informaci\'{o} Cient\'{i}fica, Campus UAB, C. Albareda s/n, 08193 Bellaterra (Barcelona), Spain\label{aff68}
\and
INFN section of Naples, Via Cinthia 6, 80126, Napoli, Italy\label{aff69}
\and
Institute for Astronomy, University of Hawaii, 2680 Woodlawn Drive, Honolulu, HI 96822, USA\label{aff70}
\and
Dipartimento di Fisica e Astronomia "Augusto Righi" - Alma Mater Studiorum Universit\`a di Bologna, Viale Berti Pichat 6/2, 40127 Bologna, Italy\label{aff71}
\and
European Space Agency/ESRIN, Largo Galileo Galilei 1, 00044 Frascati, Roma, Italy\label{aff72}
\and
Universit\'e Claude Bernard Lyon 1, CNRS/IN2P3, IP2I Lyon, UMR 5822, Villeurbanne, F-69100, France\label{aff73}
\and
Institut de Ci\`{e}ncies del Cosmos (ICCUB), Universitat de Barcelona (IEEC-UB), Mart\'{i} i Franqu\`{e}s 1, 08028 Barcelona, Spain\label{aff74}
\and
Instituci\'o Catalana de Recerca i Estudis Avan\c{c}ats (ICREA), Passeig de Llu\'{\i}s Companys 23, 08010 Barcelona, Spain\label{aff75}
\and
Institut de Ciencies de l'Espai (IEEC-CSIC), Campus UAB, Carrer de Can Magrans, s/n Cerdanyola del Vall\'es, 08193 Barcelona, Spain\label{aff76}
\and
UCB Lyon 1, CNRS/IN2P3, IUF, IP2I Lyon, 4 rue Enrico Fermi, 69622 Villeurbanne, France\label{aff77}
\and
Mullard Space Science Laboratory, University College London, Holmbury St Mary, Dorking, Surrey RH5 6NT, UK\label{aff78}
\and
Universit\'e Paris-Saclay, CNRS, Institut d'astrophysique spatiale, 91405, Orsay, France\label{aff79}
\and
INFN-Padova, Via Marzolo 8, 35131 Padova, Italy\label{aff80}
\and
Aix-Marseille Universit\'e, CNRS/IN2P3, CPPM, Marseille, France\label{aff81}
\and
INFN-Bologna, Via Irnerio 46, 40126 Bologna, Italy\label{aff82}
\and
University Observatory, LMU Faculty of Physics, Scheinerstr.~1, 81679 Munich, Germany\label{aff83}
\and
FRACTAL S.L.N.E., calle Tulip\'an 2, Portal 13 1A, 28231, Las Rozas de Madrid, Spain\label{aff84}
\and
Max Planck Institute for Extraterrestrial Physics, Giessenbachstr. 1, 85748 Garching, Germany\label{aff85}
\and
Universit\"ats-Sternwarte M\"unchen, Fakult\"at f\"ur Physik, Ludwig-Maximilians-Universit\"at M\"unchen, Scheinerstr.~1, 81679 M\"unchen, Germany\label{aff86}
\and
Dipartimento di Fisica "Aldo Pontremoli", Universit\`a degli Studi di Milano, Via Celoria 16, 20133 Milano, Italy\label{aff87}
\and
INFN-Sezione di Milano, Via Celoria 16, 20133 Milano, Italy\label{aff88}
\and
Department of Physics, Lancaster University, Lancaster, LA1 4YB, UK\label{aff89}
\and
Felix Hormuth Engineering, Goethestr. 17, 69181 Leimen, Germany\label{aff90}
\and
Technical University of Denmark, Elektrovej 327, 2800 Kgs. Lyngby, Denmark\label{aff91}
\and
Cosmic Dawn Center (DAWN), Denmark\label{aff92}
\and
NASA Goddard Space Flight Center, Greenbelt, MD 20771, USA\label{aff93}
\and
Universit\'e de Gen\`eve, D\'epartement de Physique Th\'eorique and Centre for Astroparticle Physics, 24 quai Ernest-Ansermet, CH-1211 Gen\`eve 4, Switzerland\label{aff94}
\and
Department of Physics, P.O. Box 64, University of Helsinki, 00014 Helsinki, Finland\label{aff95}
\and
Helsinki Institute of Physics, Gustaf H{\"a}llstr{\"o}min katu 2, University of Helsinki, 00014 Helsinki, Finland\label{aff96}
\and
Laboratoire d'etude de l'Univers et des phenomenes eXtremes, Observatoire de Paris, Universit\'e PSL, Sorbonne Universit\'e, CNRS, 92190 Meudon, France\label{aff97}
\and
SKAO, Jodrell Bank, Lower Withington, Macclesfield SK11 9FT, UK\label{aff98}
\and
Centre de Calcul de l'IN2P3/CNRS, 21 avenue Pierre de Coubertin 69627 Villeurbanne Cedex, France\label{aff99}
\and
Universit\"at Bonn, Argelander-Institut f\"ur Astronomie, Auf dem H\"ugel 71, 53121 Bonn, Germany\label{aff100}
\and
INFN-Sezione di Roma, Piazzale Aldo Moro, 2 - c/o Dipartimento di Fisica, Edificio G. Marconi, 00185 Roma, Italy\label{aff101}
\and
Dipartimento di Fisica e Astronomia "Augusto Righi" - Alma Mater Studiorum Universit\`a di Bologna, via Piero Gobetti 93/2, 40129 Bologna, Italy\label{aff102}
\and
Department of Physics, Institute for Computational Cosmology, Durham University, South Road, Durham, DH1 3LE, UK\label{aff103}
\and
Universit\'e Paris Cit\'e, CNRS, Astroparticule et Cosmologie, 75013 Paris, France\label{aff104}
\and
CNRS-UCB International Research Laboratory, Centre Pierre Bin\'etruy, IRL2007, CPB-IN2P3, Berkeley, USA\label{aff105}
\and
Institut d'Astrophysique de Paris, 98bis Boulevard Arago, 75014, Paris, France\label{aff106}
\and
Institute of Physics, Laboratory of Astrophysics, Ecole Polytechnique F\'ed\'erale de Lausanne (EPFL), Observatoire de Sauverny, 1290 Versoix, Switzerland\label{aff107}
\and
Telespazio UK S.L. for European Space Agency (ESA), Camino bajo del Castillo, s/n, Urbanizacion Villafranca del Castillo, Villanueva de la Ca\~nada, 28692 Madrid, Spain\label{aff108}
\and
Institut de F\'{i}sica d'Altes Energies (IFAE), The Barcelona Institute of Science and Technology, Campus UAB, 08193 Bellaterra (Barcelona), Spain\label{aff109}
\and
European Space Agency/ESTEC, Keplerlaan 1, 2201 AZ Noordwijk, The Netherlands\label{aff110}
\and
DARK, Niels Bohr Institute, University of Copenhagen, Jagtvej 155, 2200 Copenhagen, Denmark\label{aff111}
\and
Waterloo Centre for Astrophysics, University of Waterloo, Waterloo, Ontario N2L 3G1, Canada\label{aff112}
\and
Department of Physics and Astronomy, University of Waterloo, Waterloo, Ontario N2L 3G1, Canada\label{aff113}
\and
Perimeter Institute for Theoretical Physics, Waterloo, Ontario N2L 2Y5, Canada\label{aff114}
\and
Space Science Data Center, Italian Space Agency, via del Politecnico snc, 00133 Roma, Italy\label{aff115}
\and
Centre National d'Etudes Spatiales -- Centre spatial de Toulouse, 18 avenue Edouard Belin, 31401 Toulouse Cedex 9, France\label{aff116}
\and
Institute of Space Science, Str. Atomistilor, nr. 409 M\u{a}gurele, Ilfov, 077125, Romania\label{aff117}
\and
Consejo Superior de Investigaciones Cientificas, Calle Serrano 117, 28006 Madrid, Spain\label{aff118}
\and
Dipartimento di Fisica e Astronomia "G. Galilei", Universit\`a di Padova, Via Marzolo 8, 35131 Padova, Italy\label{aff119}
\and
Caltech/IPAC, 1200 E. California Blvd., Pasadena, CA 91125, USA\label{aff120}
\and
Institut de Recherche en Astrophysique et Plan\'etologie (IRAP), Universit\'e de Toulouse, CNRS, UPS, CNES, 14 Av. Edouard Belin, 31400 Toulouse, France\label{aff121}
\and
Universit\'e St Joseph; Faculty of Sciences, Beirut, Lebanon\label{aff122}
\and
Departamento de F\'isica, FCFM, Universidad de Chile, Blanco Encalada 2008, Santiago, Chile\label{aff123}
\and
Universit\"at Innsbruck, Institut f\"ur Astro- und Teilchenphysik, Technikerstr. 25/8, 6020 Innsbruck, Austria\label{aff124}
\and
Institut d'Estudis Espacials de Catalunya (IEEC),  Edifici RDIT, Campus UPC, 08860 Castelldefels, Barcelona, Spain\label{aff125}
\and
Satlantis, University Science Park, Sede Bld 48940, Leioa-Bilbao, Spain\label{aff126}
\and
Institute of Space Sciences (ICE, CSIC), Campus UAB, Carrer de Can Magrans, s/n, 08193 Barcelona, Spain\label{aff127}
\and
Department of Physics and Helsinki Institute of Physics, Gustaf H\"allstr\"omin katu 2, University of Helsinki, 00014 Helsinki, Finland\label{aff128}
\and
Departamento de F\'isica, Faculdade de Ci\^encias, Universidade de Lisboa, Edif\'icio C8, Campo Grande, PT1749-016 Lisboa, Portugal\label{aff129}
\and
Instituto de Astrof\'isica e Ci\^encias do Espa\c{c}o, Faculdade de Ci\^encias, Universidade de Lisboa, Tapada da Ajuda, 1349-018 Lisboa, Portugal\label{aff130}
\and
Cosmic Dawn Center (DAWN)\label{aff131}
\and
Niels Bohr Institute, University of Copenhagen, Jagtvej 128, 2200 Copenhagen, Denmark\label{aff132}
\and
Universidad Polit\'ecnica de Cartagena, Departamento de Electr\'onica y Tecnolog\'ia de Computadoras,  Plaza del Hospital 1, 30202 Cartagena, Spain\label{aff133}
\and
Universit\'e PSL, Observatoire de Paris, Sorbonne Universit\'e, CNRS, LERMA, 75014, Paris, France\label{aff134}
\and
Universit\'e Paris-Cit\'e, 5 Rue Thomas Mann, 75013, Paris, France\label{aff135}
\and
Aurora Technology for European Space Agency (ESA), Camino bajo del Castillo, s/n, Urbanizacion Villafranca del Castillo, Villanueva de la Ca\~nada, 28692 Madrid, Spain\label{aff136}
\and
Dipartimento di Fisica - Sezione di Astronomia, Universit\`a di Trieste, Via Tiepolo 11, 34131 Trieste, Italy\label{aff137}
\and
ICL, Junia, Universit\'e Catholique de Lille, LITL, 59000 Lille, France\label{aff138}
\and
ICSC - Centro Nazionale di Ricerca in High Performance Computing, Big Data e Quantum Computing, Via Magnanelli 2, Bologna, Italy\label{aff139}}    

\abstract{
We report the discovery of 31 new high-$z$ quasars in the redshift range $6.6 < z < 7.8$. These quasars were selected from approximately \SI{3000}{\deg\squared} of sky covered during the first 1.5 years of the Euclid Wide Survey, representing the initial results of the \Euclid high-$z$ quasar search. Our candidate selection employed multiple machine-learning and probabilistic techniques applied to the \Euclid \IE, \YE, \JE, and \HE images, supplemented by ancillary $z$-band data when available. Spectroscopic follow-up observations were carried out with Keck, Magellan, and the Large Binocular Telescope (LBT). Among the new discoveries, there are 12 quasars at $z \geq 7$, more than doubling the number of previously known quasars at $z \geq 7$. 
The newly discovered quasars exhibit $21.2 < \JE < 23.2$ ($-25.5 < M_{\rm 1450} < -23.6$), extending quasar studies to the faint end of the quasar luminosity function (QLF) at $z\gtrsim7$. The quasar with the highest-$z$, EUCL\,J172902.75$+$641018.1 at $z \approx 7.77$, sets the new redshift record for the most distant quasar ever reported. 
These discoveries demonstrate \Euclid's transformative role in high-$z$ quasar discovery and set the stage for future follow-up studies of the early galaxies hosting quasars, supermassive black hole growth, and the intergalactic medium in the epoch of reionisation.
}

%
% Provide up to five key words:
%

%    from the list in
%     https://www.aanda.org/for-authors/latex-issues/information-files#pop}
%
% Add short versions of title and author list for page headings
%
   \titlerunning{Discovery of 31 new quasars at $6.6 < z < 7.8$}
   \authorrunning{D.~Yang et al.}
   
   \maketitle
   \nolinenumbers

\section{\label{sec:Intro}Introduction}

Since the first discovery of quasars at $z>5.7$ \citep[corresponding to $<1~\mathrm{Gyr}$ of cosmic time assuming a $\Lambda$CDM model;][]{Fan2000}, subsequent searches over the past two decades have pushed the redshift frontier to $z\sim7.5$, and have provided key insights into galaxy formation and cosmology \citep[for a review see][]{Fan2023}. Specifically, the highest-redshift quasars \citep[$z \gtrsim 7.5$;][]{Banados2018,Yang2020,Wang2021} place stringent constraints on the formation and growth of supermassive black holes (SMBHs): only the most massive black hole seeds appear to be sufficient to produce these quasars within the limited cosmic time available under the standard Eddington-limited accretion scenario \citep[e.g.,][]{Inayoshi2020,Yang2021,Farina2022,Bosman2025}. Beyond this, submillimetre observations of some other high-$z$ quasars indicate that their host galaxies might be among the most massive \citep[$>10^{10}\,M_\odot$; e.g.,][]{Neeleman2021} and actively star-forming \citep[star-formation rate above 100\,$M_\odot\,\mathrm{yr}^{-1}$; e.g.,][]{Wang2024} objects known in the early Universe. At the same time, the SMBHs powering these luminous quasars appear to be over-massive relative to the local SMBH-galaxy mass relation \citep[e.g.,][]{Kormendy2013}, suggesting an early SMBH-galaxy co-evolution scenario, in which black hole growth outpaces that of the host galaxy (e.g. \citealt{Neeleman2021,Farina2022,Yue2024,Wang2024}; but also see \citealt{Lauer2007,Izumi2019,Li2022,Silverman2025}). Finally, as the most luminous non-transient sources in the Universe, high-$z$ quasars serve as powerful beacons for studying the intergalactic medium (IGM), the reionisation \citep[e.g.,][]{Davies2018,Yang2020IGM,Greig2024,Kist2025,Hennawi2025} and large-scale structure (LSS) during the epoch of reionisation \citep[EoR; e.g.,][]{Wang2023,Pizzati2024,Eilers2024}.

However, a major challenge to fully understanding high-$z$ quasars and leveraging them as probes of the early Universe is our limited knowledge of the quasar population at $z > 7$. Prior to this work, only nine quasars were found at $z>7$ since the first discovery in 2011 \citep{Mortlock2011,Banados2018,Wang2018z7,Matsuoka2019b,Matsuoka2019,Yang2020,Wang2021,Banados2025,Matsuoka2025}, while hundreds of quasars had been identified at $6 < z < 7$. 
The lack of large samples at $z>7$ hinders us from tracing the evolutionary pathways of the quasars towards their formation and from constraining the IGM and LSS at earlier epochs. While the time interval between $z= 7$ and $z = 6$ spans only of the order of 200\,Myr, key astrophysical processes can evolve significantly over such timescales. SMBHs, for instance, can grow rapidly under Eddington-limited accretion, which corresponds to an $e$-folding (Salpeter) timescale of $\sim 45\,\mathrm{Myr}$ assuming a fiducial radiative efficiency of 10\%, allowing their masses to increase by a factor of $\sim 100$ within this period. Concurrently, the IGM is expected to undergo drastic evolution, with the neutral hydrogen fraction changing from nearly unity at $z \gtrsim 7$ to $\lesssim1\%$ at $z \sim 5.5$ \citep[e.g.,][]{Planck2020,Bosman2022,Jin2023,Davies2026}.

Despite the critical importance of extending the analyses of quasars to $z\gtrsim 7$, finding them remains exceptionally challenging. Quasars at this epoch are among the rarest objects in the Universe, with a typical surface density of approximately 1 per $100\,{\rm deg}^2$ down to $J=23$ \citep[e.g.,][]{Matsuoka2023}. Moreover, at $z\gtrsim 7$ the \lya transition is redshifted to $\gtrsim\SI{1}{\micro\meter}$, which happens to match both the band gap energy at which silicon-based CCDs become insensitive, and where atmospheric OH emission significantly elevates the sky background. Thus, assembling a large sample at $z \gtrsim 7$ requires deep large area near-infrared (NIR) imaging surveys---a task that remains prohibitively difficult from the ground.

In February 2024, the \Euclid telescope \citep[\SI{1.2}{m};][]{EuclidSkyOverview} commenced the six-year Euclid Wide Survey \citep[EWS;][]{Scaramella-EP1}, which aims at mapping \SI{14000}{\deg\squared} of the extragalactic sky. At the core of the mission are two instruments, the Visible Camera \citep[VIS;][]{EuclidSkyVIS} and the Near Infrared Spectrometer and Photometer \citep[NISP;][]{EuclidSkyNISP}, which observe the sky simultaneously in the optical and NIR using a dichroic beamsplitter. \Euclid achieves unprecedented depths for a wide-field survey in both optical and NIR, reaching $5\sigma$ limits of 24.5 in the NIR and 26.2 in the optical for point sources. For comparison, the UKIRT Infrared Deep Sky Survey \citep[UKIDSS;][]{Lawrence2007} and the VISTA Kilo-degree Infrared Galaxy (VIKING) Survey \citep[][]{Edge2013Viking} cover $\sim\SI{1000}{\deg\squared}$ and reach $\lesssim 22$~mag in the NIR; the VISTA Hemisphere Survey \citep[VHS;][]{McMahon2013} covers $\sim\SI{20000}{\deg\squared}$ and reaches $\sim 21$~mag in its $J$ band; and the UKIRT Hemisphere Survey \citep[UHS;]{Dye2018} covers $\sim\SI{10500}{\deg\squared}$ and reaches $\sim 20.5$~mag in its $J$ band. The unprecedented survey area and depth of \Euclid make it a game-changer for the field of high-$z$ quasar searches. 

\citet{Barnett-EP5} predicted the discovery of over 100 quasars with $7.0 < z < 7.5$, and $\sim25$ quasars beyond $z=7.5$, including an order of 8 beyond $z = 8.0$ from the full EWS. These predicted detections extend into the regime of faint quasars ($J\sim23$) at $z \gtrsim 7$, comparable to those identified in the deepest ground-based surveys; namely, the Subaru High-$z$ Exploration of Low-Luminosity Quasars \citep[SHELLQs; e.g.,][]{Matsuoka2016} and the Canada--France High-$z$ Quasar Survey \citep[CFHQS; e.g.,][]{Willott2010}, but over vastly larger sky coverage and at significantly higher redshifts.

This paper presents the initial results of the \Euclid high-$z$ quasar search using the on-the-fly data of the EWS (see Sect.~\ref{sec:pf}) between 14 February and 11 August 2025. The imaging data and photometric catalogues used for the candidate selection are introduced in Sect.~\ref{sec:euclid}, followed by a brief overview of the candidate selection methods in Sect.~\ref{sec:selection}. The spectroscopic follow-up observation of these candidates are described in Sect.~\ref{sec:spectroscpy}. The new quasar discoveries are presented and discussed in Sect.~\ref{sec:new_quasar}. When relevant, we adopt the following cosmological parameters: $\Omega_{\rm m}=0.3111$, $\Omega_\Lambda=0.6899$, $\Omega_{\rm b}=0.0489$, $h=0.6766$, and $\sigma_{\rm 8}=0.8102$ \citep{Planck2020}. All the magnitudes are presented in the AB system \citep{Oke1983}.

\section{\label{sec:euclid}EWS and photometry catalogues}

\subsection{\label{sec:ews}EWS}

As introduced in Sect.~\ref{sec:Intro}, the EWS, conducted from 2024 to 2030, is designed to map approximately \SI{14000}{\deg\squared} of the extragalactic sky in both visible and NIR bands using the two onboard instruments: VIS and NISP. The survey employs a step-and-stare observing strategy, in which the spacecraft collects data at fixed sky positions before slewing to subsequent pointings. Each pointing covers a field of view of \SI{0.53}{\deg\squared}, with VIS and NISP operating simultaneously. Observations at each field are performed using a four-point dither pattern. The EWS reaches $5\sigma$ point-source sensitivities of 24.5\,mag in the three NIR bands, \YE (949.6\,--1212.3\,nm), \JE (1167.6\,--1567.0\,nm), and \HE (1521.5\,--2021.4\,nm), and 26.2\,mag in the broad optical band \IE (550\,--900\,nm). NISP also provides grism slitless spectroscopy across the full EWS, with a resolving power of $R>480$ over the range 1206\,--1892\,nm \citep{Schirmer-EP18} for a point source. Although this work uses only \Euclid's photometry, the spectroscopy has strong potential for the direct identification of luminous high-$z$ quasars \citep[see][]{Banados25}. However, the grism data are relatively shallow and will detect only the brightest quasars.

\begin{figure}[htbp!]
\centering
\includegraphics[angle=0,width=1.0\hsize]{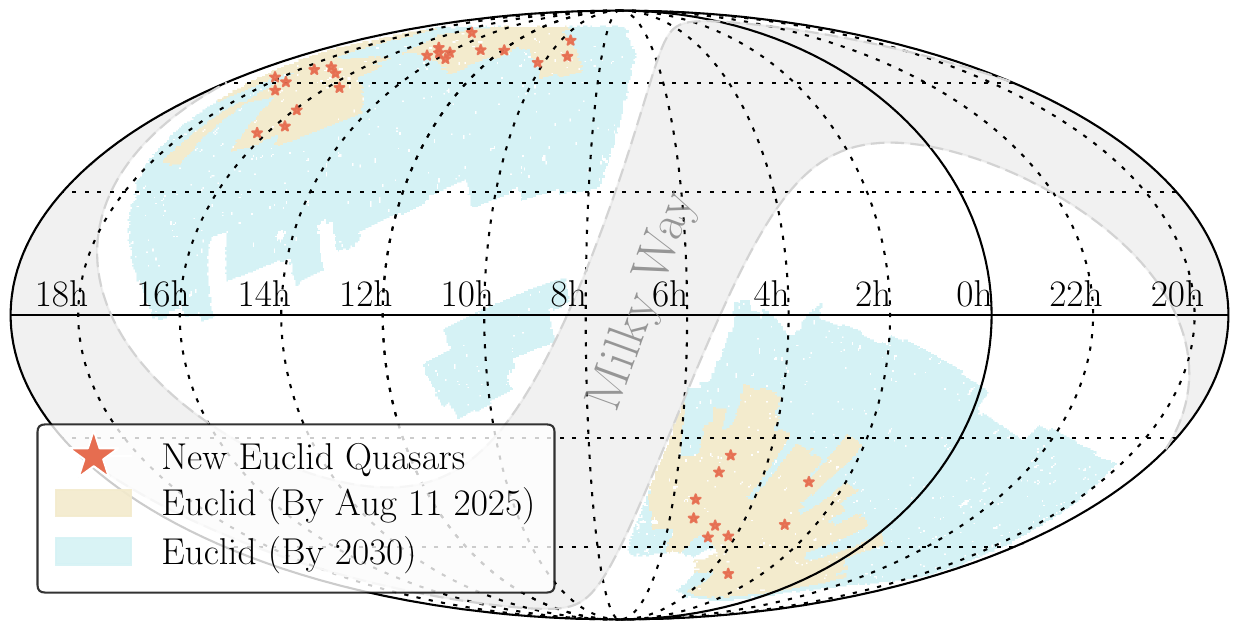}
\caption{Projection of the EWS area and the locations of the newly discovered quasars in J2000.0 equatorial coordinates. The regions observed by 11 August 2025, utilised for the quasar search presented in this work, are shown in beige. The full survey footprint expected by the end of the mission in 2030 is overlaid in cyan. The positions of newly discovered \Euclid high-$z$ quasars are marked in red.}
\label{fig:survey}
\end{figure}

The quasar search described in this work utilised EWS imaging data acquired between 14 February 2024 and 11 August 2025, as shown in Fig.~\ref{fig:survey}. The total area covered amounts to approximately \SI{3000}{\deg\squared}, with roughly \SI{1040}{\deg\squared} located in the northern hemisphere and \SI{1960}{\deg\squared} in the southern hemisphere. We note that the ground-based spectroscopic follow-up observations over these regions, particularly in the south, is ongoing and has not yet reached full completeness. A comprehensive statistical analysis of the selection function and the quasar population will be presented in forthcoming publications.

\subsection{\label{sec:ancillary}Ancillary imaging surveys}

While \Euclid imaging provides the primary basis for our quasar candidate selection, ancillary data from external surveys are also incorporated to enhance selection efficiency and robustness. Specifically, as the \IE band is a broad optical filter, deep $z$-band imaging can significantly improve the ability to detect the IGM absorption trough and to select robust high-$z$ quasar candidates at $z \gtrsim 7$. To this end, we performed forced photometry on $z$-band images obtained with the Hyper Suprime-Cam \citep[HSC;][]{Miyazaki2018HSC} as part of the Ultraviolet Near-Infrared Optical Northern Survey \citep[UNIONS;][]{Gwyn2025}, or from the Dark Energy Survey \citep[DES;][]{DES2016} when available. The 5$\sigma$ depth of these $z$-band images are 24.1 (\ang{;;2} diameter aperture) and 23.1 (\ang{;;1.95} diameter aperture), respectively. These images are processed through the \Euclid Processing Function \texttt{EXT} (see Sect.~\ref{sec:pf} for details).

Additionally, we make use of the LOFAR Two-metre Sky Survey \citep[LoTSS;][]{Shimwell2022} to prioritise candidates with associated radio emission \citep[e.g.,][]{Gloudemans2022}. We specifically utilise data from LoTSS Data Release 3, which covers $\approx89\%$ of the northern sky at a central frequency of 144\,MHz, with a median rms sensitivity of $\sim 100\,\mu$Jy\,beam$^{-1}$ and beam size of \ang{;;6}. 

\subsection{\label{sec:pf}Related Euclid processing functions and data products}

The Euclid Consortium (EC) has developed a suite of software Processing Functions (PFs) to efficiently manage and process the vast volume of data generated by \Euclid. This study utilises image and catalogue products produced at various processing levels by several of these PFs, namely \texttt{VIS}, \texttt{NIR}, \texttt{MER}, and \texttt{EXT}. In this subsection, we provide a brief overview of the relevant data products. 

\texttt{VIS}. The \texttt{VIS} PF generates calibrated optical images and source catalogues from raw \Euclid VIS exposures. It includes a calibration pipeline for instrumental corrections and a science pipeline for applying the corrections and conducting measurements. Multiple dithered exposures are co-added using \texttt{SWarp} \citep{Bertin2010swarp} to produce stacked images, from which PSF models are derived with \texttt{PSFEx} \citep{Bertin2013psfex} and source catalogues are extracted using \texttt{SExtractor} \citep{Bertin1996sextractor}. The most relevant \texttt{VIS} data products to this work are the stacked images and their associated catalogues. More details about \texttt{VIS} PF are given in \citet{Q1-TP002}.

\texttt{NIR}. The \texttt{NIR} PF processes raw NISP exposures to produce calibrated NIR images and catalogues. It includes a common pre-processing stage shared with the spectroscopic pipeline, calibration, and stacking components. Stacked images are generated from dithered exposures per field, accompanied by PSF models computed with \texttt{PSFEx}, and source catalogues extracted using \texttt{SExtractor}. For this work, we primarily use these stacked NIR images and catalogues in the \YE, \JE, and \HE bands. For more details about the \texttt{NIR} PF, we refer the readers to \citet{Q1-TP003}.

\texttt{MER}. The \texttt{MER} PF constructs multi-wavelength photometric catalogues by merging calibrated imaging data from the \texttt{VIS} and \texttt{NIR} PFs with external ground-based data (e.g. UNIONS). It produces background-subtracted mosaics per band and performs independent source detection on stacked VIS images and on a combined NIR detection image. The resulting catalogues include multi-band photometry (Kron, aperture, and template-fitting), morphological classifications, and quality flags. Photometric measurements are computed using PSF-matched images and convolution kernels. The input PSFs from \texttt{NIR}/\texttt{VIS} were propagated for each mosaiced image and evaluated at the detected source positions, together with the PSF convolution kernels. For the southern sky, instead of using \texttt{NIR}/\texttt{VIS} stacked frames, we used the \texttt{MER} mosaics to perform photometry measurements (see Sect.~\ref{sec:photometry}). For more details, we refer to \citet{Q1-TP004}.

\texttt{EXT}. The \texttt{EXT} PF ingests and calibrates external ground-based imaging data for integration with \Euclid data products. In principle, \texttt{EXT} images (e.g. $z$ band) are already incorporated into \texttt{MER}. However, in cases where the external data were available but \texttt{MER} had not yet been triggered, due to the incomplete coverage of certain external bands, we directly utilised the calibrated \texttt{EXT} single-epoch frames for forced-photometry measurements. 

After raw data from the spacecraft were downlinked, they were processed within the Euclid Data Processing System (DPS), where all data products are generated and tracked through successive PFs. Once validated, these data are pushed to the Science Archive System (SAS) in its on-the-fly (OTF) environment. The data in DPS and OTF are processed soon after acquisition to provide access as early as possible. As a result, different PF versions may be used, and the processing is therefore heterogeneous. The publicly available data are distributed through the SAS Public Data Release environment, where all products are reprocessed homogeneously using a fixed set of PFs. The data used in this work were taken directly from the DPS, and are thus not part of any data release. The motivation is to access the largest possible sky area before each follow-up confirmation campaign to search for extremely rare objects.

\section{\label{sec:selection}Photometric candidate selection}

\subsection{\label{sec:photometry}Custom photometry catalogues for candidate selection}

The candidate selection was initiated based on available \Euclid catalogues to create custom PSF photometry catalogues that fully exploit the sensitivity of \Euclid for point-like sources. With the PSF photometry, we assessed the probability of each candidate being a quasar with multiple algorithms described in Sect.~\ref{sec:three_algorithms}. Specifically, we first selected optically faint, point-like sources from the \texttt{VIS}+\texttt{NIR} catalogues and then performed PSF photometry on this pre-selected sample. Two parallel workflows were employed for the photometric measurements, depending on the availability of \texttt{MER} products at the time of data processing and spectroscopic follow-up. When \texttt{MER} products were not yet available, fluxes in the \Euclid bands were measured directly on the \texttt{VIS} and \texttt{NIR} stacked images. 
In this case, $z$-band fluxes were obtained by averaging the measurement from individual single-epoch frames delivered by the \texttt{EXT} PF (see Sect.~\ref{sec:pf}). When \texttt{MER} mosaics were available, all fluxes -- including those in the \Euclid and external bands -- were measured directly from the co-added \texttt{MER} images. A more detailed summary of the workflow is as follow:

\begin{itemize}
    \item[] \makebox[1.4cm][l]{{Step 1 ---}} Cross-matching sources across the three NISP bands using a \ang{;;0.3} matching radius to build a combined NIR source catalogue from the single-band \texttt{SExtractor} catalogues created by \texttt{NIR}.

    \item[] \makebox[1.4cm][l]{{Step 2 ---}} Applying signal-to-noise ratio (S/N) cuts in the \JE and \HE bands ($\mathrm{S/N} > 5$), and select isolated sources by requiring the angular distance to the nearest NIR-detected neighbour to exceed \ang{;;1.2}.

    \item[] \makebox[1.4cm][l]{{Step 3 ---}} Selecting point-like sources by requiring the \texttt{SExtractor} parameter $\texttt{CLASS\_STAR} > 0.6$ in both the \JE and \HE bands.

    \item[] \makebox[1.4cm][l]{{Step 4 ---}} Identifying all VIS observations overlapping with the given NISP observation.

    \item[] \makebox[1.4cm][l]{{Step 5 ---}} For each overlapping VIS pointing, cross-matching the cleaned NIR source catalogue with the corresponding VIS catalogue using a \ang{;;0.3} radius. Select dropout candidates by identifying sources with no VIS detection or with faint VIS flux relative to their \JE flux ($f_{\IE} / f_{\JE} < 0.06$, equivalent to $\IE - \JE > 3.05$).

    \item[] \makebox[1.4cm][l]{{Step 6 ---}} Discarding sources whose image in VIS is affected by pixel-masking or segmentation, to ensure reliable flux measurements.

    \item[] \makebox[1.4cm][l]{{Step 7 ---}} Measuring the PSF fluxes in each band, including the $z$ band when available, using \texttt{Photutils} \citep{Bradley2016photutils}. The local background is estimated from an annular region using a 3$\sigma$-clipped statistic. As noted previously, when \texttt{MER} products are available, fluxes are measured from the \texttt{MER} mosaics; otherwise, they are extracted from the \texttt{VIS} and \texttt{NIR} stacked images. All three PFs construct PSF models with \texttt{PSFEx} \citep{Bertin2013psfex} on a per-tile or per-pointing basis. For the \texttt{VIS} and \texttt{NIR} stacks, we use the average PSF stamp, while for the \texttt{MER} mosaics, we adopted the PSF stamp closest to each target.

    \item[] \makebox[1.4cm][l]{{Step 8 ---}} Defining the final dropout sample by applying the colour selection criteria on the PSF fluxes: $f_{\IE} / f_{\JE} < 0.06$ (equivalent to $\IE - \JE > 3.05$) and $|f_z / f_{\JE}| < 2$ (equivalent to $z - \JE > 0.75$).
\end{itemize}

These catalogues were subsequently passed to the candidate selection pipeline to produce the final targets list. 

\subsection{\label{sec:three_algorithms}Selection algorithms}

Our candidate selection method employed several different algorithms to enhance the efficiency and robustness. Candidates were independently selected from the photometric catalogues introduced in Sect.~\ref{sec:photometry} using these algorithms. Each method provides both a photometric redshift estimate ($z_{\rm phot}$) and a quantitative selection metric. Empirical thresholds on these metrics were applied to select high-confidence candidates from each method. Artefacts and spurious sources were subsequently removed through a visual inspection. Candidates were manually prioritised during spectroscopic follow-up, taking into account the outputs from all three methods. This process resulted in a final, merged list of high-priority targets. Detailed descriptions of the implementation of each algorithm and assessments of their performance will be presented in forthcoming publications. Here, we provide a brief overview of the methodologies.

The first method, Extreme Deconvolution High-Redshift Quasar selection (\textsc{XDHZQSO}) is based on the framework introduced and implemented in \citet{Nanni2022} and \cite{YangD2024}, but adopts conditional density estimation \citep{Kang2024} so that the densities of different populations vary smoothly with additional variables. Specifically, we consider only two classes, high-$z$ quasar and an all-inclusive contaminant class. We model their densities in flux-ratio space (relative to \JE) using the Extreme Deconvolution technique \citep{Bovy2011}, allowing the means and covariances of the Gaussian mixture to vary continuously depending on \JE\ magnitude and, for quasars, redshift. The quasar density is built from synthetic photometry generated with a new generative model (D.~Yang et al., in prep.), trained on a sample of 16\,198 quasars within the redshift range $0.45 < z < 3.0$, drawn from the SDSS-III BOSS and the SDSS-IV Extended BOSS (eBOSS) surveys \citep{Dawson2013,Dawson2016}. The contaminant density is estimated empirically from the flux-ratio distribution of the pre-selected sources described in Sect.~\ref{sec:photometry}, which are overwhelmingly non-quasar contaminants. This approximation is justified because contaminants outnumber true high-$z$ quasars by several orders of magnitude in this sample.

The second method employs XGBoost \citep{chen_xgboost_2016}, a widely adopted gradient-boosting algorithm for selecting quasars in large photometric catalogues \citep[e.g.,][]{calderone_boost_2024,Fu2025:2503.14141v2}. This second approach uses the same two-class formulation as the first method, considering only a high-$z$ quasar population and an all-inclusive contaminant population. Using fluxes and flux ratios as features, we construct a training set with positive and negative examples. We generated positives via an independent, machine learning-based model \citep[][]{Guarneri2025} trained on the same SDSS III and SDSS IV BOSS/eBOSS quasar set as above; negatives are drawn from the full pre-selected dataset. The classifier returns a per-source probability, $P_{\rm QSO}$, used to vet the candidate list by removing all sources with $P_{\rm QSO} < P_{\rm QSO, th} = 0.85$ and prioritise high-probability objects. A second XGBoost model is then trained via quantile regression and only on positive examples to estimate photometric redshifts and their associated uncertainties.

Our third method uses template-based SED fitting with \texttt{eazypy} \citep{EAZY} and \texttt{LePhare} \citep{LePhare}. Both codes are supplied with the same custom template library, which is based on the extragalactic template sets of \citet{Salvato2022} and \citet{Wolf2024}, and extended to include common stellar contaminants \citep{Martinez2026}. For every object we compute the statistic $R_{\chi^2} \equiv \chi^2_{\star}/{\chi^2_\mathrm{gal}}$,
where $\chi^2_{\star}$ and $\chi^2_{\mathrm{gal}}$ denote the best-fit $\chi^2$ values for the stellar and extragalactic template families, respectively. Objects with $R_{\chi^2}>3$ were retained and ordered by decreasing $R_{\chi^2}$. 

Finally, following \citet{Mortlock2012}, \cite{Barnett-EP5}, and \cite{Ezziati2025}, we implemented a Bayesian model comparison (BMC) procedure. We constructed colour models for high-$z$ quasars and the principal contaminants (MLT dwarfs and intermediate-redshift galaxies). For each source, we evaluated the class-conditional likelihoods that explicitly account for photometric uncertainties, assumed to be Gaussian-distributed. These likelihoods are combined with analytic surface-density priors that depend on redshift, apparent magnitude, and Galactic coordinates to yield the posterior quasar probability $P_q$. Candidates were then ranked by $P_q$ and we retained those with $P_q \ge 0.5$.

\section{\label{sec:spectroscpy}Spectroscopic follow-up observation}

To confirm the high-$z$ quasar candidates, spectroscopic observations with large-aperture telescopes were essential. We conducted the follow-up campaign using the Keck\,I and Keck\,II telescopes (\SI{10}{m}) at the W.~M.~Keck Observatory, the Magellan Baade telescope (\SI{6.5}{m}) at Las Campanas Observatory, and the LBT on Mount Graham. Spectroscopic follow-up was conducted over 20.5 nights across the 2024A, 2024B, 2025A, and 2025B semesters. Six different instruments were employed to confirm the new quasars reported in this work: the Low Resolution Imaging Spectrometer \citep[LRIS;][]{Oke1995LRIS} and the Multi-Object Spectrometer For Infra-Red Exploration \citep[MOSFIRE;][]{McLean2012MOSFIRE} on Keck\,I, the Keck Cosmic Web Imager \citep[KCWI;][]{Morrissey2018KCWI,McGurk2024KCRM} on Keck\,II, the Folded-port InfraRed Echellette \citep[FIRE;][]{Simcoe2010FIRE} on the Magellan Baade telescope, the Multi-object Double Spectrograph \citep[MODS;][]{Pogge2010} and the LBT Utility Camera in the Infrared \citep[LUCI;][]{Seifert2003LUCI} on LBT. A summary of the discovery observations obtained with these instruments is presented in Table~\ref{table:discovery}. Detailed descriptions of the instruments and the associated observations are provided in the following sub-sections.

\subsection{\label{sec:LRIS}Keck/LRIS}

LRIS is an optical double spectrograph on Keck\,I. For the purpose of high-$z$ quasar confirmation, we only used the red channel with a plate scale of \ang{;;0.123} per pixel. 
For this program, we used a \ang{;;1} slit, the gold-coated 600/10000 grating, and the D680 dichroic. The data were binned over 2$\times$2 pixels before readout. This configuration provided a wavelength coverage of 750\,--1060\,nm and a spectral resolution of full width at half maximum (FWHM) $\sim 150\,\kms$ ($R \sim 2000$). Each observing sequence consisted of four \SI{300}{s} exposures, and each target was typically observed through one to three such sequences. 

Observations were carried out in the 2024A semester (U259, PI: Hennawi), with a total of 2.5 nights allocated, split into five half-nights in June and July 2024. One half-night of observation was cancelled due to extreme weather conditions on the summit. The seeing was around \ang{;;0.8}--\ang{;;1} during the remaining time.

\subsection{\label{sec:KCWI}Keck/KCWI}

KCWI is an optical integral field spectrograph mounted on Keck\,II. For this program, we only used the red-channel, namely the Keck Cosmic Reionization Mapper \citep[KCRM;][]{McGurk2024KCRM}. Observations were conducted using the medium slicer configuration, which provides a field of view of 16\farcs5\,$\times$\,20\farcs4, composed of 24 slices each with a slice width of \ang{;;0.69}. The RM2 grating was employed in the red arm, covering 894\,--1070\,nm. A binning of $2 \times 2$ was applied. The spectral resolution of this setup is ${\rm FWHM} \approx \SI{107}{\kms}$ ($R\sim 2800$). Like our LRIS program, each observing sequence also included four \SI{300}{s} exposures, and each target was typically observed with one to three sequences. 

The KCWI observations were conducted across the 2024B (4 nights; U267, PI: Hennawi), 2025A (3 nights; U275, PI: Hennawi), and 2025B (1 night; U262, PI: Hennawi) semesters. Approximately two nights in 2024B were not suitable for high-$z$ quasar confirmation due to cloudy weather and/or poor seeing. In 2025A, 2.5 nights were lost to a shutter malfunction on Keck\,II, and the remaining half night was completed in June 2025. Observations in 2025B were successfully carried out in August 2025.

\subsection{\label{sec:MOSFIRE}Keck/MOSFIRE}

MOSFIRE is a NIR multi-object spectrograph on Keck\,I, covering the \textit{Y}, \textit{J}, \textit{H}, and \textit{K} bands. We primarily employed the \textit{Y} band setting with a \ang{;;1} slit, providing a wavelength coverage of 972\,--1125\,nm and a spectral resolution of ${\rm FWHM}\approx 126\ {\rm km\,s^{-1}}$ ($R\sim 2372$). Slits with widths of \ang{;;0.7}, \ang{;;1.2}, and \ang{;;1.5} were utilised depending on the seeing conditions. For candidates with photometric redshifts $z_{\rm phot} > 8.2$ (see Sect.~\ref{sec:strategy}), we switched to the \textit{J2} filter, whose wavelength coverage is 1117\,--1246\,nm. Observations were carried out in an ABAB dither pattern with individual exposure times of \SI{150}{s}. At least one dither sequence was obtained per target, corresponding to a minimum total integration time of \SI{600}{s}. 

MOSFIRE observations were carried out during the 2024B (4 nights; U418, PI: Hennawi), 2025A (5 nights; U275, N110, PI: Hennawi), and 2025B (0.5 nights; U262, PI: Hennawi) semesters. Approximately one half-night in 2024B and  1.5 nights in 2025A were compromised by poor seeing and/or substantial extinction. The 2025B observations were conducted under seeing conditions exceeding \ang{;;2}.

\subsection{\label{sec:FIRE}Magellan/FIRE}

FIRE is a NIR cross-dispersed echelle spectrograph on the Magellan Baade 6.5m telescope. For confirmation of quasar candidates, we used FIRE in its long-slit (prism) mode with a 1\farcs0 slit, providing a spectral resolution of ${\rm FWHM}\sim1000$--1660\,km\,s$^{-1}$ ($R \sim 180$--300) and covering a wavelength range of 820--2500\,nm. We obtained a typical on-source exposure time of \SI{900}{s} $\times\, 2$ per target, in an AB nodding sequence.

Observations were carried out during 2024B (4 nights; PI: J.~Yang) and 2025A (0.5 nights; PI: X.~Fan). In a few cases (e.g., due to faint targets or poor weather conditions), we increased the exposure time to \SI{900}{s} $\times\, 4$.

For one newly identified quasar, we also took 3.25 hours of Echelle mode observations for high-resolution and high-quality NIR spectra (see Fig.~\ref{fig:J0522}). The Echelle mode provides a spectral resolution of $R\sim 4800$ for a \ang{;;0.75} slit.

\subsection{\label{sec:LBT}LBT/MODS and LBT/LUCI}

MODS is an optical dual-channel spectrograph, and LUCI is a NIR imager and multi-object spectrograph on the LBT, covering the \textit{Y}, \textit{J}, \textit{H}, and \textit{K} bands. LBT Observations were carried out between February and September 2025 in 2025A (ID: MPIA-2025A-007) and 2025B (ID: MPIA-2025B-004) semesters. We used the binocular mode with the red grating for MODS and the G200 grating with zJ+zJ filters for LUCI. The MODS red grating covers a wavelength range of 580--1000\,nm, while LUCI G200 with zJ filters spans 900--1200\,nm. We employed long-slit spectroscopy with slit widths of \ang{;;1.0}--\ang{;;1.2} and utilised an ABBA dithering pattern for optimal sky subtraction. Individual exposure times were 300 or \SI{600}{s} for MODS and \SI{240}{s} for LUCI, with total integration times of 1--2 hours for most targets. 

We also performed a deep LUCI follow-up observation of EUCL J1729+6410 (see Fig.~\ref{fig:J1729}), the highest redshift quasar reported in this work. The total on-source exposure time is 2.8 hours in binocular mode with an average seeing of \ang{;;0.6}.

\subsection{\label{sec:strategy}Observation strategy and data reduction}

To estimate the required exposure times for spectroscopic confirmation, we leveraged the photometric redshifts provided by the suite of algorithms described in Sect.~\ref{sec:selection}, in combination with the fake source injection approach introduced by \citet{YangD2024}. In brief, this method involves injecting synthetic quasar spectra with a given redshift and $J$-band magnitude into real 2D spectroscopic data. We then determined the minimum exposure time necessary for each instrument to yield a sufficient S/N for visually identifying the Lyman break. These exposure time estimates only serve as reference values and lower limits, due to the uncertainties in $z_{\rm phot}$ and the variability of observing conditions. They turned out to be robust for typical quasars overall.

For candidates in the north, we typically conducted initial observations with Keck/LRIS or Keck/KCWI, followed by Keck/MOSFIRE if there was no signal detected at the position of the target in the two optical instruments and $z_{\rm phot}$ suggested a potentially higher redshift. In contrast, for southern targets, the longslit mode of Magellan/FIRE, covering the full wavelength range from \SI{800}{nm} to \SI{2500}{nm}, enabled us to perform comprehensive spectroscopic follow-up without switching between instruments.

During observations, we conducted on-the-fly data reduction with \texttt{PypeIt} \citep{Prochaska2020pypeit} for real-time decision-making. All spectra presented in this work were subsequently reduced in a uniform manner using \texttt{PypeIt}.

\section{\label{sec:new_quasar}Discovery of 31 new high-$z$ quasars}

\begin{figure*}[htbp!]
  \begin{center}
    \includegraphics[angle=0,width=1.95\columnwidth]{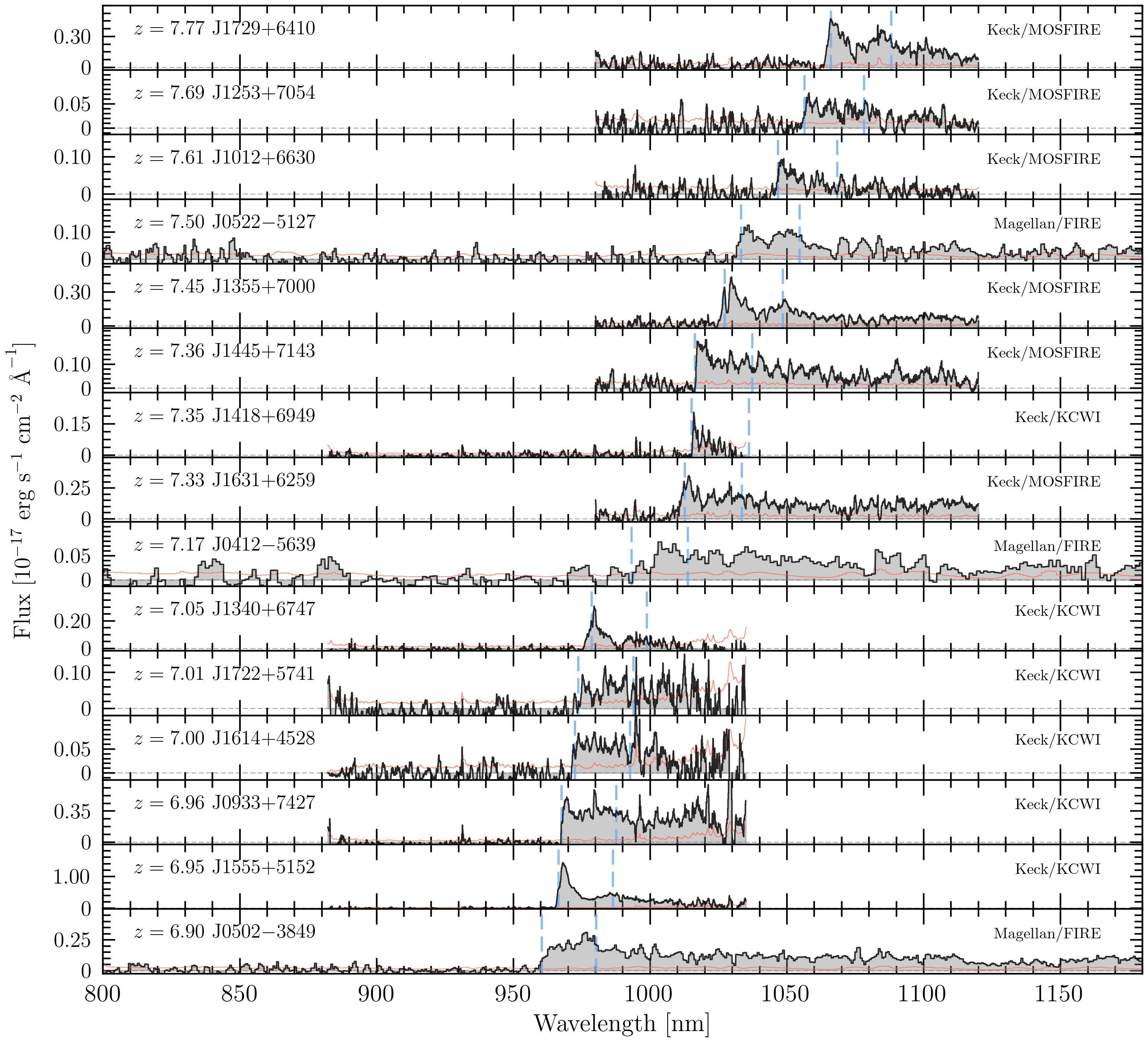}
   \end{center}
\caption{Discovery spectra of the new quasars at $z\geq 6.9$. The redshifts estimated from the \lya breaks, the short names, and the instruments for the discoveries are noted in the each panels. The spectra are fluxed, but not corrected for telluric absorption. They are also smoothed with inverse-variance weights and a 3 or 11 pixel window for Magellan or Keck, respectively. The red curves in each panel show the smoothed noise vectors. The dashed, vertical light blue lines indicate the positions of \lya and \nv lines.}
\label{fig:discovery_spectra}
\end{figure*}

We have conducted spectroscopic follow-up of 123 high-$z$ quasar candidates with the aforementioned observations. With these observations, we have identified 31 new quasars at redshifts $6.6 < z < 7.8$, whose discovery spectra are shown in Fig.~\ref{fig:discovery_spectra} ($z\geq6.9$) and Fig.~\ref{fig:discovery_spectra_2} ($z<6.9$). The cutouts of all the new quasars are displayed in Appendix~\ref{apdx:data_public}, and their 2D spectra are also shown in Appendix~\ref{apdx:discovery_spectra}. The summary of the rest of the candidates is provided in Sect.~\ref{sec:contaminant}. Notably, 12 of these quasars lie at $z \geq 7$, more than doubling the number of known quasars at $z \geq 7$ prior to \Euclid. Among them, EUCL\,J1729, at $z \approx 7.77$, is the most distant quasar known thus far. 

Table~\ref{table:new_quasars} summarises the key properties of the newly identified quasars, including their \Euclid names\footnote{The naming convention for sources identified in this work follows the standard IAU format, with the prefix `EUCL' denoting selection from \Euclid data, followed by the J2000.0 right ascension and declination coordinates in \JE band. The final digits in both the RA and Dec components of the source names reflect a positional precision that is consistent with \Euclid{’s} astrometric accuracy, which is better than 100\,mas.}, estimated redshifts, \JE-band coordinates, photometry, rest-frame ultraviolet absolute magnitudes ($M_{1450}$), and bolometric luminosities ($L_{\mathrm{bol}}$). Redshifts were estimated based on the location of the Ly$\alpha$ emission line and/or the onset of the Gunn--Peterson trough \citep{Gunn1965}. 
The typical redshift uncertainty of such visual determinations is approximately 0.05--0.1 \citep[e.g.][]{Matsuoka2016}; however, in extreme cases where Ly$\alpha$ is weak or strongly absorbed (see note~b in Table~\ref{table:discovery}), the uncertainty can increase to $\sim0.2$.
The $M_{1450}$ values were derived from the observed \JE-band magnitudes by extrapolating a power-law continuum ($F_{\lambda} \propto \lambda^{\alpha}$) with a fixed spectral slope of $\alpha = -1.7$ \citep[e.g.][]{Telfer2002}. Bolometric luminosities were converted from $M_{1450}$ using the empirical conversion from \citet{Runnoe2012}. 

\begin{figure*}[htbp!]
  \centering
    \includegraphics[width=0.49\textwidth]{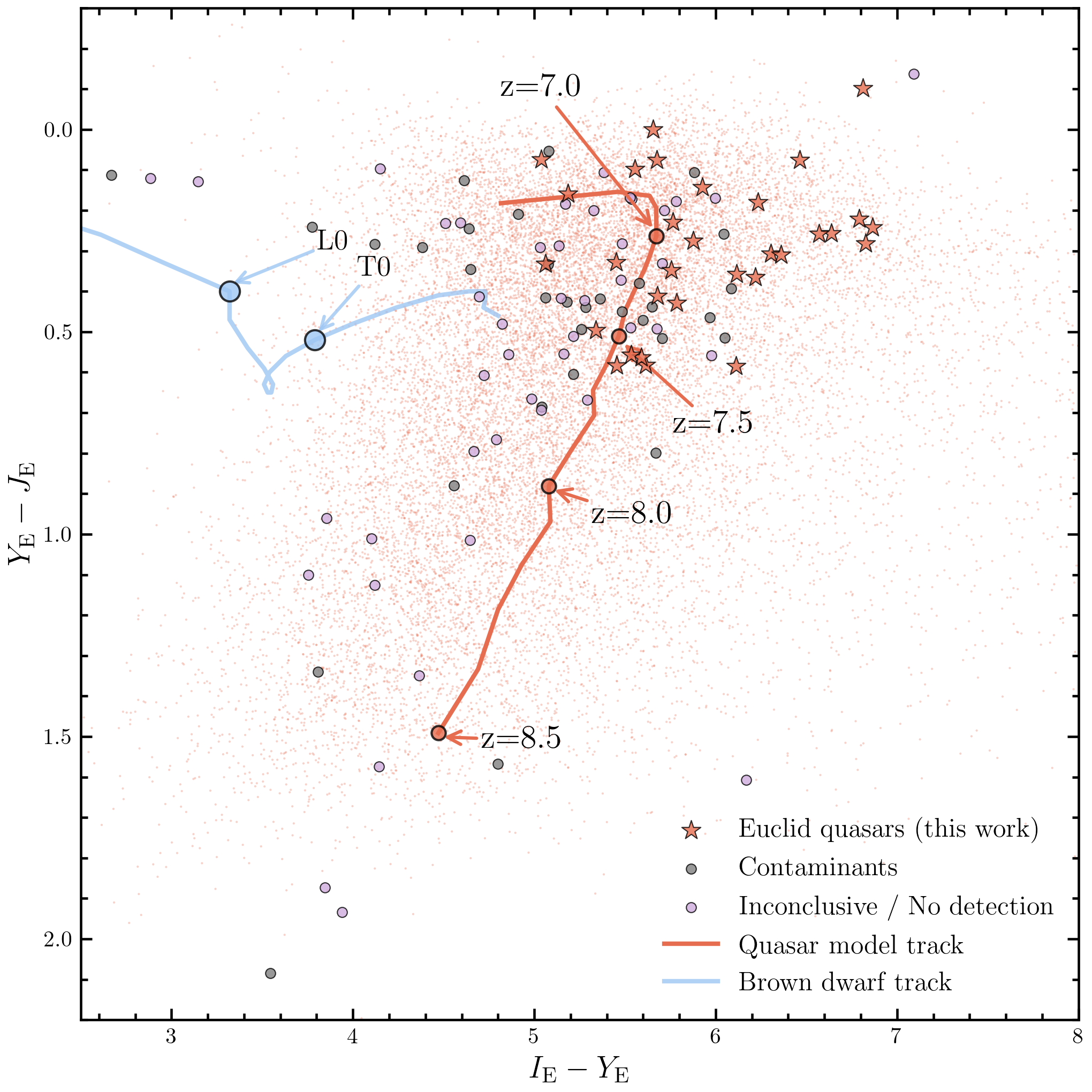}
  \hfill
    \includegraphics[width=0.49\textwidth]{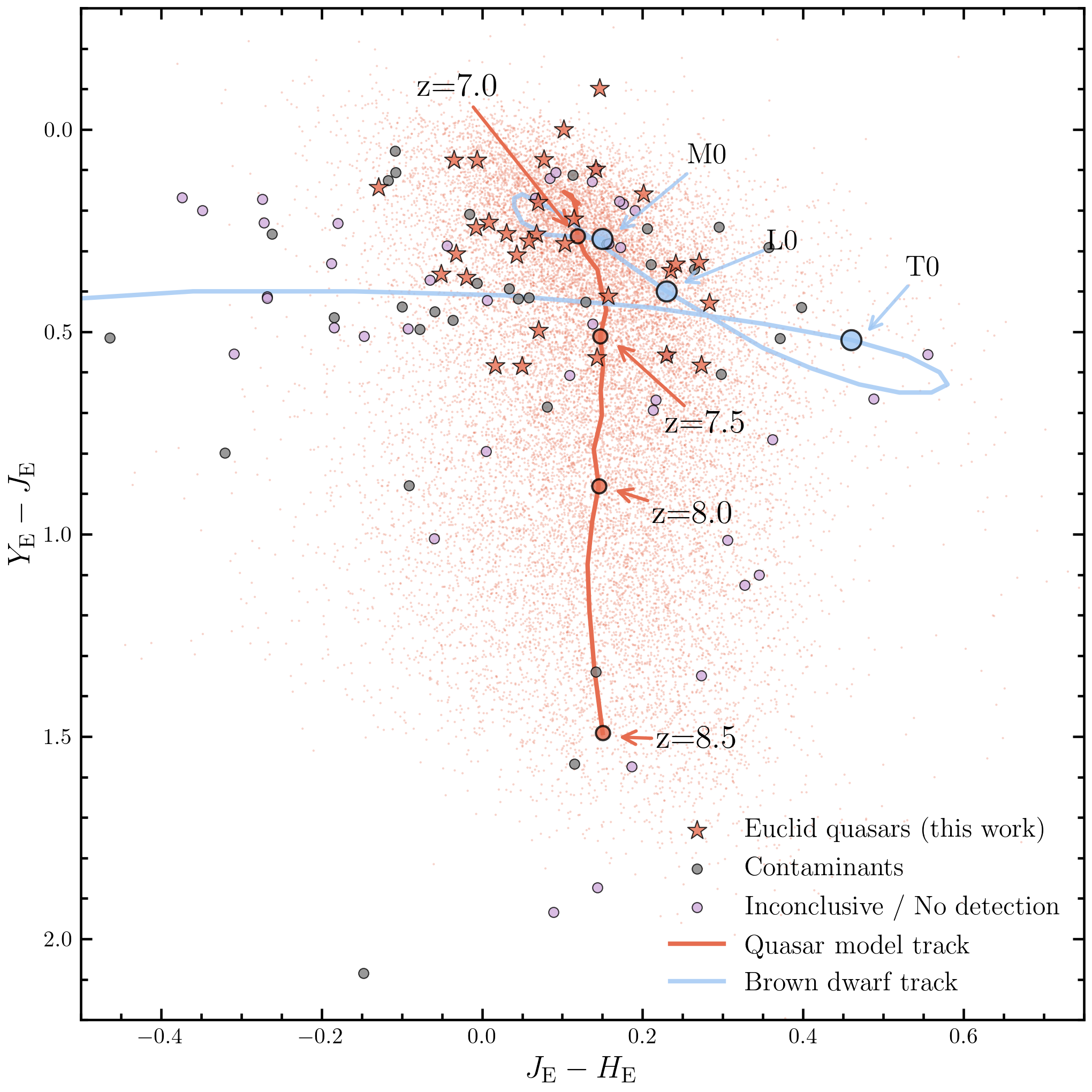}
  \caption{
    Colour-colour diagrams of newly discovered quasars and other candidates.
    \textit{Left}: $\IE - \YE$ versus $\YE - \JE$ colour-colour diagram. 
    \textit{Right}: $\JE - \HE$ versus $\YE - \JE$ colour-colour diagram. 
    The red stars mark the new \Euclid quasars presented in this work, the grey circles denote the identified contaminants, and the violet circles are those marked as inconclusive or no detection (see Sect.~\ref{sec:contaminant}).
    The red curve indicates the mean colour track of high-$z$ quasars based on the synthetic quasar samples (the red points; Yang et~al., in prep.), with redshifts of 7.0, 7.5, 8.0, and 8.5 annotated. 
    The light blue curve represents the empirical brown-dwarf colour track from \citet{Barnett-EP5}, with selected spectral types (M0, L0, T0) labelled.
  }
  \label{fig:color_color}
\end{figure*}

Figure~\ref{fig:color_color} shows the distribution of the newly discovered quasars in the colour--colour diagrams ($\IE-\YE$ versus $\YE-\JE$ and $\JE-\HE$ versus $\YE-\JE$). We overplotted the synthetic colour track of high-$z$ quasars (D.~Yang et~al., in prep.) and the empirical brown-dwarf track from \citet{Barnett-EP5}, both convolved with typical \Euclid photometric noise. The new quasars lie close to the mean quasar track while exhibiting noticeable scatter, consistent with the diversity seen in the synthetic high-$z$ quasar samples.

\begin{figure*}[htbp!]
  \begin{center}
    \includegraphics[angle=0,width=2\columnwidth]{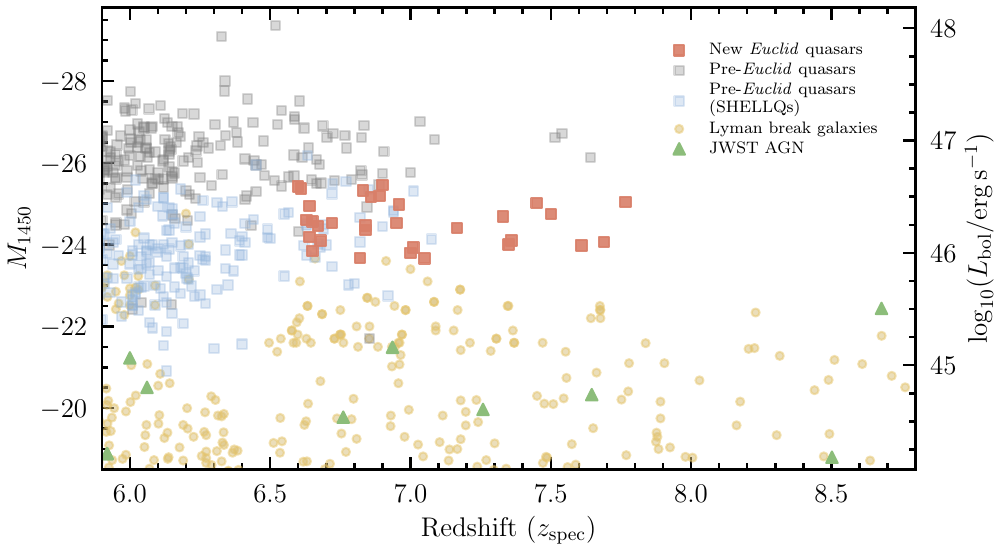}
   \end{center}
\caption{
Redshift versus absolute magnitude at 1450\,\AA, $M_{1450}$, for the new \Euclid-discovered quasars (red squares), compared to pre-\Euclid quasars (grey squares), including the SHELLQs sample (light blue squares). Yellow circles represent LBGs compiled from \cite{Matsuoka2016,Matsuoka2018ApJS,Matsuoka2018PASJ,Matsuoka2019b,Matsuoka2022,Matsuoka2025,Bouwens2022,Roberts-Borsani2024,Roberts-Borsani2025,Harikane2025}. Green triangles denote faint AGN candidates discovered with JWST with rest-frame UV spectroscopic observation \citep{Harikane2023,Kokorev2023,Larson2023,Maiolino2024,Schindler2024,Lin2025}. The right-hand axis shows the corresponding bolometric luminosity scale converted from $M_{1450}$.
}
\label{fig:z-MUV}
\end{figure*}

Figure~\ref{fig:z-MUV} presents the distribution of redshift versus $M_{1450}$ for the high-$z$ quasars discovered in this work (red squares). For comparison, we also show previously known quasars \citep[grey and green circles;][]{Fan2023}, LBGs \citep[yellow circles;][]{Matsuoka2016,Matsuoka2018ApJS,Matsuoka2018PASJ,Matsuoka2019b,Matsuoka2022,Matsuoka2025,Bouwens2022,Roberts-Borsani2024,Roberts-Borsani2025,Harikane2025} and active galactic nuclei (AGN; $L_{\rm bol}\lesssim 10^{45.5}\ {\rm erg\,s^{-1}}$) candidates from various JWST programs that have rest-frame ultraviolet (UV) spectroscopic observations \citep[green triangles;][]{Harikane2023,Kokorev2023,Larson2023,Maiolino2024,Schindler2024,Lin2025}. The discoveries of these less luminous quasars in this work probe a previously unexplored region of parameter space, reaching lower luminosities ($M_{\rm 1450}\sim-24$) at $z\gtrsim7$. The implications of these faint high-redshift quasars are discussed in Sect.~\ref{sec:discussion_uvfaint}.

\begin{figure*}[htbp!]
  \begin{center}
    \includegraphics[angle=0,width=1.95\columnwidth]{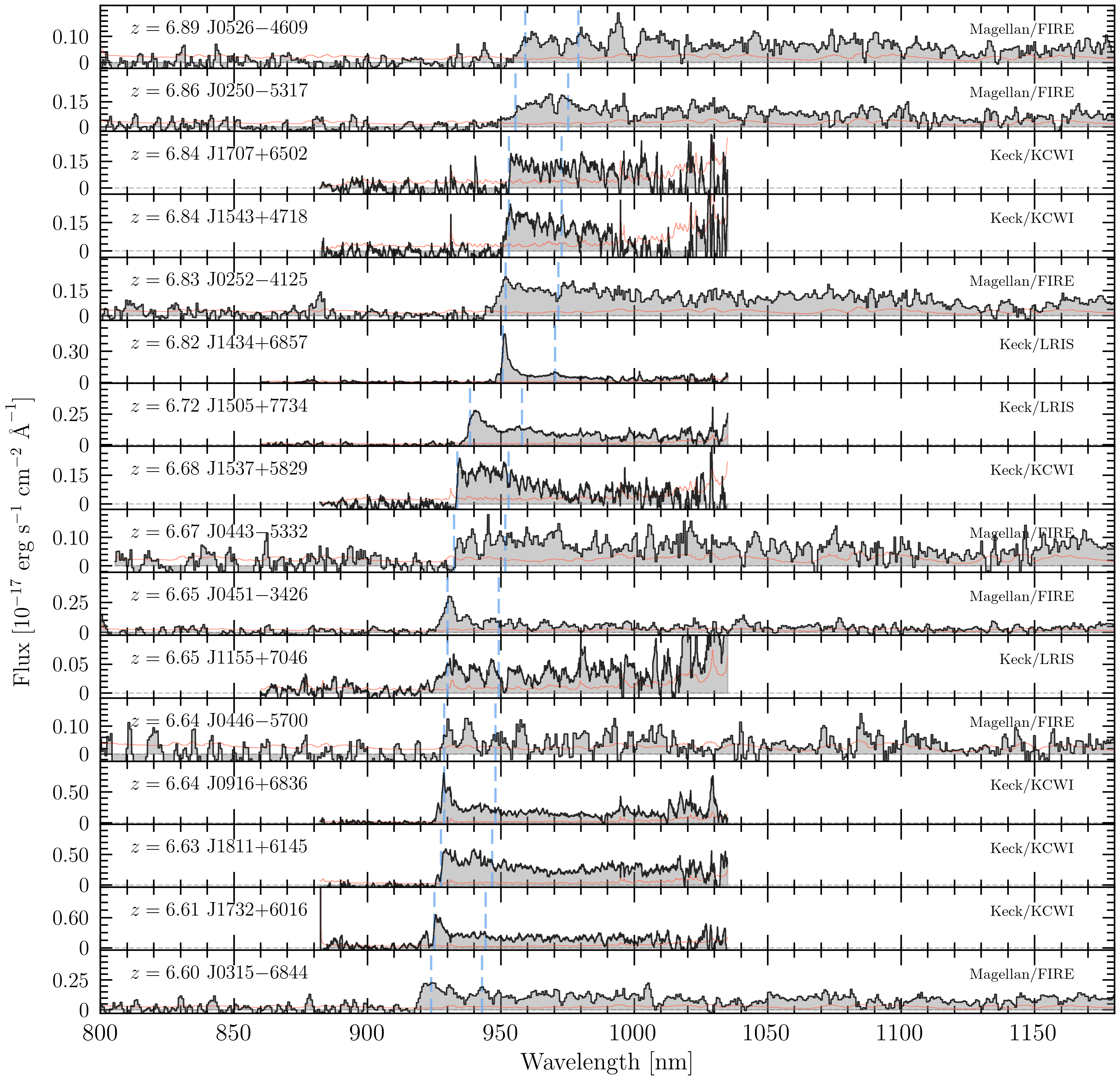}
   \end{center}
\caption{Same as Fig.~\ref{fig:discovery_spectra} but for the new quasars at $z< 6.9$.}
\label{fig:discovery_spectra_2}
\end{figure*}

\subsection{\label{sec:individual}Notes on individual quasars}

\begin{figure*}[htbp!]
  \begin{center}
    \includegraphics[angle=0,width=1.95\columnwidth]{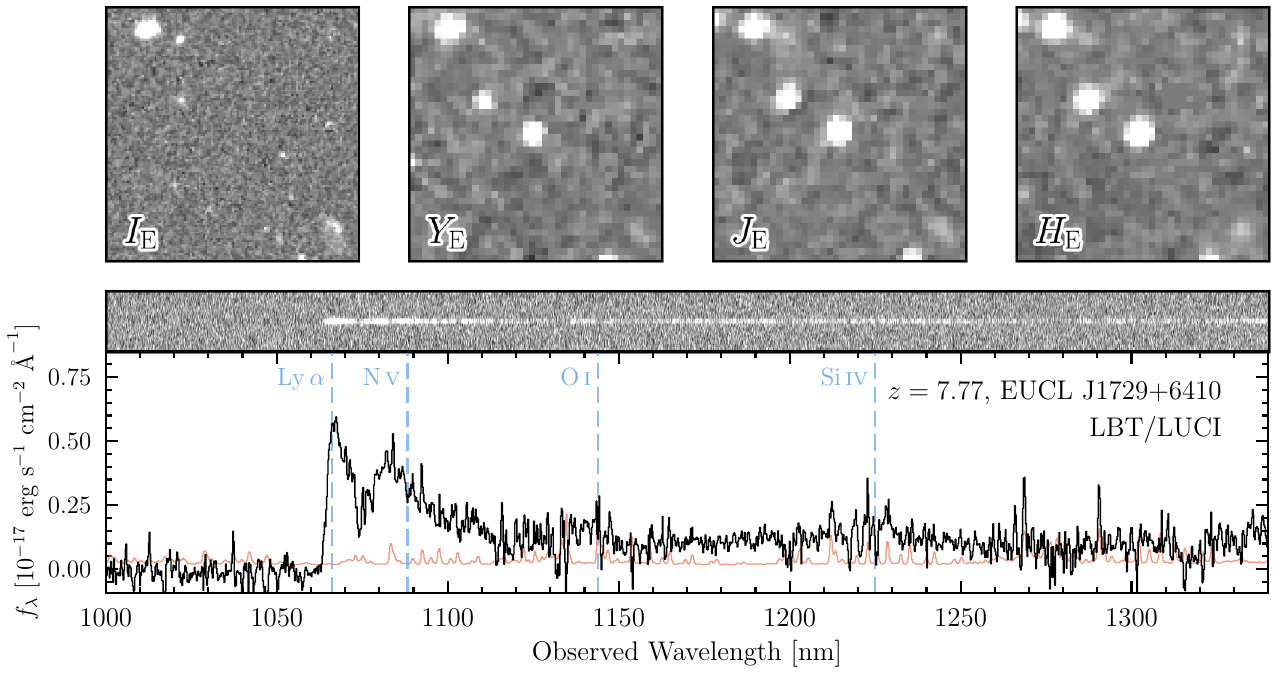}
   \end{center}
\caption{\Euclid cutouts and LBT/LUCI spectrum of EUCL\,J1729 ($z\approx7.77$), the most distant quasar discovered to date. {\it Top}: $12^{\prime\prime}\times12^{\prime\prime}$ cutouts. From left to right: \Euclid \IE, \YE, \JE, and \HE bands. {\it Bottom}: The 1D spectrum has been smoothed using a 5 pixel inverse-variance-weighted boxcar. The total exposure time is \SI{10080}{s}.}
\label{fig:J1729}
\end{figure*}

\noindent\paragraph{EUCL\,J172902.75$+$641018.1:}
EUCL\,J1729 is the most distant quasar discovered to date, with $z\approx 7.77$. Figure~\ref{fig:J1729} displays its 2D and 1D LBT/LUCI spectrum with a total integration time of \SI{10080}{s} (2.80\,h). \added{Based on visual inspection of the multiple emission features (\lya, \nv, \oi, \siiv) shown up in the spectrum, we adopt a conservative redshift uncertainty of $\Delta z \lesssim 0.05$ for this target.} The age of the Universe at its redshift is approximately \SI{662}{Myr}. The redshift increment over the previous record-holder \citep{Wang2021} is \added{approximately 0.13}, corresponding to $\sim \SI{15}{Myr}$. Notably, a broad absorption feature is present blueward of the \nv emission line, suggesting that EUCL\,J1729 is likely a broad absorption line (BAL) or mini-BAL quasar. The \oi and \siiv emission lines are also detected, with tentative, relatively narrow absorption features seen in their vicinity.

\smallskip
\noindent\paragraph{EUCL\,J125308.55$+$705432.3, EUCL\,J101255.87$+$663058.0:} EUCL\,J1253 ($z\approx 7.69$) and EUCL\,J1012 ($z\approx 7.61$) are the second and third most distant quasars identified in this work. They are also the faintest known quasars at $z \gtrsim 7.5$, with $M_{1450} \sim -24$, making them nearly an order of magnitude less luminous in the rest-frame UV than the three previously known quasars at similar redshifts (see Fig.~\ref{fig:z-MUV}). They provide a unique opportunity to probe the faint end of the QLF at the highest redshifts currently accessible. Sub-millimetre follow-up observations of EUCL\,J1253 have revealed a large gas reservoir in its host galaxy. The data and the analysis will be presented in a companion paper (Belladitta et~al., in prep.).

\smallskip
\noindent\paragraph{EUCL\,J052209.82$-$512709.2:}
We conducted an Echelle-mode observation of EUCL\,J0522 ($z\approx 7.50$) using Magellan/FIRE (see Sect.~\ref{sec:FIRE}), as it is among the brightest $z > 7$ quasars identified in this work. However, its \JE-band magnitude measured from the \texttt{NIR} stacked image is affected by hot pixels overlapping with the target, \added{which bias the Euclid $J_E$ photometry to be artificially brighter. We also note that the quasar probabilities computed by all methods were correspondingly impacted, and in particular were reduced relative to what would be obtained using the corrected photometry.} At the time of writing, we have obtained a JWST/NIRSpec spectrum of this object. Using this spectrum and the \JE-band transmission curve, we measured a corrected \JE-band magnitude of 22.17. The full analysis of the JWST/NIRSpec data will be presented in a forthcoming publication.

\begin{figure*}[htbp!]
  \begin{center}
    \includegraphics[angle=0,width=1.95\columnwidth,trim=0 0 0 0,clip]{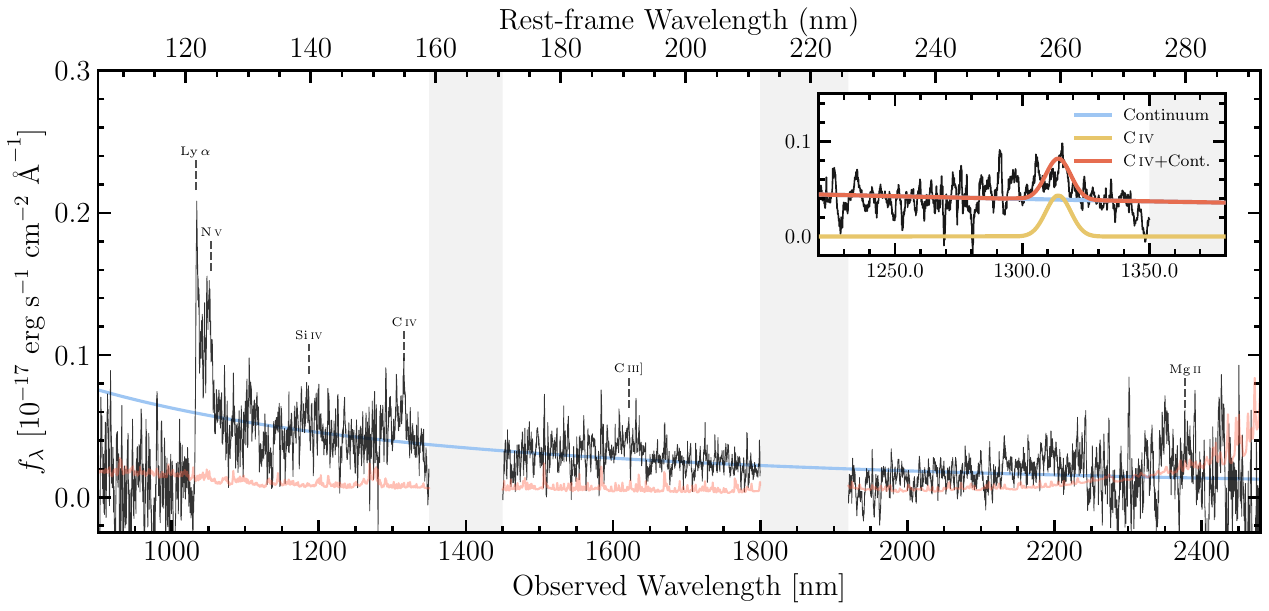}
   \end{center}
\caption{Magellan/FIRE echelle spectrum of EUCL\,J0522, one of the brightest $z\sim 7.5$ quasars in this work ($\JE\approx22.17$). The total integration time is \SI{11700}{s}. The black line shows the observed spectrum, inverse-variance smoothed with a 31 pixel boxcar. The light red curve shows the smoothed noise vector. The blue curve represents the best-fit power-law continuum. Locations of typical emission lines seen in a quasar spectrum are marked by vertical dashed lines, including Ly$\alpha$, \ion{N}{v}, \ion{Si}{iv}, \ion{C}{iv}, \ion{C}{iii]}, and \ion{Mg}{ii}. Grey shaded regions indicate wavelength ranges heavily affected by telluric absorption. The inset displays the \ion{C}{iv} region with the best-fit emission-line model generated using \texttt{Sculptor}, showing the individual \ion{C}{iv} component (yellow), continuum (blue), and combined fit (red).}
\label{fig:J0522}
\end{figure*}

Figure~\ref{fig:J0522} shows the Echelle spectrum of this quasar with a total integration time of \SI{11700}{s} (3.25\,h). Although EUCL\,J0522 is among the most luminous quasars at $z > 7$ in our sample, it remains approximately two magnitudes fainter than previously known quasars at comparable redshifts (see Fig.~\ref{fig:z-MUV}) in the rest-frame UV. The spectrum covers a broad rest-frame wavelength range, from \lya\ to \mgii, and reveals several prominent emission features, including \lya, \nv, \siiv, \civ, and \ciii. The \mgii line is not detected, likely due to insufficient S/N.

We fit the \civ emission line using a single-component Gaussian profile with the spectral fitting tool \texttt{Sculptor} \citep{Schindler2022sculptor}, following the methodology described in \citet{Schindler2020}. The fit yields a FWHM of \civ of approximately $2743 \pm 400\,\kms$. Using the single-epoch black hole mass estimator based on the \civ line from \citet{Vestergaard2006}, we derived $\log_{\rm 10}(M_{\rm BH}/M_\odot)\approx7.6\pm0.4$ \added{(the uncertainty is dominated by the scatter of the scaling relation, $\sim 0.36$ dex)}. This corresponds to an Eddington ratio of $\lambda_{\rm Edd}\approx4.5$ for EUCL\,J0522. Assuming instead an Eddington-limited accretion rate ($\lambda_{\rm Edd}=1$), the implied black hole mass would be $\log_{10}(M_{\rm BH}/M_\odot)\sim8.25$. Given the limitations of \civ as a virial mass tracer \citep[e.g.,][]{Coatman2016} and the modest S/N of the current data, a detailed analysis of the central black hole properties is deferred to future work using higher-quality data \added{with wider spectral coverage}, such as deeper JWST/NIRSpec observations.

\smallskip
\noindent\paragraph{EUCL\,J041250.73$-$563949.7, EUCL\,J134031.50$+$674715.1, EUCL\,J025029.93$-$531706.7, EUCL\,J044326.27$-$533214.9, EUCL\,J115522.89$+$704612.3:} These objects represent the most apparent examples exhibiting atypical \lya emission profiles. Such variations might suggest the presence of strong damping wing absorption, proximity damped \lya systems (pDLAs), or, in some cases, the possibility that the sources are in fact luminous galaxies. However, we caution against over-interpreting these features based solely on the current discovery spectra -- particularly those obtained with the Magellan prism mode -- due to their low spectral resolution and modest S/N. Nevertheless, the diversity among UV-selected sources at the faint end ($M_{\rm UV} \sim -24$; see Sect.~\ref{sec:discussion_uvfaint}) remains poorly constrained, and deeper spectroscopic follow-ups are essential to investigating their nature.

\begin{figure*}[htbp!]
  \begin{center}
    \includegraphics[angle=0,width=1.75\columnwidth]{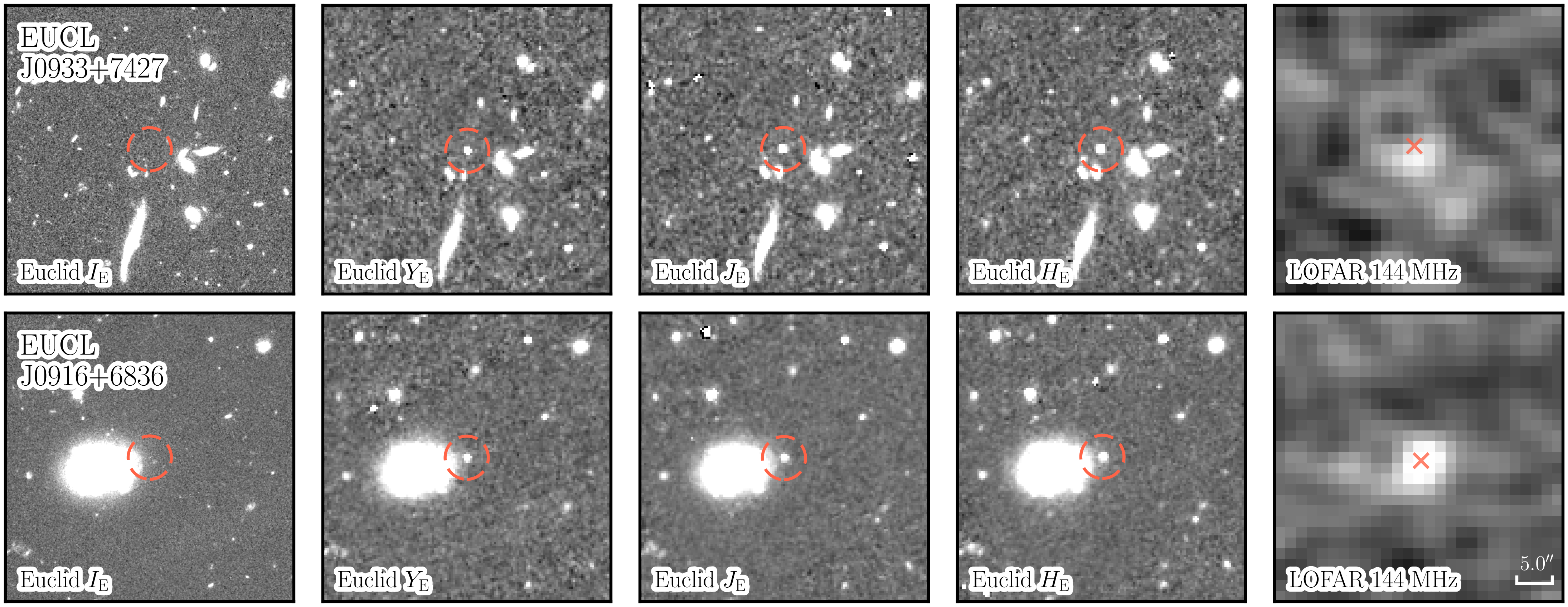}
   \end{center}
  \caption{Cutouts of two newly discovered LOFAR-detected quasars, EUCL\,J0933$+$7427 (bottom) at $z\approx6.96$, and EUCL\,J0916$+$6836 (top) at $z\approx6.6$.  From left to right: \Euclid \IE, \YE, \JE, and \HE bands, followed by the LOFAR \SI{144}{MHz} image. Each cutout covers $\ang{;;40}\times\ang{;;40}$ region. Red circles in \Euclid images and red crosses in LOFAR images both denote the \JE-band positions of the quasars. A scale bar of a \ang{;;5} length is shown in the last cutout.}
\label{fig:lofar}
\end{figure*}

\smallskip
\noindent\paragraph{EUCL\,J091639.93$+$683652.9, EUCL\,J093330.60$+$742730.0:}
To identify new quasars with significant radio emission ($>5\sigma$), we cross-matched the LoTSS DR3 catalogues with our sample using a matching radius of \ang{;;2}. Two quasars, EUCL\,J0916 ($z\approx 6.64$) and EUCL\,J0933 ($z\approx 6.96$), have positive matching results, as shown in Table~\ref{table:radio}. Their \SI{144}{MHz} images, along with their \Euclid cutouts, are presented in Fig.~\ref{fig:lofar}. It is worth noting that the association between the NIR and radio components of EUCL\,J0933 remains somewhat uncertain, as can be seen from the cutouts. 
\added{However, given the low S/N (3.5) of the radio detection and the beam size of LoTSS (\ang{;;6}), the expected astrometric uncertainty is $\sim{\rm FWHM}/(2\times{\rm S/N})\approx\ang{;;0.9}$, so the \ang{;;1.76} offset corresponds to $\sim$2$\sigma$ and is not implausible for a real association. The corresponding 2D tail probability is of order $10\%$, much higher than the probability of random association between a LoTSS detection with S/N$\geq3.5$ and a $z\geq6.5$ quasar.}
Assuming these fluxes are indeed from the quasars and a power-law index of 0.7, the radio powers at rest-frame \SI{144}{MHz} are $\logten \left(P_{144}/{\rm W\,Hz^{-1}}\right)\approx26.64$ and $26.19$, respectively. These values place them among the most radio-luminous quasars known at $z>6.5$ \citep[e.g.,][]{Gloudemans2022}, \added{and are thus more naturally explained by quasar jet activity than by star formation in the host galaxy.}

\begin{table}
\caption{Radio properties of the two \Euclid quasars with LOFAR counterparts.}
\label{table:radio}
\centering
\begin{tabular}{lccc}
\hline\hline
Target & Redshift & $S_{\rm 144\,MHz}$ (mJy) & Separation \\
\hline
EUCL\,J0916 & 6.60 & $1.53\pm0.25$ & \ang{;;0.69} \\
EUCL\,J0933 & 6.96 & $0.49\pm0.14$ & \ang{;;1.76} \\
\hline
\end{tabular}
\tablefoot{
Both quasars are detected in the LoTSS DR3 catalogue with $>5\sigma$ significance in peak flux. The quoted values are integrated 144 MHz flux densities and their associated uncertainties. The angular separation indicates the distance between the \JE-band and radio positions. The beam size of LoTSS DR3 is around \ang{;;6}.
}
\end{table}

\subsection{\label{sec:contaminant}Notes on other candidates}

We conducted spectroscopic observations for a total of 123 high-$z$ quasar candidates. Of these, 31 were confirmed as quasars, while the remaining 92 were classified into three categories: `contaminants', `inconclusive' and `no detection'. Table~\ref{table:contaminant} summarises the classification scheme and definitions applied to these non-quasar sources. In Appendix~\ref{sec:append_contaminant}, we also show the 2D/1D spectra of some targets that were identified as `contaminants', and the coordinates of all of them are provided in Table~\ref{table:contaminant_list}.

\begin{table}
\centering
\caption{Classification of the remaining 92 candidates.}\label{table:contaminant}
\begin{threeparttable}[b]
\begin{tabular}{p{1.4cm} c p{4.7cm}}
\toprule\toprule
Label & Number & Description \\
\midrule
Contaminants & 45 & A trace was detected but ruled out as a high-$z$ quasar; 28 of the 45 were visually classified as possible brown dwarfs based on their spectral features. \\
\addlinespace
Inconclusive & 25 & A trace was detected, but the data quality was insufficient to determine whether a \lya break is present. Some remain high-priority targets for future observations. \\
\addlinespace
No detection & 22 & No detectable flux was observed in the spectroscopic data. \\
\bottomrule
\end{tabular}
\end{threeparttable}
\end{table}

A substantial fraction of the `no detection' sources are likely to be contaminants based on their colours and $z_{\rm phot}$ estimates. This is supported by the fact that brown dwarfs and low-redshift red galaxies are generally more difficult to detect or classify than high-$z$ quasars at similar \JE-band magnitudes. High-$z$ quasars typically show a pronounced flux density near Ly$\alpha$ and a sharp break blueward of the line, while contaminants tend to exhibit a more gradual decline towards the blue end. Consequently, only sources with $z_{\rm phot} \gtrsim 7$ that are undetected in the optical are retained as viable IR follow-up targets. Objects undetected in both the optical and NIR, or those with $z_{\rm phot} < 7$ and no optical detection, are likely to be contaminants.

\subsection{\label{sec:stack}Composite spectra of \Euclid high-$z$ quasars}

\begin{figure*}
\sidecaption
\includegraphics[width=0.70\textwidth]{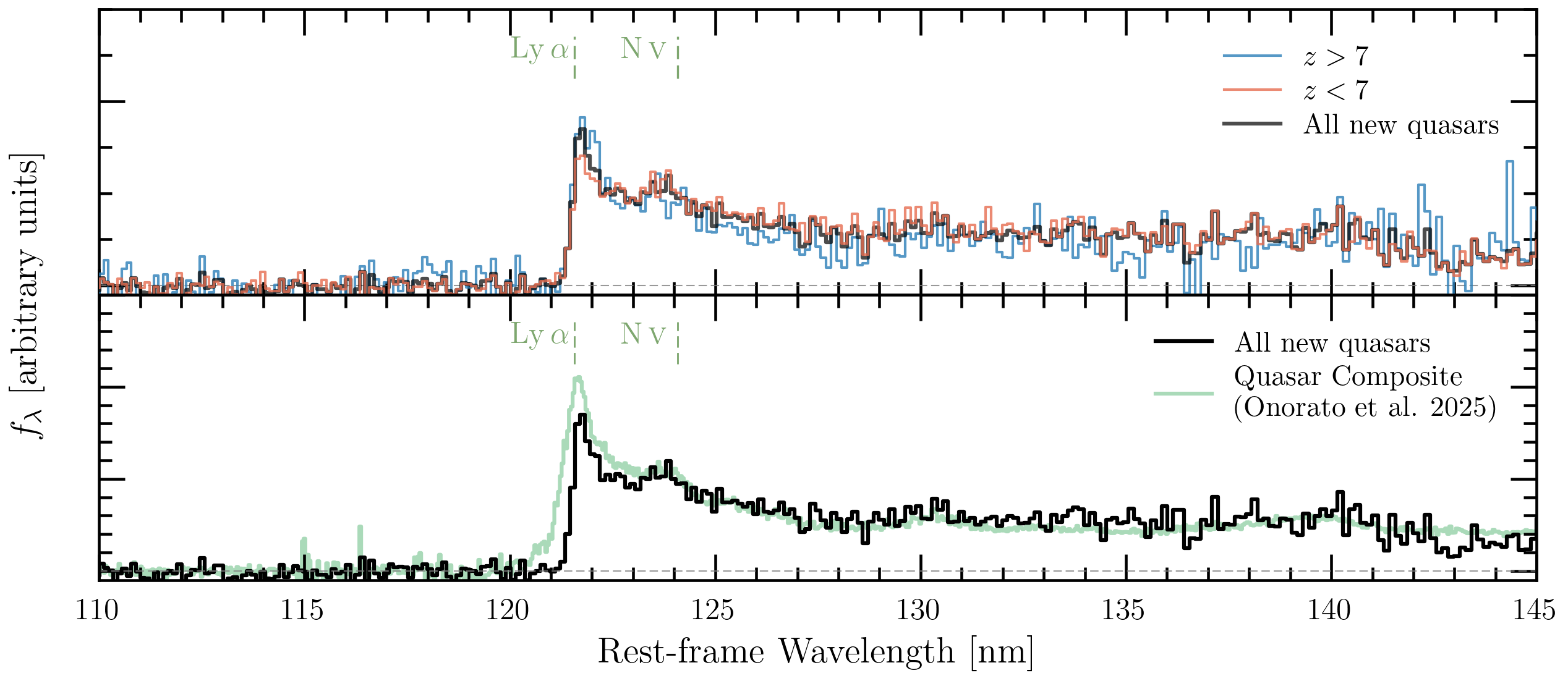}
\caption{Stacked rest-frame UV spectra of the new \Euclid quasars. The top panel shows the stacked rest-frame UV spectra of all new \Euclid quasars, the $z > 7$ sample, and the $z < 7$ sample. The bottom panel compares the stacked spectrum of all new quasars with a stacked spectrum of $z>6.5$ bright quasars from \citet{Onorato2025}. Each spectrum was shifted to the rest frame, re-binned onto a common grid with $\Delta v=300$\,km\,s$^{-1}$, and scaled to match a fiducial absolute magnitude of $M_{\rm 1450} = -24$. We show the median-stacked results here. No significant evolution is identified with these stacked spectra.}
\label{fig:stack}
\end{figure*}

The rest-frame UV/optical SEDs of quasars are generally thought to exhibit little (if any), evolution with redshift \citep[e.g.,][]{Shen2019,Yang2021,Onorato2025}. To investigate this further, we constructed composite spectra for the $z > 7$, $z < 7$, and full quasar samples in this work. Each spectrum was shifted to the rest frame using its systemic redshift and re-binned onto a common rest-frame wavelength grid with a velocity spacing of $\Delta v = 300\,\mathrm{km\,s^{-1}}$ per pixel. To normalise the spectra, we calculated the average flux density between 1250 and 1400\,\AA\ and scaled each spectrum to match the composite spectrum from \citet{Onorato2025} over the same interval. Median-stacked spectra were then generated for the $z > 7$, $z < 7$, and full samples. The resulting composite spectra are shown in Fig.~\ref{fig:stack}, alongside the \citet{Onorato2025} composite spectrum for comparison.

The $z > 7$ composite spectrum appears to exhibit stronger \lya emission relative to the $z < 7$ counterpart, which may seem to contradict expectations of stronger damping wing absorption at higher redshifts. This can be partially attributed to a higher fraction of `weak Ly$\alpha$' objects within the $z < 7$ sample. Whether the observed weakness of \lya in these sources is intrinsic or primarily driven by the limited spectral resolution remains to be explored with future follow-up data. Furthermore, the composite spectrum constructed from all quasars in this work is broadly consistent with the stacked spectrum from \citet{Onorato2025}, with the exception of a somewhat weaker \lya region, which is potentially indicative of stronger damping wing absorption due to smaller proximity zone for quasars that are less luminous in the rest-UV \citep[e.g.,][]{Ishimoto2020}, \added{and/or because the average redshift of our sample is higher than that of \citet{Onorato2025b}, which would imply a higher average IGM neutral fraction.}

\added{Finally, the good agreement outside the Ly$\alpha$ region reinforces the well-known lack of strong spectral evolution in quasar rest-frame UV spectra at high redshift, indicating a rapid establishment of the physical conditions in the broad-line region at early times.}

\section{\label{sec:discussion_uvfaint}Populating the faint end at $z\gtrsim7$}

As shown in Fig.~\ref{fig:z-MUV}, compared with the quasar samples prior to \Euclid, most of the quasars reported in this work are notably faint in the rest-frame UV. In particular, those at $z \gtrsim 7$ are generally 1--2~mag fainter than the quasars discovered prior to \Euclid\ at comparable redshifts. At $z < 7$, the new quasars exhibit luminosities similar to those found in the SHELLQs sample \citep{Matsuoka2016,Matsuoka2018ApJS,Matsuoka2018PASJ,Matsuoka2019b,Matsuoka2022}. Here we briefly discuss the significance of the high-$z$ sources in this luminosity range.

\subsection{\label{sec:qso_or_lbg}Quasars or galaxies?}

\begin{figure}[htbp!]
  \begin{center}
    \includegraphics[angle=0,width=1\columnwidth,trim=0 0 0 0,clip]{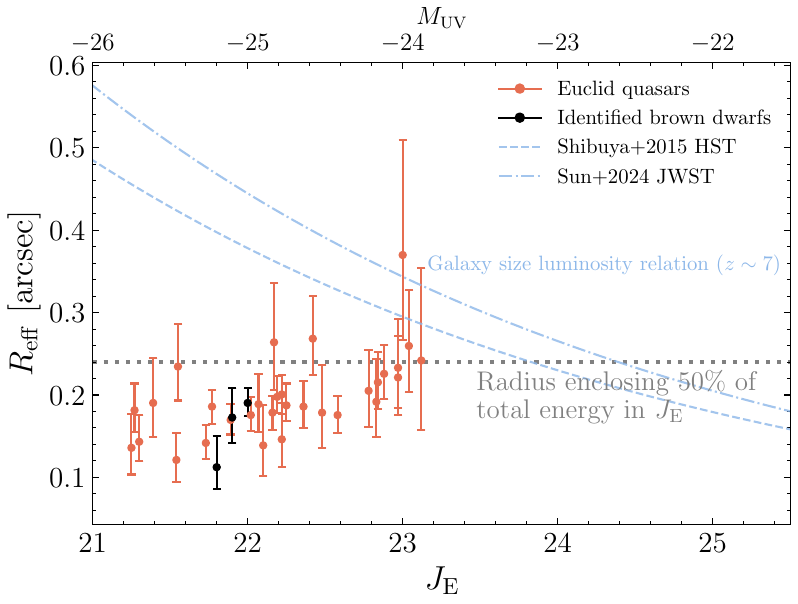}
   \end{center}
\caption{
Magnitude versus fitted size of \Euclid quasars (red orange) and identified brown dwarfs (black) in the \JE band. The effective radii ($R_{\mathrm{eff}}$) are measured from \Euclid \YE, \JE, and \HE images assuming a Sersic profile. The light blue lines represent the empirical size-luminosity relations for galaxies at $z \sim 7$ \citep{Shibuya2015,Sun2022}. The grey horizontal line represents the 50\% enclosed energy radius \citep[$R_{\rm EE50}$;][]{EuclidSkyNISP} for the PSF of the \JE band. The \Euclid quasars are consistent with being point sources.
}
\label{fig:reff}
\end{figure}

The luminosity functions of quasars and LBGs appear to intersect at $M_{1450} \sim -23$ to $-24$ \citep[e.g.,][]{Finkelstein2022,Matsuoka2023,Harikane2025,Marques-Chaves2025}. Therefore, it is not unexpected that some of the faintest objects identified as `quasars' might instead be bright high-$z$ galaxies \citep[e.g.,][]{Sobral2015,Matsuoka2022}, particularly if they lack prominent broad \lya emission (see the examples listed in Sect.~\ref{sec:individual}). While deep NIR spectroscopic observations are required to confirm their true nature, we provide a preliminary assessment by testing their consistency with being point sources. We performed Markov chain Monte Carlo (MCMC) fitting with a S\'ersic profile using \texttt{pysersic} \citep{Pasha2023pysersic} on the \YE, \JE, and \HE images independently, and we subsequently multiplied the posterior distributions of the effective radius $R_{\mathrm{eff}}$ to get the combined posterior. The distribution of \JE magnitude versus the fitted $R_{\mathrm{eff}}$ is shown in Fig.~\ref{fig:reff}.

\added{The inferred $R_{\rm eff}$ of the sources from single-component S\'ersic profile fitting are largely consistent with the \JE-band PSF: for all but one source, the posterior distribution of $R_{\rm eff}$ overlaps the \JE-band $R_{\rm EE50}$ (the radius enclosing 50\% of the total energy; grey dashed line in Fig.~\ref{fig:reff}) within the 1$\sigma$ credible interval \citep[]{EuclidSkyNISP}.} The only exception is EUCL\,J1012+6630, for which the fit is likely affected by a nearby foreground source rather than the source being intrinsically extended. \added{The Ly$\alpha$ line also appears relatively broad ($\gtrsim 1000$ km s$^{-1}$), which is consistent with a quasar interpretation.} As a comparison, we also show the size measurements for three newly identified faint brown dwarfs in our follow-up campaign. We further compare the inferred sizes with the empirical size-luminosity relations of $z \sim 7$ galaxies from \citet{Shibuya2015} and \cite{Sun2022}. \added{However, these relations do not place strong constraints on the very bright end as they're based on surveys in relatively small area.} The intersection between the galaxy size-luminosity relations with the $R_{\rm EE50}$ line at $\JE \sim 24$ suggests that distinguishing quasars from galaxies based solely on their sizes in \Euclid images becomes increasingly challenging for sources fainter than $\JE > 23$. Nevertheless, a further refinement of size measurements has the potential to exploit the diffraction-limited resolution of \Euclid, enabling even tighter constraints to be placed on the sizes of faint objects.

\subsection{Less luminous quasars as distinctive probes of the early Universe}

High-$z$ quasars at the luminous end, with $M_{\rm 1450}\lesssim-25$, are often referred to as the `tip of the iceberg' and were thought to be unrepresentative of the broader SMBH population \citep[see, e.g.,][]{Lauer2007,Li2022}. Some observational studies indeed support this concern: less luminous quasars at $z \gtrsim 6$, such as those identified by SHELLQs and the CFHQS, appear to follow SMBH-galaxy mass relations more consistent with the local Universe \citep[e.g.,][]{Willott2017,Izumi2019,Silverman2025}, in contrast to their more luminous counterparts. At $z > 7$, however, such studies have been extremely limited: only one type-I quasar with $M_{1450} > -25$ was known prior to \Euclid{} \citep{Matsuoka2019}. The discoveries reported here significantly expand this regime, enabling systematic investigation of the demographics, accretion properties, and luminosity function of low-luminosity quasars at $z > 7$.

On the other hand, quasars with faint UV luminosities may exhibit more diverse behaviour, as shown in Sect.~\ref{sec:individual}. 
Studies at $z \sim 6$ have already revealed intriguing properties in the host galaxies of similarly faint quasars: \citet{Onoue2025} detected Balmer absorption lines of stellar origin, a spectral signature commonly associated with post-starburst activity in the galaxies; \citet{Bouwens2025} reported that some low-luminosity quasars might reside in massive galaxies with stellar masses comparable to those hosting luminous quasars. Moreover, as mentioned earlier, at these luminosities the quasar and galaxy luminosity functions intersect, implying that it is also possible that some sources identified as quasars may in fact be exceptionally luminous star-forming galaxies \citep[e.g.,][at lower redshift]{Marques-Chaves2022}. While such cases complicate classification, they are themselves of considerable interest, offering insights into the most massive and actively star-forming systems in the early Universe \citep[e.g.,][]{Marques-Chaves2025}.

In addition, quasars with lower UV luminosities may also be important for reionisation studies. Their smaller proximity zones \citep[e.g.,][]{Ishimoto2020, Onorato2025b} imply more neutral surroundings, making them especially sensitive probes of the ionisation state of the IGM.

\subsection{Future exploration toward even fainter regimes}

We also note that the sensitivity limit of \Euclid has not yet been reached. Most of the quasars reported here are brighter than $\JE \sim 23$, whereas the $5\sigma$ depth of NISP imaging reaches $\mathrm{AB} \approx 24.5$. 
Identifying quasars at these fainter limits, however, is not feasible with ground-based spectroscopy: detecting a $z \gtrsim 7$ quasar with $\JE \sim 23$ typically requires $\gtrsim 1$--2 hours of integration on a 10-m telescope. For even fainter targets, utilising HST or JWST will be the only viable option.

Nevertheless, such faint samples are of irreplaceable value. JWST has recently revealed a population of EoR sources with broad Balmer emission lines, possibly indicative of active SMBHs \citep[e.g.,][]{Harikane2023,Larson2023,Kokorev2023,Lin2024,Greene2024,Maiolino2024,Matthee2024}. However, these sources differ from classical type-I quasars in important respects: they often lack strong high-ionisation lines, and show varied behaviour in variability \citep[e.g.,][]{Kokubo2024,Tee2025}, X-ray emission \citep{Ananna2024,Yue2024}, radio detection \citep[e.g.,][]{Perger2025,Gloudemans2025}, and mid-IR torus emission \citep[e.g.,][]{Akins2025}. Their number densities also exceed predictions from QLF extrapolations by factors of 10--100 \citep[e.g.,][]{Harikane2023,Pizzati2025}. Building quasar samples down to $M_{\rm UV} \gtrsim -22$ -- comparable to the brightest JWST-identified AGN -- will be essential to elucidate the connections and distinctions between quasar, galaxy, and JWST AGN populations.

\section{\label{sec:summary}Summary}

\Euclid has long been anticipated to revolutionise the search for high-$z$ quasars. In this work, we demonstrate that this potential is now being realised: within the first one and a half year of the EWS ($\sim$\SI{3000}{\deg\squared}), we have uncovered 31 new quasars spanning the redshift range $6.6 < z < 7.8$ (see Fig.~\ref{fig:discovery_spectra} and Table~\ref{table:discovery}). Notably, 12 of these quasars lie at $z \geq 7$, more than doubling the number of known quasars at these redshifts prior to \Euclid. Among them, EUCL\,J1729+6410, at $z \approx 7.77$ with $M_{1450} \approx -25.05$, currently represents the most distant quasar known (see Fig.~\ref{fig:J1729}). A large fraction of the newly discovered quasars are also on the faint end of the QLF, with $M_{\rm 1450}\sim -24$ (see Fig.~\ref{fig:z-MUV}). Furthermore, two of the newly discovered quasars show significant ($>5\sigma$) radio counterparts in the LoTSS, underscoring the promising synergy between \Euclid and LOFAR in identifying and characterising the high-$z$ quasar population at $z\sim7$ and beyond. These discoveries mark a substantial advance into the epoch of cosmic reionisation, showcasing \Euclid's unprecedented capability to extend the redshift frontier of quasar studies. Follow-up observations are already underway, although many of these sources are faint and challenging to characterise spectroscopically (see Fig.~\ref{fig:z-MUV}). Facilities such as JWST, NOEMA, and ALMA, will be crucial for future characterization.

Finally, the results presented here are broadly consistent with the forecast of \citet{Barnett-EP5}, according to which $\sim 20$ quasars should exist at $z\gtrsim7$ in 3000\,deg$^2$ down to $\JE\sim 23$, while this work has already yielded 14 such quasars among 31 new discoveries. 
\replaced{Because our search and spectroscopic follow-up are not yet complete (particularly in the south, where fewer confirmation resources have been available), this comparison is provisional; nevertheless, it could be suggestive that some existing QLFs underpredict the number density at $z\gtrsim 7$ and/or at the faint end. We caution against overinterpreting this possible tension pending a formal analysis of the selection function and luminosity function.}
{ Since our search is incomplete, particularly in the south due to less follow-up confirmation resources, this comparison may imply that existing QLFs may underpredict the space density at $z\gtrsim 7$ and/or at the faint end. However, we caution against overinterpreting these statistics, pending a formal analysis of the selection function and luminosity function. }
Finally, as the \Euclid footprint expands and confirmations proceed smoothly, the first $z>8$ quasars are likely to be identified in the near future.

\section*{Data availability}

This work has made use of Euclid data from the European Space Agency (ESA) mission Euclid. The quasar cutouts are available at \url{https://doi.org/10.57780/esa-50769fd}. The discovery spectra are available from the corresponding author upon reasonable request.

\begin{acknowledgements}
%\AckERO  
D.~Yang acknowledges helpful discussions with members of the ENIGMA group at UC Santa Barbara and Leiden University, in particular Elia Pizzati, Silvia Onorato, and Diego Gonz\'alez for their valuable feedback. D.~Yang thanks Shane Bechtel for assistance during observations, and acknowledges the early contributions of Riccardo Nanni, which helped lay the foundation for this work.
D.~Yang and J.~F.~Hennawi acknowledge support from the European Research Council (ERC) under the European Union’s Horizon 2020 research and innovation program (grant agreement No 885301). 
J.~F.~Hennawi acknowledges support from NSF grant No. 2307180. 
F.~Guarneri and J.-T.~Schindler are supported by the Deutsche Forschungsgemeinschaft (DFG,
German Research Foundation) - Project number 518006966. 
F.~Wang acknowledges support from NSF award AST-2513040.
R.~Decarli acknowledges support from the INAF GO 2022 grant ``The birth of the giants: JWST sheds light on the build-up of quasars at cosmic dawn'', INAF Minigrant 2024 ``The interstellar medium at high redshift'', and by the PRIN MUR ``2022935STW'', RFF M4.C2.1.1, CUP J53D23001570006 and C53D23000950006. 
S.~E.~I.~Bosman is supported by the Deutsche Forschungsgemeinschaft (DFG) under Emmy Noether grant number BO 5771/1-1.
Y.~Harikane acknowledges support from the Japan Society for the Promotion of Science (JSPS) Grant-in-Aid for Scientific Research (24H00245), the JSPS Core-to-Core Program (JPJSCCA20210003), and the JSPS International Leading Research (22K21349).
J.~R.~Weaver acknowledges that support for this work was provided by The Brinson Foundation through a Brinson Prize Fellowship grant.
\AckEC  
% Keck
Some of the data presented herein were obtained at Keck Observatory, which is a private 501(c)3 non-profit organization operated as a scientific partnership among the California Institute of Technology, the University of California, and the National Aeronautics and Space Administration. The Observatory was made possible by the generous financial support of the W. M. Keck Foundation. The authors wish to recognise and acknowledge the very significant cultural role and reverence that the summit of Maunakea has always had within the Native Hawaiian community. We are most fortunate to have the opportunity to conduct observations from this mountain.
% Magellan
This paper includes data gathered with the 6.5 meter Magellan Telescopes located at Las Campanas Observatory, Chile. 
% LBT
This work uses LBT data from programmes: MPIA-2025A-007 and MPIA-2025B-004. The LBT is an international collaboration among institutions in the United States, Italy and Germany. LBT Corporation partners are: The University of Arizona on behalf of the Arizona university system; Istituto Nazionale di Astrofisica, Italy; LBT Beteiligungsgesellschaft, Germany, representing the Max-Planck Society, the Astrophysical Institute Potsdam, and Heidelberg University; The Ohio State University, and The Research Corporation, on behalf of The University of Notre Dame, University of Minnesota, and University of Virginia,
% Lofar
LOFAR data products were provided by the LOFAR Surveys Key Science project (LSKSP; https://lofar-surveys.org/) and were derived from observations with the International LOFAR Telescope (ILT). LOFAR (van Haarlem et al. 2013) is the Low Frequency Array designed and constructed by ASTRON. It has observing, data processing, and data storage facilities in several countries, which are owned by various parties (each with their own funding sources), and which are collectively operated by the ILT foundation under a joint scientific policy. The efforts of the LSKSP have benefited from funding from the European Research Council, NOVA, NWO, CNRS-INSU, the SURF Co-operative, the UK Science and Technology Funding Council and the Jülich Supercomputing Centre.
% WISHES and DES
We make use of $z$-band imaging from the Wide Imaging with Subaru HSC of the Euclid Sky (WISHES) program operated by the Subaru Telescope, which is operated by the National Astronomical Observatory of Japan.
We also make use of $z$-band imaging from the Dark Energy Survey (DES), based on observations obtained at the Cerro Tololo Inter-American Observatory, National Optical Astronomy Observatory, which is operated by AURA under a cooperative agreement with the National Science Foundation.
% software
This work made use of \texttt{NumPy} \citep{Harris2020}, \texttt{SciPy} \citep{Virtanen2020}, \texttt{Astropy} \citep{astropy:2013,astropy:2018,astropy:2022}, \texttt{PypeIt} \citep{Prochaska2020pypeit}, \texttt{Sculptor} \citep{Schindler2022sculptor}, \texttt{condXD} \citep{Kang2024}, \texttt{pysersic} \citep{Pasha2023pysersic}, \texttt{Matplotlib} \citep{hunter2007}, \texttt{corner.py} \citep{Foreman-Mackey2016}, \texttt{healpy} \citep{Zonca2020}.
\end{acknowledgements}

%
% Here comes the reference list, generated via bibtex from
% your bibfile my.bib and Euclid.bib. Please make sure that
% the same paper is not referenced twice, one from your my.bib
% file, and once from Euclid.bib.
%

\bibliography{aa58883-26} % add my.bib, containing your bibentry file 

\begin{thebibliography}{139}
\expandafter\ifx\csname natexlab\endcsname\relax\def\natexlab#1{#1}\fi

\bibitem[{{Akins} {et~al.}(2025){Akins}, {Casey}, {Lambrides}, {Allen}, {Andika}, {Brinch}, {Champagne}, {Cooper}, {Ding}, {Drakos}, {Faisst}, {Finkelstein}, {Franco}, {Fujimoto}, {Gentile}, {Gillman}, {Gozaliasl}, {Harish}, {Hayward}, {Hirschmann}, {Ilbert}, {Kartaltepe}, {Kocevski}, {Koekemoer}, {Kokorev}, {Liu}, {Long}, {McCracken}, {McKinney}, {Onoue}, {Paquereau}, {Renzini}, {Rhodes}, {Robertson}, {Shuntov}, {Silverman}, {Tanaka}, {Toft}, {Trakhtenbrot}, {Valentino}, \& {Zavala}}]{Akins2025}
{Akins}, H.~B., {Casey}, C.~M., {Lambrides}, E., {et~al.} 2025, \apj, 991, 37

\bibitem[{{Ananna} {et~al.}(2024){Ananna}, {Bogd{\'a}n}, {Kov{\'a}cs}, {Natarajan}, \& {Hickox}}]{Ananna2024}
{Ananna}, T.~T., {Bogd{\'a}n}, {\'A}., {Kov{\'a}cs}, O.~E., {Natarajan}, P., \& {Hickox}, R.~C. 2024, \apjl, 969, L18

\bibitem[{{Arnouts} \& {Ilbert}(2011)}]{LePhare}
{Arnouts}, S. \& {Ilbert}, O. 2011, {LePHARE: Photometric Analysis for Redshift Estimate}, Astrophysics Source Code Library, record ascl:1108.009

\bibitem[{{Astropy Collaboration} {et~al.}(2022){Astropy Collaboration}, {Price-Whelan}, {Lim}, {Earl}, {Starkman}, {Bradley}, {Shupe}, {Patil}, {Corrales}, {Brasseur}, {N{"o}the}, {Donath}, {Tollerud}, {Morris}, {Ginsburg}, {Vaher}, {Weaver}, {Tocknell}, {Jamieson}, {van Kerkwijk}, {Robitaille}, {Merry}, {Bachetti}, {G{"u}nther}, {Aldcroft}, {Alvarado-Montes}, {Archibald}, {B{'o}di}, {Bapat}, {Barentsen}, {Baz{'a}n}, {Biswas}, {Boquien}, {Burke}, {Cara}, {Cara}, {Conroy}, {Conseil}, {Craig}, {Cross}, {Cruz}, {D'Eugenio}, {Dencheva}, {Devillepoix}, {Dietrich}, {Eigenbrot}, {Erben}, {Ferreira}, {Foreman-Mackey}, {Fox}, {Freij}, {Garg}, {Geda}, {Glattly}, {Gondhalekar}, {Gordon}, {Grant}, {Greenfield}, {Groener}, {Guest}, {Gurovich}, {Handberg}, {Hart}, {Hatfield-Dodds}, {Homeier}, {Hosseinzadeh}, {Jenness}, {Jones}, {Joseph}, {Kalmbach}, {Karamehmetoglu}, {Ka{l}uszy{'n}ski}, {Kelley}, {Kern}, {Kerzendorf}, {Koch}, {Kulumani}, {Lee}, {Ly}, {Ma}, {MacBride}, {Maljaars}, {Muna}, {Murphy}, {Norman}, {O'Steen},
  {Oman}, {Pacifici}, {Pascual}, {Pascual-Granado}, {Patil}, {Perren}, {Pickering}, {Rastogi}, {Roulston}, {Ryan}, {Rykoff}, {Sabater}, {Sakurikar}, {Salgado}, {Sanghi}, {Saunders}, {Savchenko}, {Schwardt}, {Seifert-Eckert}, {Shih}, {Jain}, {Shukla}, {Sick}, {Simpson}, {Singanamalla}, {Singer}, {Singhal}, {Sinha}, {Sip{H{o}}cz}, {Spitler}, {Stansby}, {Streicher}, {{{S}}umak}, {Swinbank}, {Taranu}, {Tewary}, {Tremblay}, {Val-Borro}, {Van Kooten}, {Vasovi{'c}}, {Verma}, {de Miranda Cardoso}, {Williams}, {Wilson}, {Winkel}, {Wood-Vasey}, {Xue}, {Yoachim}, {Zhang}, {Zonca}, \& {Astropy Project Contributors}}]{astropy:2022}
{Astropy Collaboration}, {Price-Whelan}, A.~M., {Lim}, P.~L., {et~al.} 2022, \apj, 935, 167

\bibitem[{{Astropy Collaboration} {et~al.}(2018){Astropy Collaboration}, {Price-Whelan}, {Sip{\H{o}}cz}, {G{\"u}nther}, {Lim}, {Crawford}, {Conseil}, {Shupe}, {Craig}, {Dencheva}, {Ginsburg}, {Vand erPlas}, {Bradley}, {P{\'e}rez-Su{\'a}rez}, {de Val-Borro}, {Aldcroft}, {Cruz}, {Robitaille}, {Tollerud}, {Ardelean}, {Babej}, {Bach}, {Bachetti}, {Bakanov}, {Bamford}, {Barentsen}, {Barmby}, {Baumbach}, {Berry}, {Biscani}, {Boquien}, {Bostroem}, {Bouma}, {Brammer}, {Bray}, {Breytenbach}, {Buddelmeijer}, {Burke}, {Calderone}, {Cano Rodr{\'\i}guez}, {Cara}, {Cardoso}, {Cheedella}, {Copin}, {Corrales}, {Crichton}, {D'Avella}, {Deil}, {Depagne}, {Dietrich}, {Donath}, {Droettboom}, {Earl}, {Erben}, {Fabbro}, {Ferreira}, {Finethy}, {Fox}, {Garrison}, {Gibbons}, {Goldstein}, {Gommers}, {Greco}, {Greenfield}, {Groener}, {Grollier}, {Hagen}, {Hirst}, {Homeier}, {Horton}, {Hosseinzadeh}, {Hu}, {Hunkeler}, {Ivezi{\'c}}, {Jain}, {Jenness}, {Kanarek}, {Kendrew}, {Kern}, {Kerzendorf}, {Khvalko}, {King}, {Kirkby}, {Kulkarni},
  {Kumar}, {Lee}, {Lenz}, {Littlefair}, {Ma}, {Macleod}, {Mastropietro}, {McCully}, {Montagnac}, {Morris}, {Mueller}, {Mumford}, {Muna}, {Murphy}, {Nelson}, {Nguyen}, {Ninan}, {N{\"o}the}, {Ogaz}, {Oh}, {Parejko}, {Parley}, {Pascual}, {Patil}, {Patil}, {Plunkett}, {Prochaska}, {Rastogi}, {Reddy Janga}, {Sabater}, {Sakurikar}, {Seifert}, {Sherbert}, {Sherwood-Taylor}, {Shih}, {Sick}, {Silbiger}, {Singanamalla}, {Singer}, {Sladen}, {Sooley}, {Sornarajah}, {Streicher}, {Teuben}, {Thomas}, {Tremblay}, {Turner}, {Terr{\'o}n}, {van Kerkwijk}, {de la Vega}, {Watkins}, {Weaver}, {Whitmore}, {Woillez}, {Zabalza}, \& {Astropy Contributors}}]{astropy:2018}
{Astropy Collaboration}, {Price-Whelan}, A.~M., {Sip{\H{o}}cz}, B.~M., {et~al.} 2018, \aj, 156, 123

\bibitem[{{Astropy Collaboration} {et~al.}(2013){Astropy Collaboration}, {Robitaille}, {Tollerud}, {Greenfield}, {Droettboom}, {Bray}, {Aldcroft}, {Davis}, {Ginsburg}, {Price-Whelan}, {Kerzendorf}, {Conley}, {Crighton}, {Barbary}, {Muna}, {Ferguson}, {Grollier}, {Parikh}, {Nair}, {Unther}, {Deil}, {Woillez}, {Conseil}, {Kramer}, {Turner}, {Singer}, {Fox}, {Weaver}, {Zabalza}, {Edwards}, {Azalee Bostroem}, {Burke}, {Casey}, {Crawford}, {Dencheva}, {Ely}, {Jenness}, {Labrie}, {Lim}, {Pierfederici}, {Pontzen}, {Ptak}, {Refsdal}, {Servillat}, \& {Streicher}}]{astropy:2013}
{Astropy Collaboration}, {Robitaille}, T.~P., {Tollerud}, E.~J., {et~al.} 2013, \aap, 558, A33

\bibitem[{{Ba{\~n}ados} {et~al.}(2025{\natexlab{a}}){Ba{\~n}ados}, {Le Brun}, {Belladitta}, {et~al.}}]{Banados25}
{Ba{\~n}ados}, E., {Le Brun}, V., {Belladitta}, S., {et~al.} 2025{\natexlab{a}}, \mnras, 542, 1088

\bibitem[{{Ba{\~n}ados} {et~al.}(2025{\natexlab{b}}){Ba{\~n}ados}, {Momjian}, {Connor}, {Belladitta}, {Decarli}, {Mazzucchelli}, {Venemans}, {Walter}, {Wang}, {Xie}, {Barth}, {Eilers}, {Fan}, {Khusanova}, {Schindler}, {Stern}, {Yang}, {Andika}, {Carilli}, {Farina}, {Fabian}, {Hennawi}, {Pensabene}, \& {Rojas-Ruiz}}]{Banados2025}
{Ba{\~n}ados}, E., {Momjian}, E., {Connor}, T., {et~al.} 2025{\natexlab{b}}, Nature Astronomy, 9, 293

\bibitem[{{Ba{\~n}ados} {et~al.}(2018){Ba{\~n}ados}, {Venemans}, {Mazzucchelli}, {Farina}, {Walter}, {Wang}, {Decarli}, {Stern}, {Fan}, {Davies}, {Hennawi}, {Simcoe}, {Turner}, {Rix}, {Yang}, {Kelson}, {Rudie}, \& {Winters}}]{Banados2018}
{Ba{\~n}ados}, E., {Venemans}, B.~P., {Mazzucchelli}, C., {et~al.} 2018, \nat, 553, 473

\bibitem[{{Bertin}(2010)}]{Bertin2010swarp}
{Bertin}, E. 2010, {SWarp: Resampling and Co-adding FITS Images Together}, Astrophysics Source Code Library, record ascl:1010.068

\bibitem[{{Bertin}(2013)}]{Bertin2013psfex}
{Bertin}, E. 2013, {PSFEx: Point Spread Function Extractor}, Astrophysics Source Code Library, record ascl:1301.001

\bibitem[{{Bertin} \& {Arnouts}(1996)}]{Bertin1996sextractor}
{Bertin}, E. \& {Arnouts}, S. 1996, \aaps, 117, 393

\bibitem[{{Bosman} {et~al.}(2025){Bosman}, {{\'A}lvarez-M{\'a}rquez}, {Davies}, {Protu{\v{s}}ov{\'a}}, {Hennawi}, {Yang}, {Spina}, {Colina}, {Fan}, {{\"O}stlin}, {Walter}, {Wang}, {Ward}, {Alonso Herrero}, {Barth}, {Belladitta}, {Boogaard}, {Caputi}, {Connor}, {{\v{D}}urov{\v{c}}{\'\i}kov{\'a}}, {Eilers}, {Crespo G{\'o}mez}, {Hjorth}, {Jun}, {Langeroodi}, {Liu}, {Lupi}, {Mazzucchelli}, {Pye}, {Rinaldi}, {van der Werf}, \& {Volonteri}}]{Bosman2025}
{Bosman}, S. E.~I., {{\'A}lvarez-M{\'a}rquez}, J., {Davies}, F.~B., {et~al.} 2025, ApJ, arXiv:2511.02902, submitted

\bibitem[{{Bosman} {et~al.}(2022){Bosman}, {Davies}, {Becker}, {Keating}, {Davies}, {Zhu}, {Eilers}, {D'Odorico}, {Bian}, {Bischetti}, {Cristiani}, {Fan}, {Farina}, {Haehnelt}, {Hennawi}, {Kulkarni}, {Mesinger}, {Meyer}, {Onoue}, {Pallottini}, {Qin}, {Ryan-Weber}, {Schindler}, {Walter}, {Wang}, \& {Yang}}]{Bosman2022}
{Bosman}, S. E.~I., {Davies}, F.~B., {Becker}, G.~D., {et~al.} 2022, \mnras, 514, 55

\bibitem[{{Bouwens} {et~al.}(2025){Bouwens}, {Banados}, {Decarli}, {Hennawi}, {Yang}, {Algera}, {Aravena}, {Farina}, {Gloudemans}, {Hodge}, {Inami}, {Matthee}, {Meyer}, {Naidu}, {Oesch}, {Rottgering}, {Schouws}, {Smit}, {Stefanon}, {van der Werf}, {Venemans}, {Walter}, \& {Fudamoto}}]{Bouwens2025}
{Bouwens}, R.~J., {Banados}, E., {Decarli}, R., {et~al.} 2025, \aap, submitted, arXiv:2506.24128

\bibitem[{{Bouwens} {et~al.}(2022){Bouwens}, {Smit}, {Schouws}, {Stefanon}, {Bowler}, {Endsley}, {Gonzalez}, {Inami}, {Stark}, {Oesch}, {Hodge}, {Aravena}, {da Cunha}, {Dayal}, {de Looze}, {Ferrara}, {Fudamoto}, {Graziani}, {Li}, {Nanayakkara}, {Pallottini}, {Schneider}, {Sommovigo}, {Topping}, {van der Werf}, {Algera}, {Barrufet}, {Hygate}, {Labb{\'e}}, {Riechers}, \& {Witstok}}]{Bouwens2022}
{Bouwens}, R.~J., {Smit}, R., {Schouws}, S., {et~al.} 2022, \apj, 931, 160

\bibitem[{{Bovy} {et~al.}(2011){Bovy}, {Hogg}, \& {Roweis}}]{Bovy2011}
{Bovy}, J., {Hogg}, D.~W., \& {Roweis}, S.~T. 2011, Annals of Applied Statistics, 5, 1657

\bibitem[{{Bradley} {et~al.}(2016){Bradley}, {Sipocz}, {Robitaille}, {Tollerud}, {Deil}, {Vin{\'\i}cius}, {Barbary}, {G{\"u}nther}, {Bostroem}, {Droettboom}, {Bray}, {Bratholm}, {Pickering}, {Craig}, {Pascual}, {Greco}, {Donath}, {Kerzendorf}, {Littlefair}, {Barentsen}, {D'Eugenio}, \& {Weaver}}]{Bradley2016photutils}
{Bradley}, L., {Sipocz}, B., {Robitaille}, T., {et~al.} 2016, {Photutils: Photometry tools}, Astrophysics Source Code Library, record ascl:1609.011

\bibitem[{{Brammer} {et~al.}(2008){Brammer}, {van Dokkum}, \& {Coppi}}]{EAZY}
{Brammer}, G.~B., {van Dokkum}, P.~G., \& {Coppi}, P. 2008, \apj, 686, 1503

\bibitem[{Calderone {et~al.}(2024)Calderone, Guarneri, Porru, Cristiani, Grazian, Nicastro, Bischetti, Boutsia, Cupani, D'Odorico, Feruglio, \& Fontanot}]{calderone_boost_2024}
Calderone, G., Guarneri, F., Porru, M., {et~al.} 2024, A\&A, 683, A34

\bibitem[{Chen \& Guestrin(2016)}]{chen_xgboost_2016}
Chen, T. \& Guestrin, C. 2016, in Proceedings of the 22nd {ACM} {SIGKDD} {International} {Conference} on {Knowledge} {Discovery} and {Data} {Mining}, {KDD} '16 (New York, NY, USA: Association for Computing Machinery), 785--794

\bibitem[{{Coatman} {et~al.}(2016){Coatman}, {Hewett}, {Banerji}, \& {Richards}}]{Coatman2016}
{Coatman}, L., {Hewett}, P.~C., {Banerji}, M., \& {Richards}, G.~T. 2016, \mnras, 461, 647

\bibitem[{{Dark Energy Survey Collaboration} {et~al.}(2016){Dark Energy Survey Collaboration}, {Abbott}, {Abdalla}, {Aleksi{\'c}}, {Allam}, {Amara}, {Bacon}, {Balbinot}, {Banerji}, {Bechtol}, {Benoit-L{\'e}vy}, {Bernstein}, {Bertin}, {Blazek}, {Bonnett}, {Bridle}, {Brooks}, {Brunner}, {Buckley-Geer}, {Burke}, {Caminha}, {Capozzi}, {Carlsen}, {Carnero-Rosell}, {Carollo}, {Carrasco-Kind}, {Carretero}, {Castander}, {Clerkin}, {Collett}, {Conselice}, {Crocce}, {Cunha}, {D'Andrea}, {da Costa}, {Davis}, {Desai}, {Diehl}, {Dietrich}, {Dodelson}, {Doel}, {Drlica-Wagner}, {Estrada}, {Etherington}, {Evrard}, {Fabbri}, {Finley}, {Flaugher}, {Foley}, {Fosalba}, {Frieman}, {Garc{\'\i}a-Bellido}, {Gaztanaga}, {Gerdes}, {Giannantonio}, {Goldstein}, {Gruen}, {Gruendl}, {Guarnieri}, {Gutierrez}, {Hartley}, {Honscheid}, {Jain}, {James}, {Jeltema}, {Jouvel}, {Kessler}, {King}, {Kirk}, {Kron}, {Kuehn}, {Kuropatkin}, {Lahav}, {Li}, {Lima}, {Lin}, {Maia}, {Makler}, {Manera}, {Maraston}, {Marshall}, {Martini}, {McMahon},
  {Melchior}, {Merson}, {Miller}, {Miquel}, {Mohr}, {Morice-Atkinson}, {Naidoo}, {Neilsen}, {Nichol}, {Nord}, {Ogando}, {Ostrovski}, {Palmese}, {Papadopoulos}, {Peiris}, {Peoples}, {Percival}, {Plazas}, {Reed}, {Refregier}, {Romer}, {Roodman}, {Ross}, {Rozo}, {Rykoff}, {Sadeh}, {Sako}, {S{\'a}nchez}, {Sanchez}, {Santiago}, {Scarpine}, {Schubnell}, {Sevilla-Noarbe}, {Sheldon}, {Smith}, {Smith}, {Soares-Santos}, {Sobreira}, {Soumagnac}, {Suchyta}, {Sullivan}, {Swanson}, {Tarle}, {Thaler}, {Thomas}, {Thomas}, {Tucker}, {Vieira}, {Vikram}, {Walker}, {Wechsler}, {Weller}, {Wester}, {Whiteway}, {Wilcox}, {Yanny}, {Zhang}, \& {Zuntz}}]{DES2016}
{Dark Energy Survey Collaboration}, {Abbott}, T., {Abdalla}, F.~B., {et~al.} 2016, \mnras, 460, 1270

\bibitem[{{Davies} {et~al.}(2026){Davies}, {Bosman}, {D'Odorico}, {Campo}, {Mesinger}, {Qin}, {Becker}, {Ba{\~n}ados}, {Chen}, {Cristiani}, {Fan}, {Gallerani}, {Haehnelt}, {Keating}, {Lai}, {Ryan-Weber}, {Wang}, {Yang}, \& {Zhu}}]{Davies2026}
{Davies}, F.~B., {Bosman}, S. E.~I., {D'Odorico}, V., {et~al.} 2026, \mnras, 545, staf1862

\bibitem[{{Davies} {et~al.}(2018){Davies}, {Hennawi}, {Ba{\~n}ados}, {Luki{\'c}}, {Decarli}, {Fan}, {Farina}, {Mazzucchelli}, {Rix}, {Venemans}, {Walter}, {Wang}, \& {Yang}}]{Davies2018}
{Davies}, F.~B., {Hennawi}, J.~F., {Ba{\~n}ados}, E., {et~al.} 2018, \apj, 864, 142

\bibitem[{{Dawson} {et~al.}(2016){Dawson}, {Kneib}, {Percival}, {Alam}, {Albareti}, {Anderson}, {Armengaud}, {Aubourg}, {Bailey}, {Bautista}, {Berlind}, {Bershady}, {Beutler}, {Bizyaev}, {Blanton}, {Blomqvist}, {Bolton}, {Bovy}, {Brandt}, {Brinkmann}, {Brownstein}, {Burtin}, {Busca}, {Cai}, {Chuang}, {Clerc}, {Comparat}, {Cope}, {Croft}, {Cruz-Gonzalez}, {da Costa}, {Cousinou}, {Darling}, {de la Macorra}, {de la Torre}, {Delubac}, {du Mas des Bourboux}, {Dwelly}, {Ealet}, {Eisenstein}, {Eracleous}, {Escoffier}, {Fan}, {Finoguenov}, {Font-Ribera}, {Frinchaboy}, {Gaulme}, {Georgakakis}, {Green}, {Guo}, {Guy}, {Ho}, {Holder}, {Huehnerhoff}, {Hutchinson}, {Jing}, {Jullo}, {Kamble}, {Kinemuchi}, {Kirkby}, {Kitaura}, {Klaene}, {Laher}, {Lang}, {Laurent}, {Le Goff}, {Li}, {Liang}, {Lima}, {Lin}, {Lin}, {Lin}, {Long}, {Lundgren}, {MacDonald}, {Geimba Maia}, {Malanushenko}, {Malanushenko}, {Mariappan}, {McBride}, {McGreer}, {M{\'e}nard}, {Merloni}, {Meza}, {Montero-Dorta}, {Muna}, {Myers}, {Nandra}, {Naugle},
  {Newman}, {Noterdaeme}, {Nugent}, {Ogando}, {Olmstead}, {Oravetz}, {Oravetz}, {Padmanabhan}, {Palanque-Delabrouille}, {Pan}, {Parejko}, {P{\^a}ris}, {Peacock}, {Petitjean}, {Pieri}, {Pisani}, {Prada}, {Prakash}, {Raichoor}, {Reid}, {Rich}, {Ridl}, {Rodriguez-Torres}, {Carnero Rosell}, {Ross}, {Rossi}, {Ruan}, {Salvato}, {Sayres}, {Schneider}, {Schlegel}, {Seljak}, {Seo}, {Sesar}, {Shandera}, {Shu}, {Slosar}, {Sobreira}, {Streblyanska}, {Suzuki}, {Taylor}, {Tao}, {Tinker}, {Tojeiro}, {Vargas-Maga{\~n}a}, {Wang}, {Weaver}, {Weinberg}, {White}, {Wood-Vasey}, {Yeche}, {Zhai}, {Zhao}, {Zhao}, {Zheng}, {Ben Zhu}, \& {Zou}}]{Dawson2016}
{Dawson}, K.~S., {Kneib}, J.-P., {Percival}, W.~J., {et~al.} 2016, \aj, 151, 44

\bibitem[{{Dawson} {et~al.}(2013){Dawson}, {Schlegel}, {Ahn}, {Anderson}, {Aubourg}, {Bailey}, {Barkhouser}, {Bautista}, {Beifiori}, {Berlind}, {Bhardwaj}, {Bizyaev}, {Blake}, {Blanton}, {Blomqvist}, {Bolton}, {Borde}, {Bovy}, {Brandt}, {Brewington}, {Brinkmann}, {Brown}, {Brownstein}, {Bundy}, {Busca}, {Carithers}, {Carnero}, {Carr}, {Chen}, {Comparat}, {Connolly}, {Cope}, {Croft}, {Cuesta}, {da Costa}, {Davenport}, {Delubac}, {de Putter}, {Dhital}, {Ealet}, {Ebelke}, {Eisenstein}, {Escoffier}, {Fan}, {Filiz Ak}, {Finley}, {Font-Ribera}, {G{\'e}nova-Santos}, {Gunn}, {Guo}, {Haggard}, {Hall}, {Hamilton}, {Harris}, {Harris}, {Ho}, {Hogg}, {Holder}, {Honscheid}, {Huehnerhoff}, {Jordan}, {Jordan}, {Kauffmann}, {Kazin}, {Kirkby}, {Klaene}, {Kneib}, {Le Goff}, {Lee}, {Long}, {Loomis}, {Lundgren}, {Lupton}, {Maia}, {Makler}, {Malanushenko}, {Malanushenko}, {Mandelbaum}, {Manera}, {Maraston}, {Margala}, {Masters}, {McBride}, {McDonald}, {McGreer}, {McMahon}, {Mena}, {Miralda-Escud{\'e}}, {Montero-Dorta},
  {Montesano}, {Muna}, {Myers}, {Naugle}, {Nichol}, {Noterdaeme}, {Nuza}, {Olmstead}, {Oravetz}, {Oravetz}, {Owen}, {Padmanabhan}, {Palanque-Delabrouille}, {Pan}, {Parejko}, {P{\^a}ris}, {Percival}, {P{\'e}rez-Fournon}, {P{\'e}rez-R{\`a}fols}, {Petitjean}, {Pfaffenberger}, {Pforr}, {Pieri}, {Prada}, {Price-Whelan}, {Raddick}, {Rebolo}, {Rich}, {Richards}, {Rockosi}, {Roe}, {Ross}, {Ross}, {Rossi}, {Rubi{\~n}o-Martin}, {Samushia}, {S{\'a}nchez}, {Sayres}, {Schmidt}, {Schneider}, {Sc{\'o}ccola}, {Seo}, {Shelden}, {Sheldon}, {Shen}, {Shu}, {Slosar}, {Smee}, {Snedden}, {Stauffer}, {Steele}, {Strauss}, {Streblyanska}, {Suzuki}, {Swanson}, {Tal}, {Tanaka}, {Thomas}, {Tinker}, {Tojeiro}, {Tremonti}, {Vargas Maga{\~n}a}, {Verde}, {Viel}, {Wake}, {Watson}, {Weaver}, {Weinberg}, {Weiner}, {West}, {White}, {Wood-Vasey}, {Yeche}, {Zehavi}, {Zhao}, \& {Zheng}}]{Dawson2013}
{Dawson}, K.~S., {Schlegel}, D.~J., {Ahn}, C.~P., {et~al.} 2013, \aj, 145, 10

\bibitem[{{Dye} {et~al.}(2018){Dye}, {Lawrence}, {Read}, {Fan}, {Kerr}, {Varricatt}, {Furnell}, {Edge}, {Irwin}, {Hambly}, {Lucas}, {Almaini}, {Chambers}, {Green}, {Hewett}, {Liu}, {McGreer}, {Best}, {Zhang}, {Sutorius}, {Froebrich}, {Magnier}, {Hasinger}, {Lederer}, {Bold}, \& {Tedds}}]{Dye2018}
{Dye}, S., {Lawrence}, A., {Read}, M.~A., {et~al.} 2018, \mnras, 473, 5113

\bibitem[{{Edge} {et~al.}(2013){Edge}, {Sutherland}, {Kuijken}, {Driver}, {McMahon}, {Eales}, \& {Emerson}}]{Edge2013Viking}
{Edge}, A., {Sutherland}, W., {Kuijken}, K., {et~al.} 2013, The Messenger, 154, 32

\bibitem[{{Eilers} {et~al.}(2024){Eilers}, {Mackenzie}, {Pizzati}, {Matthee}, {Hennawi}, {Zhang}, {Bordoloi}, {Kashino}, {Lilly}, {Naidu}, {Simcoe}, {Yue}, {Frenk}, {Helly}, {Schaller}, \& {Schaye}}]{Eilers2024}
{Eilers}, A.-C., {Mackenzie}, R., {Pizzati}, E., {et~al.} 2024, \apj, 974, 275

\bibitem[{{Euclid Collaboration: Barnett} {et~al.}(2019){Euclid Collaboration: Barnett}, {Warren}, {Mortlock}, {et~al.}}]{Barnett-EP5}
{Euclid Collaboration: Barnett}, R., {Warren}, S.~J., {Mortlock}, D.~J., {et~al.} 2019, \aap, 631, A85

\bibitem[{{Euclid Collaboration: Cropper} {et~al.}(2025){Euclid Collaboration: Cropper}, {Al-Bahlawan}, {Amiaux}, {et~al.}}]{EuclidSkyVIS}
{Euclid Collaboration: Cropper}, M., {Al-Bahlawan}, A., {Amiaux}, J., {et~al.} 2025, A\&A, 697, A2

\bibitem[{{Euclid Collaboration: Jahnke} {et~al.}(2025){Euclid Collaboration: Jahnke}, {Gillard}, {Schirmer}, {et~al.}}]{EuclidSkyNISP}
{Euclid Collaboration: Jahnke}, K., {Gillard}, W., {Schirmer}, M., {et~al.} 2025, A\&A, 697, A3

\bibitem[{{Euclid Collaboration: McCracken} {et~al.}(2025){Euclid Collaboration: McCracken}, {Benson}, {Dolding}, {et~al.}}]{Q1-TP002}
{Euclid Collaboration: McCracken}, H.~J., {Benson}, K., {Dolding}, C., {et~al.} 2025, A\&A, in press (Euclid Q1 SI), \url{https://doi.org/10.1051/0004-6361/202554594}, arXiv:2503.15303

\bibitem[{{Euclid Collaboration: Mellier} {et~al.}(2025){Euclid Collaboration: Mellier}, {Abdurro'uf}, {Acevedo~Barroso}, {et~al.}}]{EuclidSkyOverview}
{Euclid Collaboration: Mellier}, Y., {Abdurro'uf}, {Acevedo~Barroso}, J., {et~al.} 2025, A\&A, 697, A1

\bibitem[{{Euclid Collaboration: Polenta} {et~al.}(2025){Euclid Collaboration: Polenta}, {Frailis}, {Alavi}, {et~al.}}]{Q1-TP003}
{Euclid Collaboration: Polenta}, G., {Frailis}, M., {Alavi}, A., {et~al.} 2025, A\&A, in press (Euclid Q1 SI), \url{https://doi.org/10.1051/0004-6361/202554657}, arXiv:2503.15304

\bibitem[{{Euclid Collaboration: Romelli} {et~al.}(2025){Euclid Collaboration: Romelli}, {K\"ummel}, {Dole}, {et~al.}}]{Q1-TP004}
{Euclid Collaboration: Romelli}, E., {K\"ummel}, M., {Dole}, H., {et~al.} 2025, A\&A, in press (Euclid Q1 SI), \url{https://doi.org/10.1051/0004-6361/202554586}, arXiv:2503.15305

\bibitem[{{Euclid Collaboration: Scaramella} {et~al.}(2022){Euclid Collaboration: Scaramella}, {Amiaux}, {Mellier}, {et~al.}}]{Scaramella-EP1}
{Euclid Collaboration: Scaramella}, R., {Amiaux}, J., {Mellier}, Y., {et~al.} 2022, \aap, 662, A112

\bibitem[{{Euclid Collaboration: Schirmer} {et~al.}(2022){Euclid Collaboration: Schirmer}, {Jahnke}, {Seidel}, {et~al.}}]{Schirmer-EP18}
{Euclid Collaboration: Schirmer}, M., {Jahnke}, K., {Seidel}, G., {et~al.} 2022, \aap, 662, A92

\bibitem[{{Ezziati} {et~al.}(2025){Ezziati}, {Pello}, {Cuby}, {Pudlo}, {Dup{\'e}}, {Lambert}, {Cuillandre}, {Ilbert}, {de la Torre}, {Arnouts}, {Jullo}, \& {Yang}}]{Ezziati2025}
{Ezziati}, M., {Pello}, R., {Cuby}, J.~G., {et~al.} 2025, \aap, 701, A282

\bibitem[{{Fan} {et~al.}(2023){Fan}, {Ba{\~n}ados}, \& {Simcoe}}]{Fan2023}
{Fan}, X., {Ba{\~n}ados}, E., \& {Simcoe}, R.~A. 2023, \araa, 61, 373

\bibitem[{{Fan} {et~al.}(2000){Fan}, {White}, {Davis}, {Becker}, {Strauss}, {Haiman}, {Schneider}, {Gregg}, {Gunn}, {Knapp}, {Lupton}, {Anderson}, {Anderson}, {Annis}, {Bahcall}, {Boroski}, {Brunner}, {Chen}, {Connolly}, {Csabai}, {Doi}, {Fukugita}, {Hennessy}, {Hindsley}, {Ichikawa}, {Ivezi{\'c}}, {Loveday}, {Meiksin}, {McKay}, {Munn}, {Newberg}, {Nichol}, {Okamura}, {Pier}, {Sekiguchi}, {Shimasaku}, {Stoughton}, {Szalay}, {Szokoly}, {Thakar}, {Vogeley}, \& {York}}]{Fan2000}
{Fan}, X., {White}, R.~L., {Davis}, M., {et~al.} 2000, \aj, 120, 1167

\bibitem[{{Farina} {et~al.}(2022){Farina}, {Schindler}, {Walter}, {Ba{\~n}ados}, {Davies}, {Decarli}, {Eilers}, {Fan}, {Hennawi}, {Mazzucchelli}, {Meyer}, {Trakhtenbrot}, {Volonteri}, {Wang}, {Worseck}, {Yang}, {Gutcke}, {Venemans}, {Bosman}, {Costa}, {De Rosa}, {Drake}, \& {Onoue}}]{Farina2022}
{Farina}, E.~P., {Schindler}, J.-T., {Walter}, F., {et~al.} 2022, \apj, 941, 106

\bibitem[{{Finkelstein} \& {Bagley}(2022)}]{Finkelstein2022}
{Finkelstein}, S.~L. \& {Bagley}, M.~B. 2022, \apj, 938, 25

\bibitem[{{Foreman-Mackey}(2016)}]{Foreman-Mackey2016}
{Foreman-Mackey}, D. 2016, The Journal of Open Source Software, 1, 24

\bibitem[{{Fu} {et~al.}(2025){Fu}, {Wu}, {Bouwens}, {Caputi}, {Pang}, {Zhu}, {Yang}, {Qin}, {Wang}, {Wolf}, {Li}, {Joshi}, {Zhang}, {Huo}, \& {Ai}}]{Fu2025:2503.14141v2}
{Fu}, Y., {Wu}, X.-B., {Bouwens}, R.~J., {et~al.} 2025, \apjs, 279, 54

\bibitem[{{Gloudemans} {et~al.}(2025){Gloudemans}, {Duncan}, {Eilers}, {Farina}, {Harikane}, {Inayoshi}, {Lambrides}, \& {Vardoulaki}}]{Gloudemans2025}
{Gloudemans}, A.~J., {Duncan}, K.~J., {Eilers}, A.-C., {et~al.} 2025, \apj, 986, 130

\bibitem[{{Gloudemans} {et~al.}(2022){Gloudemans}, {Duncan}, {Saxena}, {Harikane}, {Hill}, {Zeimann}, {R{\"o}ttgering}, {Yang}, {Best}, {Ba{\~n}ados}, {Drabent}, {Hardcastle}, {Hennawi}, {Lansbury}, {Magliocchetti}, {Miley}, {Nanni}, {Shimwell}, {Smith}, {Venemans}, \& {Wagenveld}}]{Gloudemans2022}
{Gloudemans}, A.~J., {Duncan}, K.~J., {Saxena}, A., {et~al.} 2022, \aap, 668, A27

\bibitem[{{Greene} {et~al.}(2024){Greene}, {Labbe}, {Goulding}, {Furtak}, {Chemerynska}, {Kokorev}, {Dayal}, {Volonteri}, {Williams}, {Wang}, {Setton}, {Burgasser}, {Bezanson}, {Atek}, {Brammer}, {Cutler}, {Feldmann}, {Fujimoto}, {Glazebrook}, {de Graaff}, {Khullar}, {Leja}, {Marchesini}, {Maseda}, {Matthee}, {Miller}, {Naidu}, {Nanayakkara}, {Oesch}, {Pan}, {Papovich}, {Price}, {van Dokkum}, {Weaver}, {Whitaker}, \& {Zitrin}}]{Greene2024}
{Greene}, J.~E., {Labbe}, I., {Goulding}, A.~D., {et~al.} 2024, \apj, 964, 39

\bibitem[{{Greig} {et~al.}(2024){Greig}, {Mesinger}, {Ba{\~n}ados}, {Becker}, {Bosman}, {Chen}, {Davies}, {D'Odorico}, {Eilers}, {Gallerani}, {Haehnelt}, {Keating}, {Lai}, {Qin}, {Ryan-Weber}, {Satyavolu}, {Wang}, {Yang}, \& {Zhu}}]{Greig2024}
{Greig}, B., {Mesinger}, A., {Ba{\~n}ados}, E., {et~al.} 2024, \mnras, 530, 3208

\bibitem[{{Guarneri} {et~al.}(2026){Guarneri}, {Schindler}, {Meyer}, {Yang}, {Hennawi}, {Lucie-Smith}, {Bosman}, \& {Davies}}]{Guarneri2025}
{Guarneri}, F., {Schindler}, J.~T., {Meyer}, R.~A., {et~al.} 2026, \aap, in press, arXiv:2510.23206

\bibitem[{{Gunn} \& {Peterson}(1965)}]{Gunn1965}
{Gunn}, J.~E. \& {Peterson}, B.~A. 1965, \apj, 142, 1633

\bibitem[{{Gwyn} {et~al.}(2025){Gwyn}, {McConnachie}, {Cuillandre}, {Chambers}, {Magnier}, {de Boer}, {Hudson}, {Oguri}, {Furusawa}, {Hildebrandt}, {Carlberg}, {Ellison}, {Furusawa}, {Gavazzi}, {Ibata}, {Mellier}, {Osato}, {Aussel}, {Baumont}, {Bayer}, {Boulade}, {C{\^o}t{\'e}}, {Chemaly}, {Daley}, {Duc}, {Durret}, {Ellien}, {Fabbro}, {Ferreira}, {Fitriana}, {Le Floc'h}, {Fudamoto}, {Gao}, {Goh}, {Goto}, {Guerrini}, {Guinot}, {H{\'e}nault-Brunet}, {Hammer}, {Harikane}, {Hayashi}, {Heesters}, {Ichikawa}, {Kilbinger}, {Kuzma}, {Li}, {Liaudat}, {Lin}, {M{\"u}ller}, {Martin}, {Matsuoka}, {Medina}, {Miyatake}, {Miyazaki}, {Mpetha}, {Nagao}, {Navarro}, {Niwano}, {Ogami}, {Okabe}, {Onoue}, {Paek}, {Parker}, {Patton}, {Peters}, {Prunet}, {S{\'a}nchez-Janssen}, {Schultheis}, {Sestito}, {Smith}, {Starck}, {Starkenburg}, {Stone}, {Storfer}, {Suzuki}, {Erben}, {Taibi}, {Thomas}, {Toba}, {Uchiyama}, {Valls-Gabaud}, {Venn}, {Van Waerbeke}, {Wainscoat}, {Wilkinson}, {Wittje}, {Yoshida}, {Zhang}, \& {Zhong}}]{Gwyn2025}
{Gwyn}, S., {McConnachie}, A.~W., {Cuillandre}, J.-C., {et~al.} 2025, \aj, 170, 324

\bibitem[{{Harikane} {et~al.}(2025){Harikane}, {Inoue}, {Ellis}, {Ouchi}, {Nakazato}, {Yoshida}, {Ono}, {Sun}, {Sato}, {Ferrami}, {Fujimoto}, {Kashikawa}, {McLeod}, {P{\'e}rez-Gonz{\'a}lez}, {Sawicki}, {Sugahara}, {Xu}, {Yamanaka}, {Carnall}, {Cullen}, {Dunlop}, {Egami}, {Grogin}, {Isobe}, {Koekemoer}, {Laporte}, {Lee}, {Magee}, {Matsuo}, {Matsuoka}, {Mawatari}, {Nakajima}, {Nakane}, {Tamura}, {Umeda}, \& {Yanagisawa}}]{Harikane2025}
{Harikane}, Y., {Inoue}, A.~K., {Ellis}, R.~S., {et~al.} 2025, \apj, 980, 138

\bibitem[{{Harikane} {et~al.}(2023){Harikane}, {Zhang}, {Nakajima}, {Ouchi}, {Isobe}, {Ono}, {Hatano}, {Xu}, \& {Umeda}}]{Harikane2023}
{Harikane}, Y., {Zhang}, Y., {Nakajima}, K., {et~al.} 2023, \apj, 959, 39

\bibitem[{{Harris} {et~al.}(2020){Harris}, {Millman}, {van der Walt}, {Gommers}, {Virtanen}, {Cournapeau}, {Wieser}, {Taylor}, {Berg}, {Smith}, {Kern}, {Picus}, {Hoyer}, {van Kerkwijk}, {Brett}, {Haldane}, {del R{\'\i}o}, {Wiebe}, {Peterson}, {G{\'e}rard-Marchant}, {Sheppard}, {Reddy}, {Weckesser}, {Abbasi}, {Gohlke}, \& {Oliphant}}]{Harris2020}
{Harris}, C.~R., {Millman}, K.~J., {van der Walt}, S.~J., {et~al.} 2020, \nat, 585, 357

\bibitem[{Hennawi {et~al.}(2025)Hennawi, Kist, Davies, \& Tamanas}]{Hennawi2025}
Hennawi, J.~F., Kist, T., Davies, F.~B., \& Tamanas, J. 2025, MNRAS, 539, 2621

\bibitem[{{Hunter}(2007)}]{hunter2007}
{Hunter}, J.~D. 2007, Computing in Science and Engineering, 9, 90

\bibitem[{{Inayoshi} {et~al.}(2020){Inayoshi}, {Visbal}, \& {Haiman}}]{Inayoshi2020}
{Inayoshi}, K., {Visbal}, E., \& {Haiman}, Z. 2020, \araa, 58, 27

\bibitem[{{Ishimoto} {et~al.}(2020){Ishimoto}, {Kashikawa}, {Onoue}, {Matsuoka}, {Izumi}, {Strauss}, {Fujimoto}, {Imanishi}, {Ito}, {Iwasawa}, {Kawaguchi}, {Lee}, {Liang}, {Lu}, {Momose}, {Toba}, \& {Uchiyama}}]{Ishimoto2020}
{Ishimoto}, R., {Kashikawa}, N., {Onoue}, M., {et~al.} 2020, \apj, 903, 60

\bibitem[{{Izumi} {et~al.}(2019){Izumi}, {Onoue}, {Matsuoka}, {Nagao}, {Strauss}, {Imanishi}, {Kashikawa}, {Fujimoto}, {Kohno}, {Toba}, {Umehata}, {Goto}, {Ueda}, {Shirakata}, {Silverman}, {Greene}, {Harikane}, {Hashimoto}, {Ikarashi}, {Iono}, {Iwasawa}, {Lee}, {Minezaki}, {Nakanishi}, {Tamura}, {Tang}, \& {Taniguchi}}]{Izumi2019}
{Izumi}, T., {Onoue}, M., {Matsuoka}, Y., {et~al.} 2019, \pasj, 71, 111

\bibitem[{{Jin} {et~al.}(2023){Jin}, {Yang}, {Fan}, {Wang}, {Ba{\~n}ados}, {Bian}, {Davies}, {Eilers}, {Farina}, {Hennawi}, {Pacucci}, {Venemans}, \& {Walter}}]{Jin2023}
{Jin}, X., {Yang}, J., {Fan}, X., {et~al.} 2023, \apj, 942, 59

\bibitem[{{Kang} {et~al.}(2025){Kang}, {Hennawi}, {Schindler}, {Tamanas}, \& {Nanni}}]{Kang2024}
{Kang}, Y., {Hennawi}, J.~F., {Schindler}, J.-T., {Tamanas}, J., \& {Nanni}, R. 2025, \mnras, 541, 2815

\bibitem[{{Kist} {et~al.}(2025){Kist}, {Hennawi}, \& {Davies}}]{Kist2025}
{Kist}, T., {Hennawi}, J.~F., \& {Davies}, F.~B. 2025, \mnras, 538, 2704

\bibitem[{{Kokorev} {et~al.}(2023){Kokorev}, {Fujimoto}, {Labbe}, {Greene}, {Bezanson}, {Dayal}, {Nelson}, {Atek}, {Brammer}, {Caputi}, {Chemerynska}, {Cutler}, {Feldmann}, {Fudamoto}, {Furtak}, {Goulding}, {de Graaff}, {Leja}, {Marchesini}, {Miller}, {Nanayakkara}, {Oesch}, {Pan}, {Price}, {Setton}, {Smit}, {Stefanon}, {Wang}, {Weaver}, {Whitaker}, {Williams}, \& {Zitrin}}]{Kokorev2023}
{Kokorev}, V., {Fujimoto}, S., {Labbe}, I., {et~al.} 2023, \apjl, 957, L7

\bibitem[{{Kokubo} \& {Harikane}(2025)}]{Kokubo2024}
{Kokubo}, M. \& {Harikane}, Y. 2025, \apj, 995, 24

\bibitem[{{Kormendy} \& {Ho}(2013)}]{Kormendy2013}
{Kormendy}, J. \& {Ho}, L.~C. 2013, \araa, 51, 511

\bibitem[{{Larson} {et~al.}(2023){Larson}, {Finkelstein}, {Kocevski}, {Hutchison}, {Trump}, {Arrabal Haro}, {Bromm}, {Cleri}, {Dickinson}, {Fujimoto}, {Kartaltepe}, {Koekemoer}, {Papovich}, {Pirzkal}, {Tacchella}, {Zavala}, {Bagley}, {Behroozi}, {Champagne}, {Cole}, {Jung}, {Morales}, {Yang}, {Zhang}, {Zitrin}, {Amor{\'\i}n}, {Burgarella}, {Casey}, {Ch{\'a}vez Ortiz}, {Cox}, {Chworowsky}, {Fontana}, {Gawiser}, {Grazian}, {Grogin}, {Harish}, {Hathi}, {Hirschmann}, {Holwerda}, {Juneau}, {Leung}, {Lucas}, {McGrath}, {P{\'e}rez-Gonz{\'a}lez}, {Rigby}, {Seill{\'e}}, {Simons}, {de La Vega}, {Weiner}, {Wilkins}, {Yung}, \& {Ceers Team}}]{Larson2023}
{Larson}, R.~L., {Finkelstein}, S.~L., {Kocevski}, D.~D., {et~al.} 2023, \apjl, 953, L29

\bibitem[{{Lauer} {et~al.}(2007){Lauer}, {Tremaine}, {Richstone}, \& {Faber}}]{Lauer2007}
{Lauer}, T.~R., {Tremaine}, S., {Richstone}, D., \& {Faber}, S.~M. 2007, \apj, 670, 249

\bibitem[{{Lawrence} {et~al.}(2007){Lawrence}, {Warren}, {Almaini}, {Edge}, {Hambly}, {Jameson}, {Lucas}, {Casali}, {Adamson}, {Dye}, {Emerson}, {Foucaud}, {Hewett}, {Hirst}, {Hodgkin}, {Irwin}, {Lodieu}, {McMahon}, {Simpson}, {Smail}, {Mortlock}, \& {Folger}}]{Lawrence2007}
{Lawrence}, A., {Warren}, S.~J., {Almaini}, O., {et~al.} 2007, \mnras, 379, 1599

\bibitem[{{Li} {et~al.}(2022){Li}, {Silverman}, {Izumi}, {He}, {Akiyama}, {Inayoshi}, {Matsuoka}, {Onoue}, \& {Toba}}]{Li2022}
{Li}, J., {Silverman}, J.~D., {Izumi}, T., {et~al.} 2022, \apjl, 931, L11

\bibitem[{{Lin} {et~al.}(2026){Lin}, {Fan}, {Wang}, {Sun}, {Champagne}, {Egami}, {Kakiichi}, {Lyu}, {Tee}, {Yang}, {Bian}, {Bosman}, {Cai}, {Casey}, {Decarli}, {Faisst}, {Finkelstein}, {Fujimoto}, {Harish}, {Ilbert}, {Inoue}, {Jin}, {Kartaltepe}, {Kocevski}, {Li}, {Liu}, {Liu}, {Schindler}, {Shuntov}, {Tanaka}, {Vestergaard}, {Wu}, {Zhang}, \& {Zhang}}]{Lin2025}
{Lin}, X., {Fan}, X., {Wang}, F., {et~al.} 2026, \apj, 996, 93

\bibitem[{{Lin} {et~al.}(2024){Lin}, {Wang}, {Fan}, {Cai}, {Champagne}, {Sun}, {Volonteri}, {Yang}, {Hennawi}, {Ba{\~n}ados}, {Barth}, {Eilers}, {Farina}, {Liu}, {Jin}, {Jun}, {Lupi}, {Kakiichi}, {Mazzucchelli}, {Onoue}, {Pan}, {Pizzati}, {Rojas-Ruiz}, {Schindler}, {Trakhtenbrot}, {Shen}, {Trebitsch}, {Zhuang}, {Endsley}, {Meyer}, {Li}, {Li}, {Pudoka}, {Tee}, {Wu}, \& {Zhang}}]{Lin2024}
{Lin}, X., {Wang}, F., {Fan}, X., {et~al.} 2024, \apj, 974, 147

\bibitem[{{Maiolino} {et~al.}(2024){Maiolino}, {Scholtz}, {Curtis-Lake}, {Carniani}, {Baker}, {de Graaff}, {Tacchella}, {{\"U}bler}, {D'Eugenio}, {Witstok}, {Curti}, {Arribas}, {Bunker}, {Charlot}, {Chevallard}, {Eisenstein}, {Egami}, {Ji}, {Jones}, {Lyu}, {Rawle}, {Robertson}, {Rujopakarn}, {Perna}, {Sun}, {Venturi}, {Williams}, \& {Willott}}]{Maiolino2024}
{Maiolino}, R., {Scholtz}, J., {Curtis-Lake}, E., {et~al.} 2024, \aap, 691, A145

\bibitem[{{Marques-Chaves} {et~al.}(2022){Marques-Chaves}, {Schaerer}, {{\'A}lvarez-M{\'a}rquez}, {Verhamme}, {Ceverino}, {Chisholm}, {Colina}, {Dessauges-Zavadsky}, {P{\'e}rez-Fournon}, {Saldana-Lopez}, {Upadhyaya}, \& {Vanzella}}]{Marques-Chaves2022}
{Marques-Chaves}, R., {Schaerer}, D., {{\'A}lvarez-M{\'a}rquez}, J., {et~al.} 2022, \mnras, 517, 2972

\bibitem[{{Marques-Chaves} {et~al.}(2025){Marques-Chaves}, {Schaerer}, {Dessauges-Zavadsky}, {{\'A}lvarez-M{\'a}rquez}, {Hashimoto}, {Colina}, {Inoue}, {Blanco-Prieto}, {Nakazato}, {Costantin}, {Arribas}, {Bakx}, {Ceverino}, {Crespo G{\'o}mez}, {Fudamoto}, {Hagimoto}, {Hamada}, {Matsuoka}, {Mawatari}, {Onoue}, {Osone}, {Ren}, {Sugahara}, {Terui}, \& {Yoshida}}]{Marques-Chaves2025}
{Marques-Chaves}, R., {Schaerer}, D., {Dessauges-Zavadsky}, M., {et~al.} 2025, \aap, submitted, arXiv:2510.12411

\bibitem[{{Mart{\'\i}nez-Ram{\'\i}rez} {et~al.}(2026){Mart{\'\i}nez-Ram{\'\i}rez}, {Wolf}, {Belladitta}, {Ba{\~n}ados}, {Bauer}, {Hviding}, {Stern}, {Mazzucchelli}, {Meyer}, {Treister}, \& {Loiacono}}]{Martinez2026}
{Mart{\'\i}nez-Ram{\'\i}rez}, L.~N., {Wolf}, J., {Belladitta}, S., {et~al.} 2026, \aap, 708, A117

\bibitem[{{Matsuoka} {et~al.}(2022){Matsuoka}, {Iwasawa}, {Onoue}, {Izumi}, {Kashikawa}, {Strauss}, {Imanishi}, {Nagao}, {Akiyama}, {Silverman}, {Asami}, {Bosch}, {Furusawa}, {Goto}, {Gunn}, {Harikane}, {Ikeda}, {Ishimoto}, {Kawaguchi}, {Kato}, {Kikuta}, {Kohno}, {Komiyama}, {Lee}, {Lupton}, {Minezaki}, {Miyazaki}, {Murayama}, {Nishizawa}, {Oguri}, {Ono}, {Ouchi}, {Price}, {Sameshima}, {Sugiyama}, {Tait}, {Takada}, {Takahashi}, {Takata}, {Tanaka}, {Toba}, {Utsumi}, {Wang}, \& {Yamashita}}]{Matsuoka2022}
{Matsuoka}, Y., {Iwasawa}, K., {Onoue}, M., {et~al.} 2022, \apjs, 259, 18

\bibitem[{{Matsuoka} {et~al.}(2025){Matsuoka}, {Iwasawa}, {Onoue}, {Izumi}, {Strauss}, {Akiyama}, {Aoki}, {Arita}, {Ding}, {Imanishi}, {Kashikawa}, {Kawaguchi}, {Kikuta}, {Kohno}, {Lee}, {Nagao}, {Phillips}, {Sawamura}, {Silverman}, {Takahashi}, \& {Toba}}]{Matsuoka2025}
{Matsuoka}, Y., {Iwasawa}, K., {Onoue}, M., {et~al.} 2025, \apjs, 280, 68

\bibitem[{{Matsuoka} {et~al.}(2019{\natexlab{a}}){Matsuoka}, {Iwasawa}, {Onoue}, {Kashikawa}, {Strauss}, {Lee}, {Imanishi}, {Nagao}, {Akiyama}, {Asami}, {Bosch}, {Furusawa}, {Goto}, {Gunn}, {Harikane}, {Ikeda}, {Izumi}, {Kawaguchi}, {Kato}, {Kikuta}, {Kohno}, {Komiyama}, {Koyama}, {Lupton}, {Minezaki}, {Miyazaki}, {Murayama}, {Niida}, {Nishizawa}, {Noboriguchi}, {Oguri}, {Ono}, {Ouchi}, {Price}, {Sameshima}, {Schulze}, {Silverman}, {Sugiyama}, {Tait}, {Takada}, {Takata}, {Tanaka}, {Tang}, {Toba}, {Utsumi}, {Wang}, \& {Yamashita}}]{Matsuoka2019b}
{Matsuoka}, Y., {Iwasawa}, K., {Onoue}, M., {et~al.} 2019{\natexlab{a}}, \apj, 883, 183

\bibitem[{{Matsuoka} {et~al.}(2018{\natexlab{a}}){Matsuoka}, {Iwasawa}, {Onoue}, {Kashikawa}, {Strauss}, {Lee}, {Imanishi}, {Nagao}, {Akiyama}, {Asami}, {Bosch}, {Furusawa}, {Goto}, {Gunn}, {Harikane}, {Ikeda}, {Izumi}, {Kawaguchi}, {Kato}, {Kikuta}, {Kohno}, {Komiyama}, {Lupton}, {Minezaki}, {Miyazaki}, {Morokuma}, {Murayama}, {Niida}, {Nishizawa}, {Oguri}, {Ono}, {Ouchi}, {Price}, {Sameshima}, {Schulze}, {Shirakata}, {Silverman}, {Sugiyama}, {Tait}, {Takada}, {Takata}, {Tanaka}, {Tang}, {Toba}, {Utsumi}, {Wang}, \& {Yamashita}}]{Matsuoka2018ApJS}
{Matsuoka}, Y., {Iwasawa}, K., {Onoue}, M., {et~al.} 2018{\natexlab{a}}, \apjs, 237, 5

\bibitem[{{Matsuoka} {et~al.}(2023){Matsuoka}, {Onoue}, {Iwasawa}, {Strauss}, {Kashikawa}, {Izumi}, {Nagao}, {Imanishi}, {Akiyama}, {Silverman}, {Asami}, {Bosch}, {Furusawa}, {Goto}, {Gunn}, {Harikane}, {Ikeda}, {Inayoshi}, {Ishimoto}, {Kawaguchi}, {Kikuta}, {Kohno}, {Komiyama}, {Lee}, {Lupton}, {Minezaki}, {Miyazaki}, {Murayama}, {Nishizawa}, {Oguri}, {Ono}, {Oogi}, {Ouchi}, {Price}, {Sameshima}, {Sugiyama}, {Tait}, {Takada}, {Takahashi}, {Takata}, {Tanaka}, {Toba}, {Wang}, \& {Yamashita}}]{Matsuoka2023}
{Matsuoka}, Y., {Onoue}, M., {Iwasawa}, K., {et~al.} 2023, \apjl, 949, L42

\bibitem[{{Matsuoka} {et~al.}(2018{\natexlab{b}}){Matsuoka}, {Onoue}, {Kashikawa}, {Iwasawa}, {Strauss}, {Nagao}, {Imanishi}, {Lee}, {Akiyama}, {Asami}, {Bosch}, {Foucaud}, {Furusawa}, {Goto}, {Gunn}, {Harikane}, {Ikeda}, {Izumi}, {Kawaguchi}, {Kikuta}, {Kohno}, {Komiyama}, {Lupton}, {Minezaki}, {Miyazaki}, {Morokuma}, {Murayama}, {Niida}, {Nishizawa}, {Oguri}, {Ono}, {Ouchi}, {Price}, {Sameshima}, {Schulze}, {Shirakata}, {Silverman}, {Sugiyama}, {Tait}, {Takada}, {Takata}, {Tanaka}, {Tang}, {Toba}, {Utsumi}, \& {Wang}}]{Matsuoka2018PASJ}
{Matsuoka}, Y., {Onoue}, M., {Kashikawa}, N., {et~al.} 2018{\natexlab{b}}, \pasj, 70, S35

\bibitem[{{Matsuoka} {et~al.}(2016){Matsuoka}, {Onoue}, {Kashikawa}, {Iwasawa}, {Strauss}, {Nagao}, {Imanishi}, {Niida}, {Toba}, {Akiyama}, {Asami}, {Bosch}, {Foucaud}, {Furusawa}, {Goto}, {Gunn}, {Harikane}, {Ikeda}, {Kawaguchi}, {Kikuta}, {Komiyama}, {Lupton}, {Minezaki}, {Miyazaki}, {Morokuma}, {Murayama}, {Nishizawa}, {Ono}, {Ouchi}, {Price}, {Sameshima}, {Silverman}, {Sugiyama}, {Tait}, {Takada}, {Takata}, {Tanaka}, {Tang}, \& {Utsumi}}]{Matsuoka2016}
{Matsuoka}, Y., {Onoue}, M., {Kashikawa}, N., {et~al.} 2016, \apj, 828, 26

\bibitem[{{Matsuoka} {et~al.}(2019{\natexlab{b}}){Matsuoka}, {Onoue}, {Kashikawa}, {Strauss}, {Iwasawa}, {Lee}, {Imanishi}, {Nagao}, {Akiyama}, {Asami}, {Bosch}, {Furusawa}, {Goto}, {Gunn}, {Harikane}, {Ikeda}, {Izumi}, {Kawaguchi}, {Kato}, {Kikuta}, {Kohno}, {Komiyama}, {Koyama}, {Lupton}, {Minezaki}, {Miyazaki}, {Murayama}, {Niida}, {Nishizawa}, {Noboriguchi}, {Oguri}, {Ono}, {Ouchi}, {Price}, {Sameshima}, {Schulze}, {Shirakata}, {Silverman}, {Sugiyama}, {Tait}, {Takada}, {Takata}, {Tanaka}, {Tang}, {Toba}, {Utsumi}, {Wang}, \& {Yamashita}}]{Matsuoka2019}
{Matsuoka}, Y., {Onoue}, M., {Kashikawa}, N., {et~al.} 2019{\natexlab{b}}, \apjl, 872, L2

\bibitem[{{Matthee} {et~al.}(2024){Matthee}, {Naidu}, {Brammer}, {Chisholm}, {Eilers}, {Goulding}, {Greene}, {Kashino}, {Labbe}, {Lilly}, {Mackenzie}, {Oesch}, {Weibel}, {Wuyts}, {Xiao}, {Bordoloi}, {Bouwens}, {van Dokkum}, {Illingworth}, {Kramarenko}, {Maseda}, {Mason}, {Meyer}, {Nelson}, {Reddy}, {Shivaei}, {Simcoe}, \& {Yue}}]{Matthee2024}
{Matthee}, J., {Naidu}, R.~P., {Brammer}, G., {et~al.} 2024, \apj, 963, 129

\bibitem[{{McGurk} {et~al.}(2024){McGurk}, {Matuszewski}, {Neill}, {Martin}, {Bertz}, {Rockosi}, \& {Kassis}}]{McGurk2024KCRM}
{McGurk}, R.~C., {Matuszewski}, M., {Neill}, J.~D., {et~al.} 2024, in Society of Photo-Optical Instrumentation Engineers (SPIE) Conference Series, Vol. 13096, Ground-based and Airborne Instrumentation for Astronomy X, ed. J.~J. {Bryant}, K.~{Motohara}, \& J.~R.~D. {Vernet}, 1309647

\bibitem[{{McLean} {et~al.}(2012){McLean}, {Steidel}, {Epps}, {Konidaris}, {Matthews}, {Adkins}, {Aliado}, {Brims}, {Canfield}, {Cromer}, {Fucik}, {Kulas}, {Mace}, {Magnone}, {Rodriguez}, {Rudie}, {Trainor}, {Wang}, {Weber}, \& {Weiss}}]{McLean2012MOSFIRE}
{McLean}, I.~S., {Steidel}, C.~C., {Epps}, H.~W., {et~al.} 2012, in Society of Photo-Optical Instrumentation Engineers (SPIE) Conference Series, Vol. 8446, Ground-based and Airborne Instrumentation for Astronomy IV, ed. I.~S. {McLean}, S.~K. {Ramsay}, \& H.~{Takami}, 84460J

\bibitem[{{McMahon} {et~al.}(2013){McMahon}, {Banerji}, {Gonzalez}, {Koposov}, {Bejar}, {Lodieu}, {Rebolo}, \& {VHS Collaboration}}]{McMahon2013}
{McMahon}, R.~G., {Banerji}, M., {Gonzalez}, E., {et~al.} 2013, The Messenger, 154, 35

\bibitem[{{Miyazaki} {et~al.}(2018){Miyazaki}, {Komiyama}, {Kawanomoto}, {Doi}, {Furusawa}, {Hamana}, {Hayashi}, {Ikeda}, {Kamata}, {Karoji}, {Koike}, {Kurakami}, {Miyama}, {Morokuma}, {Nakata}, {Namikawa}, {Nakaya}, {Nariai}, {Obuchi}, {Oishi}, {Okada}, {Okura}, {Tait}, {Takata}, {Tanaka}, {Tanaka}, {Terai}, {Tomono}, {Uraguchi}, {Usuda}, {Utsumi}, {Yamada}, {Yamanoi}, {Aihara}, {Fujimori}, {Mineo}, {Miyatake}, {Oguri}, {Uchida}, {Tanaka}, {Yasuda}, {Takada}, {Murayama}, {Nishizawa}, {Sugiyama}, {Chiba}, {Futamase}, {Wang}, {Chen}, {Ho}, {Liaw}, {Chiu}, {Ho}, {Lai}, {Lee}, {Jeng}, {Iwamura}, {Armstrong}, {Bickerton}, {Bosch}, {Gunn}, {Lupton}, {Loomis}, {Price}, {Smith}, {Strauss}, {Turner}, {Suzuki}, {Miyazaki}, {Muramatsu}, {Yamamoto}, {Endo}, {Ezaki}, {Ito}, {Kawaguchi}, {Sofuku}, {Taniike}, {Akutsu}, {Dojo}, {Kasumi}, {Matsuda}, {Imoto}, {Miwa}, {Suzuki}, {Takeshi}, \& {Yokota}}]{Miyazaki2018HSC}
{Miyazaki}, S., {Komiyama}, Y., {Kawanomoto}, S., {et~al.} 2018, \pasj, 70, S1

\bibitem[{{Morrissey} {et~al.}(2018){Morrissey}, {Matuszewski}, {Martin}, {Neill}, {Epps}, {Fucik}, {Weber}, {Darvish}, {Adkins}, {Allen}, {Bartos}, {Belicki}, {Cabak}, {Callahan}, {Cowley}, {Crabill}, {Deich}, {Delecroix}, {Doppman}, {Hilyard}, {James}, {Kaye}, {Kokorowski}, {Kwok}, {Lanclos}, {Milner}, {Moore}, {O'Sullivan}, {Parihar}, {Park}, {Phillips}, {Rizzi}, {Rockosi}, {Rodriguez}, {Salaun}, {Seaman}, {Sheikh}, {Weiss}, \& {Zarzaca}}]{Morrissey2018KCWI}
{Morrissey}, P., {Matuszewski}, M., {Martin}, D.~C., {et~al.} 2018, \apj, 864, 93

\bibitem[{{Mortlock} {et~al.}(2012){Mortlock}, {Patel}, {Warren}, {Hewett}, {Venemans}, {McMahon}, \& {Simpson}}]{Mortlock2012}
{Mortlock}, D.~J., {Patel}, M., {Warren}, S.~J., {et~al.} 2012, \mnras, 419, 390

\bibitem[{{Mortlock} {et~al.}(2011){Mortlock}, {Warren}, {Venemans}, {Patel}, {Hewett}, {McMahon}, {Simpson}, {Theuns}, {Gonz{\'a}les-Solares}, {Adamson}, {Dye}, {Hambly}, {Hirst}, {Irwin}, {Kuiper}, {Lawrence}, \& {R{\"o}ttgering}}]{Mortlock2011}
{Mortlock}, D.~J., {Warren}, S.~J., {Venemans}, B.~P., {et~al.} 2011, \nat, 474, 616

\bibitem[{{Nanni} {et~al.}(2022){Nanni}, {Hennawi}, {Wang}, {Yang}, {Schindler}, \& {Fan}}]{Nanni2022}
{Nanni}, R., {Hennawi}, J.~F., {Wang}, F., {et~al.} 2022, \mnras, 515, 3224

\bibitem[{{Neeleman} {et~al.}(2021){Neeleman}, {Novak}, {Venemans}, {Walter}, {Decarli}, {Kaasinen}, {Schindler}, {Ba{\~n}ados}, {Carilli}, {Drake}, {Fan}, \& {Rix}}]{Neeleman2021}
{Neeleman}, M., {Novak}, M., {Venemans}, B.~P., {et~al.} 2021, \apj, 911, 141

\bibitem[{{Oke} {et~al.}(1995){Oke}, {Cohen}, {Carr}, {Cromer}, {Dingizian}, {Harris}, {Labrecque}, {Lucinio}, {Schaal}, {Epps}, \& {Miller}}]{Oke1995LRIS}
{Oke}, J.~B., {Cohen}, J.~G., {Carr}, M., {et~al.} 1995, \pasp, 107, 375

\bibitem[{{Oke} \& {Gunn}(1983)}]{Oke1983}
{Oke}, J.~B. \& {Gunn}, J.~E. 1983, \apj, 266, 713

\bibitem[{{Onorato} {et~al.}(2026){Onorato}, {Hennawi}, {Pizzati}, {Venemans}, \& {Eilers}}]{Onorato2025b}
{Onorato}, S., {Hennawi}, J.~F., {Pizzati}, E., {Venemans}, B.~P., \& {Eilers}, A.-C. 2026, \mnras, 547, stag388

\bibitem[{{Onorato} {et~al.}(2025){Onorato}, {Hennawi}, {Schindler}, {Yang}, {Wang}, {Barth}, {Ba{\~n}ados}, {Eilers}, {Bosman}, {Davies}, {Venemans}, {Mazzucchelli}, {Belladitta}, {Vito}, {Farina}, {Andika}, {Fan}, {Walter}, {Decarli}, {Onoue}, \& {Nanni}}]{Onorato2025}
{Onorato}, S., {Hennawi}, J.~F., {Schindler}, J.-T., {et~al.} 2025, \mnras, 540, 1308

\bibitem[{{Onoue} {et~al.}(2025){Onoue}, {Ding}, {Silverman}, {Matsuoka}, {Izumi}, {Strauss}, {Ward}, {Phillips}, {Ito}, {Andika}, {Aoki}, {Arita}, {Baba}, {Bieri}, {Bosman}, {Eilers}, {Fujimoto}, {Habouzit}, {Haiman}, {Imanishi}, {Inayoshi}, {Iwasawa}, {Jahnke}, {Kashikawa}, {Kawaguchi}, {Kohno}, {Lee}, {Li}, {Lupi}, {Lyu}, {Nagao}, {Overzier}, {Schindler}, {Schramm}, {Scoggins}, {Shimasaku}, {Toba}, {Trakhtenbrot}, {Trebitsch}, {Treu}, {Umehata}, {Venemans}, {Vestergaard}, {Volonteri}, {Walter}, {Wang}, {Yang}, \& {Zhang}}]{Onoue2025}
{Onoue}, M., {Ding}, X., {Silverman}, J.~D., {et~al.} 2025, Nature Astronomy, 9, 1541

\bibitem[{{Pasha} \& {Miller}(2023)}]{Pasha2023pysersic}
{Pasha}, I. \& {Miller}, T.~B. 2023, The Journal of Open Source Software, 8, 5703

\bibitem[{{Perger} {et~al.}(2025){Perger}, {Fogasy}, {Frey}, \& {Gab{\'a}nyi}}]{Perger2025}
{Perger}, K., {Fogasy}, J., {Frey}, S., \& {Gab{\'a}nyi}, K.~{\'E}. 2025, \aap, 693, L2

\bibitem[{{Pizzati} {et~al.}(2025){Pizzati}, {Hennawi}, {Schaye}, {Eilers}, {Huang}, {Schindler}, \& {Wang}}]{Pizzati2025}
{Pizzati}, E., {Hennawi}, J.~F., {Schaye}, J., {et~al.} 2025, \mnras, 539, 2910

\bibitem[{{Pizzati} {et~al.}(2024){Pizzati}, {Hennawi}, {Schaye}, {Schaller}, {Eilers}, {Wang}, {Frenk}, {Elbers}, {Helly}, {Mackenzie}, {Matthee}, {Bordoloi}, {Kashino}, {Naidu}, \& {Yue}}]{Pizzati2024}
{Pizzati}, E., {Hennawi}, J.~F., {Schaye}, J., {et~al.} 2024, \mnras, 534, 3155

\bibitem[{{Planck Collaboration} {et~al.}(2020){Planck Collaboration}, {Aghanim}, {Akrami}, {Ashdown}, {Aumont}, {Baccigalupi}, {Ballardini}, {Banday}, {Barreiro}, {Bartolo}, {Basak}, {Battye}, {Benabed}, {Bernard}, {Bersanelli}, {Bielewicz}, {Bock}, {Bond}, {Borrill}, {Bouchet}, {Boulanger}, {Bucher}, {Burigana}, {Butler}, {Calabrese}, {Cardoso}, {Carron}, {Challinor}, {Chiang}, {Chluba}, {Colombo}, {Combet}, {Contreras}, {Crill}, {Cuttaia}, {de Bernardis}, {de Zotti}, {Delabrouille}, {Delouis}, {Di Valentino}, {Diego}, {Dor{\'e}}, {Douspis}, {Ducout}, {Dupac}, {Dusini}, {Efstathiou}, {Elsner}, {En{\ss}lin}, {Eriksen}, {Fantaye}, {Farhang}, {Fergusson}, {Fernandez-Cobos}, {Finelli}, {Forastieri}, {Frailis}, {Fraisse}, {Franceschi}, {Frolov}, {Galeotta}, {Galli}, {Ganga}, {G{\'e}nova-Santos}, {Gerbino}, {Ghosh}, {Gonz{\'a}lez-Nuevo}, {G{\'o}rski}, {Gratton}, {Gruppuso}, {Gudmundsson}, {Hamann}, {Handley}, {Hansen}, {Herranz}, {Hildebrandt}, {Hivon}, {Huang}, {Jaffe}, {Jones}, {Karakci}, {Keih{\"a}nen},
  {Keskitalo}, {Kiiveri}, {Kim}, {Kisner}, {Knox}, {Krachmalnicoff}, {Kunz}, {Kurki-Suonio}, {Lagache}, {Lamarre}, {Lasenby}, {Lattanzi}, {Lawrence}, {Le Jeune}, {Lemos}, {Lesgourgues}, {Levrier}, {Lewis}, {Liguori}, {Lilje}, {Lilley}, {Lindholm}, {L{\'o}pez-Caniego}, {Lubin}, {Ma}, {Mac{\'\i}as-P{\'e}rez}, {Maggio}, {Maino}, {Mandolesi}, {Mangilli}, {Marcos-Caballero}, {Maris}, {Martin}, {Martinelli}, {Mart{\'\i}nez-Gonz{\'a}lez}, {Matarrese}, {Mauri}, {McEwen}, {Meinhold}, {Melchiorri}, {Mennella}, {Migliaccio}, {Millea}, {Mitra}, {Miville-Desch{\^e}nes}, {Molinari}, {Montier}, {Morgante}, {Moss}, {Natoli}, {N{\o}rgaard-Nielsen}, {Pagano}, {Paoletti}, {Partridge}, {Patanchon}, {Peiris}, {Perrotta}, {Pettorino}, {Piacentini}, {Polastri}, {Polenta}, {Puget}, {Rachen}, {Reinecke}, {Remazeilles}, {Renzi}, {Rocha}, {Rosset}, {Roudier}, {Rubi{\~n}o-Mart{\'\i}n}, {Ruiz-Granados}, {Salvati}, {Sandri}, {Savelainen}, {Scott}, {Shellard}, {Sirignano}, {Sirri}, {Spencer}, {Sunyaev}, {Suur-Uski}, {Tauber}, {Tavagnacco},
  {Tenti}, {Toffolatti}, {Tomasi}, {Trombetti}, {Valenziano}, {Valiviita}, {Van Tent}, {Vibert}, {Vielva}, {Villa}, {Vittorio}, {Wandelt}, {Wehus}, {White}, {White}, {Zacchei}, \& {Zonca}}]{Planck2020}
{Planck Collaboration}, {Aghanim}, N., {Akrami}, Y., {et~al.} 2020, \aap, 641, A6

\bibitem[{{Pogge} {et~al.}(2010){Pogge}, {Atwood}, {Brewer}, {Byard}, {Derwent}, {Gonzalez}, {Martini}, {Mason}, {O'Brien}, {Osmer}, {Pappalardo}, {Steinbrecher}, {Teiga}, \& {Zhelem}}]{Pogge2010}
{Pogge}, R.~W., {Atwood}, B., {Brewer}, D.~F., {et~al.} 2010, in Society of Photo-Optical Instrumentation Engineers (SPIE) Conference Series, Vol. 7735, Ground-based and Airborne Instrumentation for Astronomy III, ed. I.~S. {McLean}, S.~K. {Ramsay}, \& H.~{Takami}, 77350A

\bibitem[{{Prochaska} {et~al.}(2020){Prochaska}, {Hennawi}, {Westfall}, {Cooke}, {Wang}, {Hsyu}, {Davies}, {Farina}, \& {Pelliccia}}]{Prochaska2020pypeit}
{Prochaska}, J., {Hennawi}, J., {Westfall}, K., {et~al.} 2020, The Journal of Open Source Software, 5, 2308

\bibitem[{{Roberts-Borsani} {et~al.}(2025){Roberts-Borsani}, {Bagley}, {Rojas-Ruiz}, {Treu}, {Morishita}, {Finkelstein}, {Trenti}, {Arrabal Haro}, {Ba{\~n}ados}, {Ch{\'a}vez Ortiz}, {Chworowsky}, {Hutchison}, {Larson}, {Leethochawalit}, {Leung}, {Mason}, {Somerville}, {Stiavelli}, {Yung}, {Kassin}, \& {Soto}}]{Roberts-Borsani2025}
{Roberts-Borsani}, G., {Bagley}, M., {Rojas-Ruiz}, S., {et~al.} 2025, \apj, 983, 18

\bibitem[{{Roberts-Borsani} {et~al.}(2024){Roberts-Borsani}, {Treu}, {Shapley}, {Fontana}, {Pentericci}, {Castellano}, {Morishita}, {Bergamini}, \& {Rosati}}]{Roberts-Borsani2024}
{Roberts-Borsani}, G., {Treu}, T., {Shapley}, A., {et~al.} 2024, \apj, 976, 193

\bibitem[{{Runnoe} {et~al.}(2012){Runnoe}, {Brotherton}, \& {Shang}}]{Runnoe2012}
{Runnoe}, J.~C., {Brotherton}, M.~S., \& {Shang}, Z. 2012, \mnras, 422, 478

\bibitem[{{Salvato} {et~al.}(2022){Salvato}, {Wolf}, {Dwelly}, {Georgakakis}, {Brusa}, {Merloni}, {Liu}, {Toba}, {Nandra}, {Lamer}, {Buchner}, {Schneider}, {Freund}, {Rau}, {Schwope}, {Nishizawa}, {Klein}, {Arcodia}, {Comparat}, {Musiimenta}, {Nagao}, {Brunner}, {Malyali}, {Finoguenov}, {Anderson}, {Shen}, {Ibarra-Medel}, {Trump}, {Brandt}, {Urry}, {Rivera}, {Krumpe}, {Urrutia}, {Miyaji}, {Ichikawa}, {Schneider}, {Fresco}, {Boller}, {Haase}, {Brownstein}, {Lane}, {Bizyaev}, \& {Nitschelm}}]{Salvato2022}
{Salvato}, M., {Wolf}, J., {Dwelly}, T., {et~al.} 2022, \aap, 661, A3

\bibitem[{{Schindler}(2022)}]{Schindler2022sculptor}
{Schindler}, J.-T. 2022, {Sculptor: Interactive modeling of astronomical spectra}, Astrophysics Source Code Library, record ascl:2202.018

\bibitem[{{Schindler} {et~al.}(2020){Schindler}, {Farina}, {Ba{\~n}ados}, {Eilers}, {Hennawi}, {Onoue}, {Venemans}, {Walter}, {Wang}, {Davies}, {Decarli}, {Rosa}, {Drake}, {Fan}, {Mazzucchelli}, {Rix}, {Worseck}, \& {Yang}}]{Schindler2020}
{Schindler}, J.-T., {Farina}, E.~P., {Ba{\~n}ados}, E., {et~al.} 2020, \apj, 905, 51

\bibitem[{{Schindler} {et~al.}(2025){Schindler}, {Hennawi}, {Davies}, {Bosman}, {Endsley}, {Wang}, {Yang}, {Barth}, {Eilers}, {Fan}, {Kakiichi}, {Maseda}, {Pizzati}, \& {Nanni}}]{Schindler2024}
{Schindler}, J.-T., {Hennawi}, J.~F., {Davies}, F.~B., {et~al.} 2025, Nature Astronomy, 9, 1732

\bibitem[{{Seifert} {et~al.}(2003){Seifert}, {Appenzeller}, {Baumeister}, {Bizenberger}, {Bomans}, {Dettmar}, {Grimm}, {Herbst}, {Hofmann}, {Juette}, {Laun}, {Lehmitz}, {Lemke}, {Lenzen}, {Mandel}, {Polsterer}, {Rohloff}, {Schuetze}, {Seltmann}, {Thatte}, {Weiser}, \& {Xu}}]{Seifert2003LUCI}
{Seifert}, W., {Appenzeller}, I., {Baumeister}, H., {et~al.} 2003, in Society of Photo-Optical Instrumentation Engineers (SPIE) Conference Series, Vol. 4841, Instrument Design and Performance for Optical/Infrared Ground-based Telescopes, ed. M.~{Iye} \& A.~F.~M. {Moorwood}, 962--973

\bibitem[{{Shen} {et~al.}(2019){Shen}, {Wu}, {Jiang}, {Ba{\~n}ados}, {Fan}, {Ho}, {Riechers}, {Strauss}, {Venemans}, {Vestergaard}, {Walter}, {Wang}, {Willott}, {Wu}, \& {Yang}}]{Shen2019}
{Shen}, Y., {Wu}, J., {Jiang}, L., {et~al.} 2019, \apj, 873, 35

\bibitem[{{Shibuya} {et~al.}(2015){Shibuya}, {Ouchi}, \& {Harikane}}]{Shibuya2015}
{Shibuya}, T., {Ouchi}, M., \& {Harikane}, Y. 2015, \apjs, 219, 15

\bibitem[{{Shimwell} {et~al.}(2022){Shimwell}, {Hardcastle}, {Tasse}, {Best}, {R{\"o}ttgering}, {Williams}, {Botteon}, {Drabent}, {Mechev}, {Shulevski}, {van Weeren}, {Bester}, {Br{\"u}ggen}, {Brunetti}, {Callingham}, {Chy{\.z}y}, {Conway}, {Dijkema}, {Duncan}, {de Gasperin}, {Hale}, {Haverkorn}, {Hugo}, {Jackson}, {Mevius}, {Miley}, {Morabito}, {Morganti}, {Offringa}, {Oonk}, {Rafferty}, {Sabater}, {Smith}, {Schwarz}, {Smirnov}, {O'Sullivan}, {Vedantham}, {White}, {Albert}, {Alegre}, {Asabere}, {Bacon}, {Bonafede}, {Bonnassieux}, {Brienza}, {Bilicki}, {Bonato}, {Calistro Rivera}, {Cassano}, {Cochrane}, {Croston}, {Cuciti}, {Dallacasa}, {Danezi}, {Dettmar}, {Di Gennaro}, {Edler}, {En{\ss}lin}, {Emig}, {Franzen}, {Garc{\'\i}a-Vergara}, {Grange}, {G{\"u}rkan}, {Hajduk}, {Heald}, {Heesen}, {Hoang}, {Hoeft}, {Horellou}, {Iacobelli}, {Jamrozy}, {Jeli{\'c}}, {Kondapally}, {Kukreti}, {Kunert-Bajraszewska}, {Magliocchetti}, {Mahatma}, {Ma{\l}ek}, {Mandal}, {Massaro}, {Meyer-Zhao}, {Mingo}, {Mostert}, {Nair},
  {Nakoneczny}, {Nikiel-Wroczy{\'n}ski}, {Orr{\'u}}, {Pajdosz-{\'S}mierciak}, {Pasini}, {Prandoni}, {van Piggelen}, {Rajpurohit}, {Retana-Montenegro}, {Riseley}, {Rowlinson}, {Saxena}, {Schrijvers}, {Sweijen}, {Siewert}, {Timmerman}, {Vaccari}, {Vink}, {West}, {Wo{\l}owska}, {Zhang}, \& {Zheng}}]{Shimwell2022}
{Shimwell}, T.~W., {Hardcastle}, M.~J., {Tasse}, C., {et~al.} 2022, \aap, 659, A1

\bibitem[{{Silverman} {et~al.}(2025){Silverman}, {Li}, {Ding}, {Onoue}, {Strauss}, {Matsuoka}, {Izumi}, {Jahnke}, {Treu}, {Volonteri}, {Phillips}, {Andika}, {Aoki}, {Arita}, {Baba}, {Bosman}, {Eilers}, {Fan}, {Fujimoto}, {Habouzit}, {Haiman}, {Imanishi}, {Inayoshi}, {Iwasawa}, {Kashikawa}, {Kawaguchi}, {Lee}, {Lupi}, {Nagao}, {Schindler}, {Schramm}, {Shimasaku}, {Toba}, {Trakhtenbrot}, {Umehata}, {Vestergaard}, {Walter}, {Wang}, \& {Yang}}]{Silverman2025}
{Silverman}, J.~D., {Li}, J., {Ding}, X., {et~al.} 2025, \apjl, 995, L67

\bibitem[{{Simcoe} {et~al.}(2010){Simcoe}, {Burgasser}, {Bochanski}, {Schechter}, {Bernstein}, {Bigelow}, {Pipher}, {Forrest}, {McMurtry}, {Smith}, \& {Fishner}}]{Simcoe2010FIRE}
{Simcoe}, R.~A., {Burgasser}, A.~J., {Bochanski}, J.~J., {et~al.} 2010, in Society of Photo-Optical Instrumentation Engineers (SPIE) Conference Series, Vol. 7735, Ground-based and Airborne Instrumentation for Astronomy III, ed. I.~S. {McLean}, S.~K. {Ramsay}, \& H.~{Takami}, 773514

\bibitem[{{Sobral} {et~al.}(2015){Sobral}, {Matthee}, {Darvish}, {Schaerer}, {Mobasher}, {R{\"o}ttgering}, {Santos}, \& {Hemmati}}]{Sobral2015}
{Sobral}, D., {Matthee}, J., {Darvish}, B., {et~al.} 2015, \apj, 808, 139

\bibitem[{{Sun} {et~al.}(2024){Sun}, {Ho}, {Zhuang}, {Ma}, {Chen}, \& {Li}}]{Sun2022}
{Sun}, W., {Ho}, L.~C., {Zhuang}, M.-Y., {et~al.} 2024, \apj, 960, 104

\bibitem[{{Tee} {et~al.}(2025){Tee}, {Fan}, {Wang}, \& {Yang}}]{Tee2025}
{Tee}, W.~L., {Fan}, X., {Wang}, F., \& {Yang}, J. 2025, \apjl, 983, L26

\bibitem[{{Telfer} {et~al.}(2002){Telfer}, {Zheng}, {Kriss}, \& {Davidsen}}]{Telfer2002}
{Telfer}, R.~C., {Zheng}, W., {Kriss}, G.~A., \& {Davidsen}, A.~F. 2002, \apj, 565, 773

\bibitem[{{Vestergaard} \& {Peterson}(2006)}]{Vestergaard2006}
{Vestergaard}, M. \& {Peterson}, B.~M. 2006, \apj, 641, 689

\bibitem[{{Virtanen} {et~al.}(2020){Virtanen}, {Gommers}, {Oliphant}, {Haberland}, {Reddy}, {Cournapeau}, {Burovski}, {Peterson}, {Weckesser}, {Bright}, {van der Walt}, {Brett}, {Wilson}, {Millman}, {Mayorov}, {Nelson}, {Jones}, {Kern}, {Larson}, {Carey}, {Polat}, {Feng}, {Moore}, {VanderPlas}, {Laxalde}, {Perktold}, {Cimrman}, {Henriksen}, {Quintero}, {Harris}, {Archibald}, {Ribeiro}, {Pedregosa}, {van Mulbregt}, \& {SciPy 1. 0 Contributors}}]{Virtanen2020}
{Virtanen}, P., {Gommers}, R., {Oliphant}, T.~E., {et~al.} 2020, Nature Medicine, 17, 261

\bibitem[{{Wang} {et~al.}(2021){Wang}, {Yang}, {Fan}, {Hennawi}, {Barth}, {Banados}, {Bian}, {Boutsia}, {Connor}, {Davies}, {Decarli}, {Eilers}, {Farina}, {Green}, {Jiang}, {Li}, {Mazzucchelli}, {Nanni}, {Schindler}, {Venemans}, {Walter}, {Wu}, \& {Yue}}]{Wang2021}
{Wang}, F., {Yang}, J., {Fan}, X., {et~al.} 2021, \apjl, 907, L1

\bibitem[{{Wang} {et~al.}(2024){Wang}, {Yang}, {Fan}, {Venemans}, {Decarli}, {Ba{\~n}ados}, {Walter}, {Barth}, {Bian}, {Davies}, {Eilers}, {Farina}, {Hennawi}, {Li}, {Mazzucchelli}, {Wang}, {Wu}, \& {Yue}}]{Wang2024}
{Wang}, F., {Yang}, J., {Fan}, X., {et~al.} 2024, \apj, 968, 9

\bibitem[{{Wang} {et~al.}(2018){Wang}, {Yang}, {Fan}, {Yue}, {Wu}, {Schindler}, {Bian}, {Li}, {Farina}, {Ba{\~n}ados}, {Davies}, {Decarli}, {Green}, {Jiang}, {Hennawi}, {Huang}, {Mazzucchelli}, {McGreer}, {Venemans}, {Walter}, \& {Beletsky}}]{Wang2018z7}
{Wang}, F., {Yang}, J., {Fan}, X., {et~al.} 2018, \apjl, 869, L9

\bibitem[{{Wang} {et~al.}(2023){Wang}, {Yang}, {Hennawi}, {Fan}, {Sun}, {Champagne}, {Costa}, {Habouzit}, {Endsley}, {Li}, {Lin}, {Meyer}, {Schindler}, {Wu}, {Ba{\~n}ados}, {Barth}, {Bhowmick}, {Bieri}, {Blecha}, {Bosman}, {Cai}, {Colina}, {Connor}, {Davies}, {Decarli}, {De Rosa}, {Drake}, {Egami}, {Eilers}, {Evans}, {Farina}, {Haiman}, {Jiang}, {Jin}, {Jun}, {Kakiichi}, {Khusanova}, {Kulkarni}, {Li}, {Liu}, {Loiacono}, {Lupi}, {Mazzucchelli}, {Onoue}, {Pudoka}, {Rojas-Ruiz}, {Shen}, {Strauss}, {Tee}, {Trakhtenbrot}, {Trebitsch}, {Venemans}, {Volonteri}, {Walter}, {Xie}, {Yue}, {Zhang}, {Zhang}, \& {Zou}}]{Wang2023}
{Wang}, F., {Yang}, J., {Hennawi}, J.~F., {et~al.} 2023, \apjl, 951, L4

\bibitem[{{Willott} {et~al.}(2017){Willott}, {Bergeron}, \& {Omont}}]{Willott2017}
{Willott}, C.~J., {Bergeron}, J., \& {Omont}, A. 2017, \apj, 850, 108

\bibitem[{{Willott} {et~al.}(2010){Willott}, {Delorme}, {Reyl{\'e}}, {Albert}, {Bergeron}, {Crampton}, {Delfosse}, {Forveille}, {Hutchings}, {McLure}, {Omont}, \& {Schade}}]{Willott2010}
{Willott}, C.~J., {Delorme}, P., {Reyl{\'e}}, C., {et~al.} 2010, \aj, 139, 906

\bibitem[{{Wolf} {et~al.}(2024){Wolf}, {Salvato}, {Belladitta}, {Arcodia}, {Ciroi}, {Di Mille}, {Sbarrato}, {Buchner}, {H{\"a}mmerich}, {Wilms}, {Collmar}, {Dwelly}, {Merloni}, {Urrutia}, \& {Nandra}}]{Wolf2024}
{Wolf}, J., {Salvato}, M., {Belladitta}, S., {et~al.} 2024, \aap, 691, A30

\bibitem[{{Yang} {et~al.}(2024){Yang}, {Schindler}, {Nanni}, {Hennawi}, {Ba{\~n}ados}, {Fan}, {Gloudemans}, {Mazzucchelli}, {Rottgering}, {Venemans}, {Wang}, \& {Yang}}]{YangD2024}
{Yang}, D.-M., {Schindler}, J.-T., {Nanni}, R., {et~al.} 2024, \mnras, 528, 2679

\bibitem[{{Yang} {et~al.}(2021){Yang}, {Wang}, {Fan}, {Barth}, {Hennawi}, {Nanni}, {Bian}, {Davies}, {Farina}, {Schindler}, {Ba{\~n}ados}, {Decarli}, {Eilers}, {Green}, {Guo}, {Jiang}, {Li}, {Venemans}, {Walter}, {Wu}, \& {Yue}}]{Yang2021}
{Yang}, J., {Wang}, F., {Fan}, X., {et~al.} 2021, \apj, 923, 262

\bibitem[{{Yang} {et~al.}(2020{\natexlab{a}}){Yang}, {Wang}, {Fan}, {Hennawi}, {Davies}, {Yue}, {Banados}, {Wu}, {Venemans}, {Barth}, {Bian}, {Boutsia}, {Decarli}, {Farina}, {Green}, {Jiang}, {Li}, {Mazzucchelli}, \& {Walter}}]{Yang2020}
{Yang}, J., {Wang}, F., {Fan}, X., {et~al.} 2020{\natexlab{a}}, \apjl, 897, L14

\bibitem[{{Yang} {et~al.}(2020{\natexlab{b}}){Yang}, {Wang}, {Fan}, {Hennawi}, {Davies}, {Yue}, {Eilers}, {Farina}, {Wu}, {Bian}, {Pacucci}, \& {Lee}}]{Yang2020IGM}
{Yang}, J., {Wang}, F., {Fan}, X., {et~al.} 2020{\natexlab{b}}, \apj, 904, 26

\bibitem[{{Yue} {et~al.}(2024){Yue}, {Eilers}, {Ananna}, {Panagiotou}, {Kara}, \& {Miyaji}}]{Yue2024}
{Yue}, M., {Eilers}, A.-C., {Ananna}, T.~T., {et~al.} 2024, \apjl, 974, L26

\bibitem[{{Zonca} {et~al.}(2020){Zonca}, {Singer}, {Lenz}, {Reinecke}, {Rosset}, {Hivon}, \& {Gorski}}]{Zonca2020}
{Zonca}, A., {Singer}, L., {Lenz}, D., {et~al.} 2020, {healpy: Python wrapper for HEALPix}, Astrophysics Source Code Library, record ascl:2008.022

\end{thebibliography}

%
% Now you can add appendices.
% In this example, the appendices are in one column mode.
% If that is not requires, comment out \onecolumn
% Note that appendices in A\&A come {\it after\/} the references.

\begin{appendix}
  \onecolumn %If you don't want single column for the Appendix, please
             %comment this out

\section{Discovery observations and new quasar properties}
\label{app:discovery_observations}

\begin{table*}[htp!]
\centering
\caption{Journal of discovery observations.}
\label{table:discovery}
\begin{threeparttable}
\begin{tabular}{lccclc}
\toprule\toprule
Target & Telescope & Instrument & Exposure (s) & Date (UTC) & Program ID \\
\midrule
EUCL\,J150537.32$+$773441.0 & Keck & LRIS & 1200 & 2024-07-01 & U259 \\
\addlinespace
\multirow{2}{*}{EUCL\,J141841.29$+$694930.6} & Keck & LRIS & 2400 & 2024-07-01 & U259 \\
                                          & Keck & KCWI & 2400 & 2025-01-21 & U267 \\
\addlinespace
\multirow{3}{*}{EUCL\,J125308.55$+$705432.3} & Keck & LRIS & 1200 & 2024-07-01\tnote{a} & U259 \\
                                          & Keck & MOSFIRE & 3900 & 2024-12-30\tnote{b} & U418 \\
                                          & Keck & MOSFIRE & 4200 & 2025-01-08 & U418 \\
\addlinespace
\multirow{3}{*}{EUCL\,J144545.82$+$714342.3} & Keck & LRIS & 1200 & 2024-07-02\tnote{a} & U259 \\
                                          & Keck & KCWI & 2400 & 2025-01-24\tnote{a} & U267 \\
                                          & Keck & MOSFIRE & 3000 & 2025-01-25 & U418 \\
\addlinespace
EUCL\,J143448.06$+$685733.2 & Keck & LRIS & 2400 & 2024-07-03 & U259 \\
\addlinespace
EUCL\,J115522.89$+$704612.3 & Keck & LRIS & 3600 & 2024-07-03 & U259 \\
\addlinespace
EUCL\,J031547.94$-$684405.6 & Magellan & FIRE & 1800 & 2024-11-16 &  \\
\addlinespace
EUCL\,J025210.37$-$412533.8 & Magellan & FIRE & 1800 & 2024-11-16 &  \\
\addlinespace
EUCL\,J025029.93$-$531706.7 & Magellan & FIRE & 1800 & 2024-11-16 &  \\
\addlinespace
\multirow{3}{*}{EUCL\,J041250.73$-$563949.7} & Magellan & FIRE & 1800 & 2024-11-16 &  \\
                                          & Magellan & FIRE & 1800 & 2024-12-20\tnote{a,c} &  \\
                                          & Magellan & FIRE & 3600 & 2025-03-16\tnote{b} &  \\
\addlinespace
EUCL\,J045148.19$-$342622.2 & Magellan & FIRE & 1800 & 2024-11-17 &  \\
\addlinespace
EUCL\,J052645.38$-$460902.7 & Magellan & FIRE & 1800 & 2024-11-17 &  \\
\addlinespace
EUCL\,J050240.93$-$384906.1 & Magellan & FIRE & 1800 & 2024-11-17 &  \\
\addlinespace
EUCL\,J052209.82$-$512709.2\tnote{d} & Magellan & FIRE & 3600 & 2024-11-18 &  \\
\addlinespace
EUCL\,J044326.27$-$533214.9 & Magellan & FIRE & 1800 & 2024-12-20 &  \\
\addlinespace
EUCL\,J091639.93$+$683652.9 & Keck & KCWI & 1200 & 2024-12-23 & U267 \\
\addlinespace
EUCL\,J093330.60$+$742730.0 & Keck & KCWI & 1200 & 2024-12-23 & U267 \\
\addlinespace
\multirow{2}{*}{EUCL\,J135534.55$+$700055.5} & Keck & KCWI & 5100 & 2024-12-23\tnote{b} & U267 \\
                                          & Keck & MOSFIRE & 3300 & 2024-12-29 & U418 \\
\addlinespace
\multirow{2}{*}{EUCL\,J134031.50$+$674715.1} & Keck & KCWI & 1200 & 2024-12-23\tnote{c} & U267 \\
                                          & Keck & KCWI & 1200 & 2025-01-24 & U267 \\
\addlinespace
EUCL\,J173243.48$+$601612.6 & Keck & KCWI & 600 & 2025-01-22\tnote{b} & U267 \\
\addlinespace
\multirow{3}{*}{EUCL\,J101255.87$+$663058.0} & Keck & KCWI & 1800 & 2025-01-22\tnote{a} & U267 \\
                                            & Keck & MOSFIRE & 2400 & 2025-06-16 & U275 \\
                                            & Keck & MOSFIRE & 5400 & 2025-06-17 & U275 \\
\addlinespace
\multirow{5}{*}{EUCL\,J172258.31$+$574133.5} & Keck & KCWI & 900 & 2025-01-23\tnote{a,c} & U267 \\
                                            & Keck & KCWI & 2100 & 2025-01-24\tnote{b} & U267 \\
                                            & Keck & MOSFIRE & 1200 & 2025-01-25\tnote{b} & U267 \\
                                            & Keck & KCWI & 1800 & 2025-01-28 & U267 \\
                                            & Keck & KCWI & 3600 & 2025-06-18 & U275 \\
\bottomrule
\end{tabular}
% \tablefoot{All dates are given in Coordinated Universal Time (UTC). The list follows the chronological order of the first observation of each quasar.}
\begin{tablenotes}
    % \centering
    \item {\bf Notes.} All dates are given in Coordinated Universal Time (UTC). The list follows the chronological order of the first observation of each quasar.
    \item [a] No detection in this observation.
    \item [b] Some exposures were taken in twilight.
    \item [c] Cloudy conditions, corresponding to $>0.5$~mag of extinction, or poor seeing, corresponding to $>$\ang{;;1.5}.
    \item [d] An echelle-mode observation was conducted for this target.
\end{tablenotes}
\end{threeparttable}
\end{table*}

\begin{table*}
\centering
\caption{Journal of discovery observations continued.}
\label{table:discovery_continued}
\begin{threeparttable}
\begin{tabular}{lccclc}
\toprule\toprule
Target & Telescope & Instrument & Exposure (s) & Date (UTC) & Program ID \\
\midrule
EUCL\,J044657.86$-$570046.6 & Magellan & FIRE & 1800 & 2025-03-16 &  \\
\addlinespace
EUCL\,J181159.50$+$614554.2 & Keck & KCWI & 900 & 2025-06-18 & U275 \\
\addlinespace
EUCL\,J163141.37$+$625920.6 & Keck & MOSFIRE & 3900 & 2025-07-18 & N110 \\
\addlinespace
\multirow{2}{*}{EUCL\,J172902.75$+$641018.1} & Keck & MOSFIRE & 8100 & 2025-08-17\tnote{b} & U262 \\
                                            & LBT & LUCI & 10080 & 2025-09-24 &  \\
\addlinespace
EUCL\,J161430.31$+$452847.1 & Keck & KCWI & 2400 & 2025-08-28 & U262 \\
\addlinespace
\multirow{2}{*}{EUCL\,J155552.09$+$515258.7} & Keck & MOSFIRE & 3600 & 2025-07-22 & N110 \\
                                            & Keck & KCWI & 1200 & 2025-08-28 & U262 \\
\addlinespace
\multirow{3}{*}{EUCL\,J170707.89$+$650252.7} & Keck & MOSFIRE & 3600 & 2025-07-17\tnote{a,b} & N110 \\
                                            & Keck & MOSFIRE & 3600 & 2025-07-18 & N110 \\
                                            & Keck & KCWI & 600 & 2025-08-28 & U262 \\
\addlinespace
\multirow{2}{*}{EUCL\,J154359.34$+$471810.4} & Keck & MOSFIRE & 3000 & 2025-08-17\tnote{b} & U262 \\
                                            & Keck & KCWI & 1200 & 2025-08-29 & U262 \\
\addlinespace
\multirow{2}{*}{EUCL\,J153722.76$+$582927.2} & Keck & MOSFIRE & 1200 & 2025-07-23 & N110 \\
                                            & Keck & KCWI & 1200 & 2025-08-28 & U262 \\
\bottomrule
\end{tabular}
\begin{tablenotes}
    \item {\bf Notes.} All dates are given in Coordinated Universal Time (UTC). This table continues Table~\ref{table:discovery}.
    \item [a] No detection in this observation.
    \item [b] Cloudy conditions, corresponding to $>0.5$~mag of extinction, and/or poor seeing, corresponding to $>$\ang{;;1.5}.
\end{tablenotes}
\end{threeparttable}
\end{table*}

\begin{sidewaystable*}[htp!]
\centering
\caption{Properties of the newly discovered quasars.}
\label{table:new_quasars}
\begin{threeparttable}[b]
    \begin{tabular}{lcccccccccc}
    \toprule\toprule
Name & RA (J2000) & Dec (J2000) & Redshift & $\IE$ & $z$\tnote{a} & \YE & \JE & \HE & $M_{1450}$ & $\log (L_{\rm bol}/L_\odot)$ \\
    \midrule
EUCL\,J172902.75$+$641018.1 & 17:29:02.75 & +64:10:18.1 & 7.77 & ${>28.28}$ & ${...}$ & ${22.61\pm 0.04}$ & ${22.02\pm 0.03}$ & ${21.97\pm 0.03}$ & --25.05 & 46.45 \\
EUCL\,J125308.55$+$705432.3 & 12:53:08.55 & +70:54:32.3 & 7.69 & ${>28.57}$ & ${>24.91}$ & ${23.56\pm 0.09}$ & ${22.97\pm 0.07}$ & ${22.96\pm 0.06}$ & --24.06 & 46.10 \\
EUCL\,J101255.87$+$663058.0 & 10:12:55.87 & +66:30:58.0 & 7.61 & ${>28.40}$ & ${>24.67}$ & ${23.50\pm 0.11}$ & ${23.00\pm 0.09}$ & ${22.93\pm 0.07}$ & --23.98 & 46.07 \\
EUCL\,J052209.82$-$512709.2 & 05:22:09.82 & $-$51:27:09.2 & 7.50 & ${>28.31}$ & ${>24.47}$ & ${22.54\pm 0.05}$ & ${22.17}$\tnote{b} & ${22.19\pm 0.04}$ & --24.75 & 46.35 \\
EUCL\,J135534.55$+$700055.5 & 13:55:34.55 & +70:00:55.5 & 7.45 & ${>28.55}$ & ${>24.66}$ & ${22.17\pm 0.03}$ & ${21.89\pm 0.03}$ & ${21.78\pm 0.03}$ & --25.01 & 46.44 \\
EUCL\,J144545.82$+$714342.3 & 14:45:45.82 & +71:43:42.3 & 7.36 & ${>28.52}$ & ${>24.86}$ & ${23.13\pm 0.09}$ & ${22.78\pm 0.06}$ & ${22.55\pm 0.06}$ & --24.10 & 46.11 \\
EUCL\,J141841.29$+$694930.6 & 14:18:41.29 & +69:49:30.6 & 7.35 & ${>28.64}$ & ${>24.91}$ & ${23.47\pm 0.08}$ & ${22.88\pm 0.04}$ & ${22.61\pm 0.05}$ & --24.00 & 46.07 \\
EUCL\,J163141.37$+$625920.6 & 16:31:41.37 & +62:59:20.6 & 7.33 & ${>28.22}$ & ${>24.63}$ & ${22.54\pm 0.04}$ & ${22.19\pm 0.03}$ & ${22.24\pm 0.03}$ & --24.69 & 46.32 \\
EUCL\,J041250.73$-$563949.7 & 04:12:50.73 & $-$56:39:49.7 & 7.17\tnote{c} & ${>27.37}$ & ${>24.33}$ & ${22.75\pm 0.05}$ & ${22.42\pm 0.03}$ & ${22.18\pm 0.03}$ & --24.41 & 46.22 \\
EUCL\,J134031.51$+$674715.0 & 13:40:31.51 & +67:47:15.0 & 7.05 & ${>28.02}$ & ${>24.96}$ & ${23.28\pm 0.14}$ & ${23.12\pm 0.10}$ & ${22.92\pm 0.09}$ & --23.67 & 45.95 \\
EUCL\,J172258.31$+$574133.5 & 17:22:58.31 & +57:41:33.5 & 7.01 & ${>28.49}$ & ${>24.73}$ & ${23.25\pm 0.07}$ & ${22.84\pm 0.05}$ & ${22.68\pm 0.05}$ & --23.93 & 46.05 \\
EUCL\,J161430.31$+$452847.1 & 16:14:30.31 & +45:28:47.1 & 7.00 & ${>28.31}$ & ${...}$ & ${23.30\pm 0.08}$ & ${22.97\pm 0.08}$ & ${22.70\pm 0.08}$ & --23.80 & 46.00 \\
EUCL\,J093330.60$+$742730.0 & 09:33:30.60 & +74:27:30.0 & 6.96 & ${>28.44}$ & ${>23.21}$ & ${22.01\pm 0.03}$ & ${21.77\pm 0.03}$ & ${21.78\pm 0.03}$ & --24.98 & 46.43 \\
EUCL\,J155552.09$+$515258.7 & 15:55:52.09 & +51:52:58.7 & 6.95 & ${>28.32}$ & ${...}$ & ${22.30\pm 0.04}$ & ${22.22\pm 0.04}$ & ${22.26\pm 0.03}$ & --24.53 & 46.27 \\
EUCL\,J050240.93$-$384906.1 & 05:02:40.93 & $-$38:49:06.1 & 6.90 & ${>27.73}$ & ${22.90\pm 0.08}$ & ${21.53\pm 0.03}$ & ${21.27\pm 0.02}$ & ${21.24\pm 0.02}$ & --25.46 & 46.61 \\
EUCL\,J052645.38$-$460902.7 & 05:26:45.38 & $-$46:09:02.7 & 6.89 & ${>28.12}$ & ${23.49\pm 0.13}$ & ${21.76\pm 0.02}$ & ${21.54\pm 0.02}$ & ${21.43\pm 0.02}$ & --25.19 & 46.51 \\
EUCL\,J025029.93$-$531706.7 & 02:50:29.93 & $-$53:17:06.7 & 6.86 & ${>27.94}$ & ${23.42\pm 0.15}$ & ${21.81\pm 0.03}$ & ${21.55\pm 0.02}$ & ${21.49\pm 0.02}$ & --25.17 & 46.50 \\
EUCL\,J170707.89$+$650252.7 & 17:07:07.89 & +65:02:52.7 & 6.84 & ${>28.42}$ & ${>24.98}$ & ${22.56\pm 0.04}$ & ${22.25\pm 0.03}$ & ${22.28\pm 0.03}$ & --24.46 & 46.24 \\
EUCL\,J154359.34$+$471810.4 & 15:43:59.34 & +47:18:10.4 & 6.84 & ${>28.07}$ & ${...}$ & ${22.63\pm 0.05}$ & ${22.36\pm 0.04}$ & ${22.30\pm 0.03}$ & --24.35 & 46.20 \\
EUCL\,J025210.37$-$412533.8 & 02:52:10.37 & $-$41:25:33.8 & 6.83 & ${>27.98}$ & ${23.19\pm 0.10}$ & ${21.70\pm 0.03}$ & ${21.39\pm 0.02}$ & ${21.34\pm 0.02}$ & --25.32 & 46.55 \\
EUCL\,J143448.06$+$685733.2 & 14:34:48.06 & +68:57:33.2 & 6.82 & ${>28.25}$ & ${>24.82}$ & ${23.04\pm 0.06}$ & ${23.04\pm 0.07}$ & ${22.94\pm 0.06}$ & --23.67 & 45.95 \\
EUCL\,J150537.32$+$773441.0 & 15:05:37.32 & +77:34:41.0 & 6.72 & ${28.13\pm 0.34}$ & ${>24.36}$ & ${22.20\pm 0.04}$ & ${22.06\pm 0.03}$ & ${22.19\pm 0.04}$ & --24.52 & 46.26 \\
EUCL\,J153722.76$+$582927.2 & 15:37:22.76 & +58:29:27.2 & 6.68 & ${>28.29}$ & ${...}$ & ${23.14\pm 0.06}$ & ${22.58\pm 0.04}$ & ${22.43\pm 0.03}$ & --24.10 & 46.11 \\
EUCL\,J044326.27$-$533214.9 & 04:43:26.27 & $-$53:32:14.9 & 6.67 & ${>28.20}$ & ${23.60\pm 0.18}$ & ${22.40\pm 0.03}$ & ${22.22\pm 0.03}$ & ${22.15\pm 0.02}$ & --24.45 & 46.24 \\
EUCL\,J045148.19$-$342622.2 & 04:51:48.19 & $-$34:26:22.2 & 6.65 & ${>27.76}$ & ${23.05\pm 0.10}$ & ${22.18\pm 0.03}$ & ${22.10\pm 0.03}$ & ${22.11\pm 0.03}$ & --24.57 & 46.28 \\
EUCL\,J115522.89$+$704612.3 & 11:55:22.89 & +70:46:12.3 & 6.65 & ${>28.48}$ & ${>24.70}$ & ${23.39\pm 0.10}$ & ${22.83\pm 0.06}$ & ${22.60\pm 0.05}$ & --23.84 & 46.02 \\
EUCL\,J044657.86$-$570046.6 & 04:46:57.86 & $-$57:00:46.6 & 6.64 & ${>28.03}$ & ${24.46\pm 0.23}$ & ${22.71\pm 0.05}$ & ${22.48\pm 0.04}$ & ${22.47\pm 0.05}$ & --24.19 & 46.14 \\
EUCL\,J091639.93$+$683652.9 & 09:16:39.93 & +68:36:52.9 & 6.64 & ${27.38\pm 0.13}$ & ${23.55\pm 0.17}$ & ${21.83\pm 0.03}$ & ${21.73\pm 0.03}$ & ${21.59\pm 0.03}$ & --24.94 & 46.42 \\
EUCL\,J181159.50$+$614554.2 & 18:11:59.50 & +61:45:54.2 & 6.63 & ${>28.34}$ & ${>24.68}$ & ${21.97\pm 0.05}$ & ${22.07\pm 0.06}$ & ${21.92\pm 0.04}$ & --24.60 & 46.29 \\
EUCL\,J173243.48$+$601612.6 & 17:32:43.48 & +60:16:12.6 & 6.61 & ${26.41\pm 0.05}$ & ${23.88\pm 0.19}$ & ${21.37\pm 0.04}$ & ${21.30\pm 0.04}$ & ${21.22\pm 0.03}$ & --25.37 & 46.57 \\
EUCL\,J031547.94$-$684405.6 & 03:15:47.94 & $-$68:44:05.6 & 6.60 & ${27.47\pm 0.27}$ & ${...}$ & ${21.69\pm 0.03}$ & ${21.26\pm 0.02}$ & ${20.97\pm 0.02}$ & --25.42 & 46.59 \\
    \bottomrule
    \end{tabular}
\begin{tablenotes}
    \item \textbf{Notes.} The quasars are listed in order of decreasing redshift. Redshifts are estimated visually from the Ly$\alpha$ break, with typical uncertainties of 0.05--0.1 \citep[e.g.][but see note~c for cases where weak or absorbed Ly$\alpha$ increases the uncertainty to $\sim$0.2]{Matsuoka2016}. Broad-band magnitudes are derived from PSF-fitting photometry (Sect.~\ref{sec:photometry}); $3\sigma$ limits are reported for non-detections. The absolute magnitude $M_{1450}$ is computed from the observed \JE-band magnitude assuming a spectral slope $\alpha=-1.7$ \citep[e.g.][]{Telfer2002}, and the bolometric luminosity $L_{\mathrm{bol}}$ from the scaling relation of \citet{Runnoe2012}.
    \item[a] \added{Entries marked with ``${...}$'' indicate no ancillary z-band imaging is available at the source position.}
    \item[b] The \JE band photometry derived from the NIR stacked image is affected by hot pixels. Here we quote the magnitude measured from the JWST/NIRSpec spectrum of this target. 
    \item[c] The redshift is determined with the \mgii line in the JWST spectrum of this target.
\end{tablenotes}
\end{threeparttable}
\end{sidewaystable*}

\clearpage
\section{Multi-band cutouts}\label{apdx:data_public}

\begin{figure*}[htbp!]
  \begin{center}
    \includegraphics[angle=0,width=0.95\columnwidth]{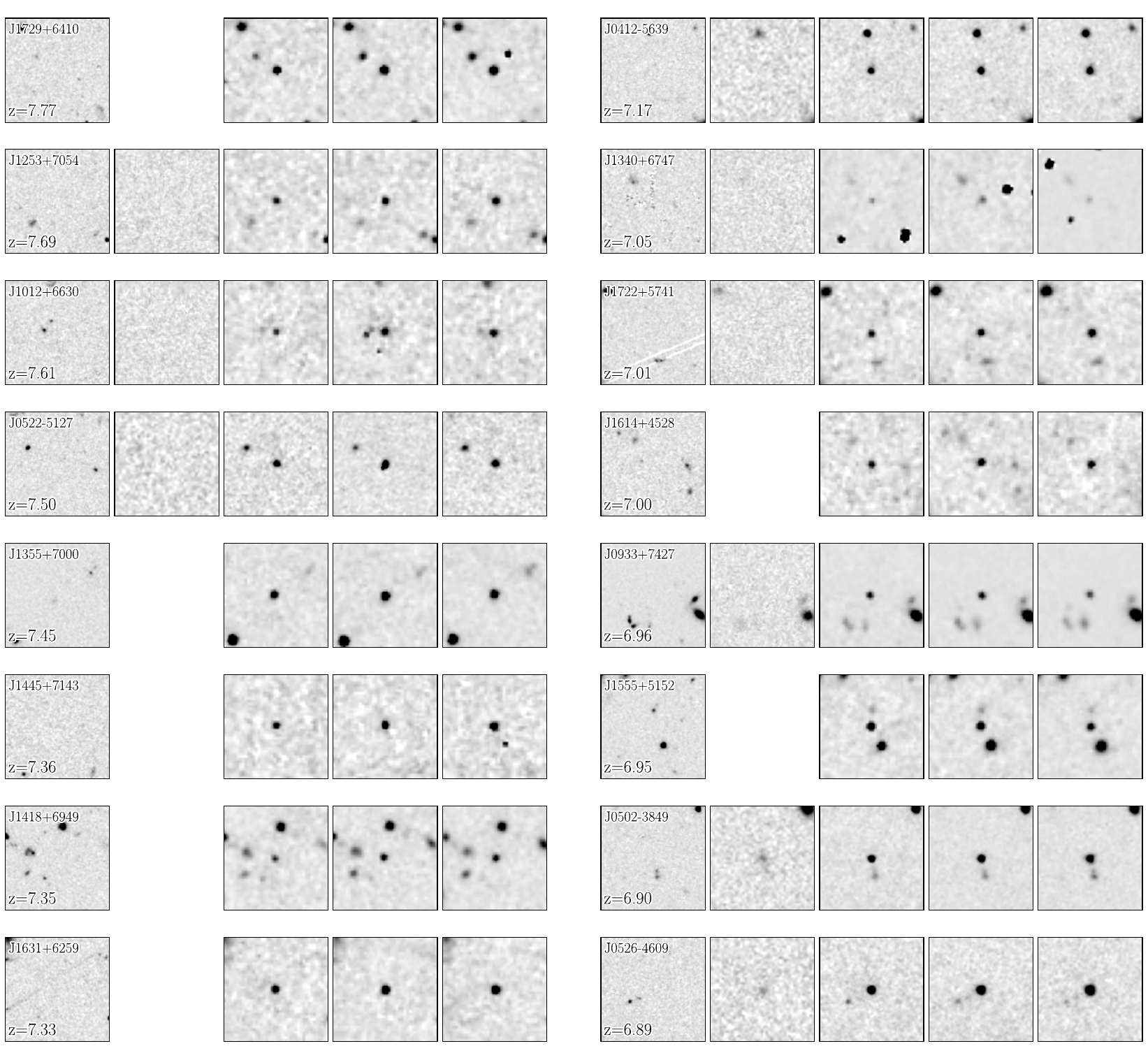}
   \end{center}
\caption{Cutouts of the new quasars ($12^{\prime\prime}\times12^{\prime\prime}$) reported in this work. From left to right: \IE, $z$, \YE, \JE, and \HE bands. Target names and their redshift are listed in the \IE-band cutouts.}
\label{fig:all_cutouts}
\end{figure*}

\begin{figure*}[htbp!]
  \begin{center}
    \includegraphics[angle=0,width=0.95\columnwidth]{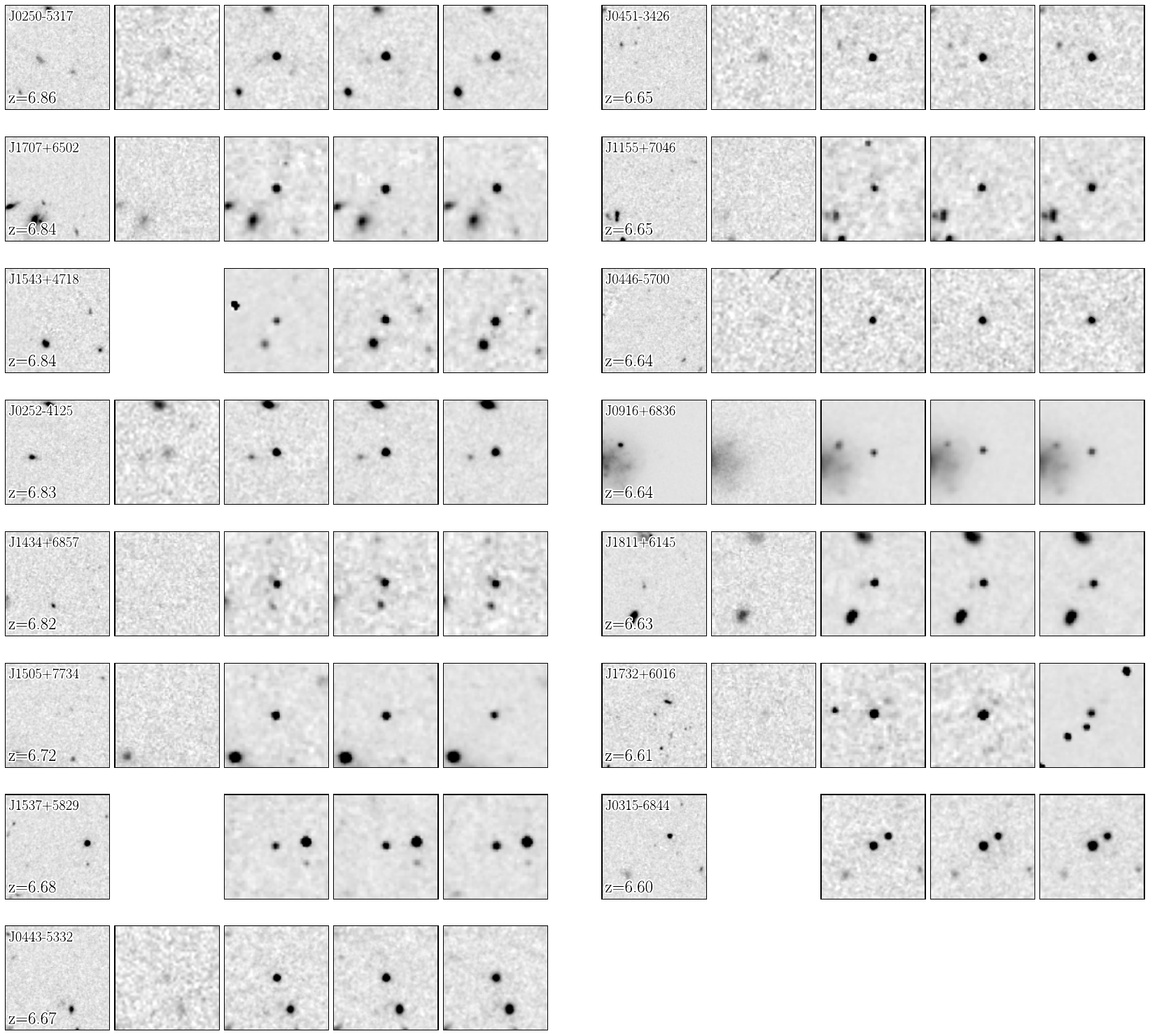}
   \end{center}
\caption{Same as Fig.~\ref{fig:all_cutouts} but for the rest of the quasars.}
\label{fig:all_cutouts_2}
\end{figure*}

Figures~\ref{fig:all_cutouts} and \ref{fig:all_cutouts_2} display the $12^{\prime\prime}\times12^{\prime\prime}$ cutouts of \IE, $z$, \YE, \JE, \HE bands of the new quasars reported in this work. These cutouts are available at \url{https://doi.org/10.57780/esa-50769fd}. The discovery spectra are available from the corresponding author upon reasonable request.

\clearpage
\section{Reduced 2D/1D discovery spectra}\label{apdx:discovery_spectra}

\begin{figure*}[htbp!]
  \begin{center}
    \includegraphics[angle=0,width=0.92\columnwidth]{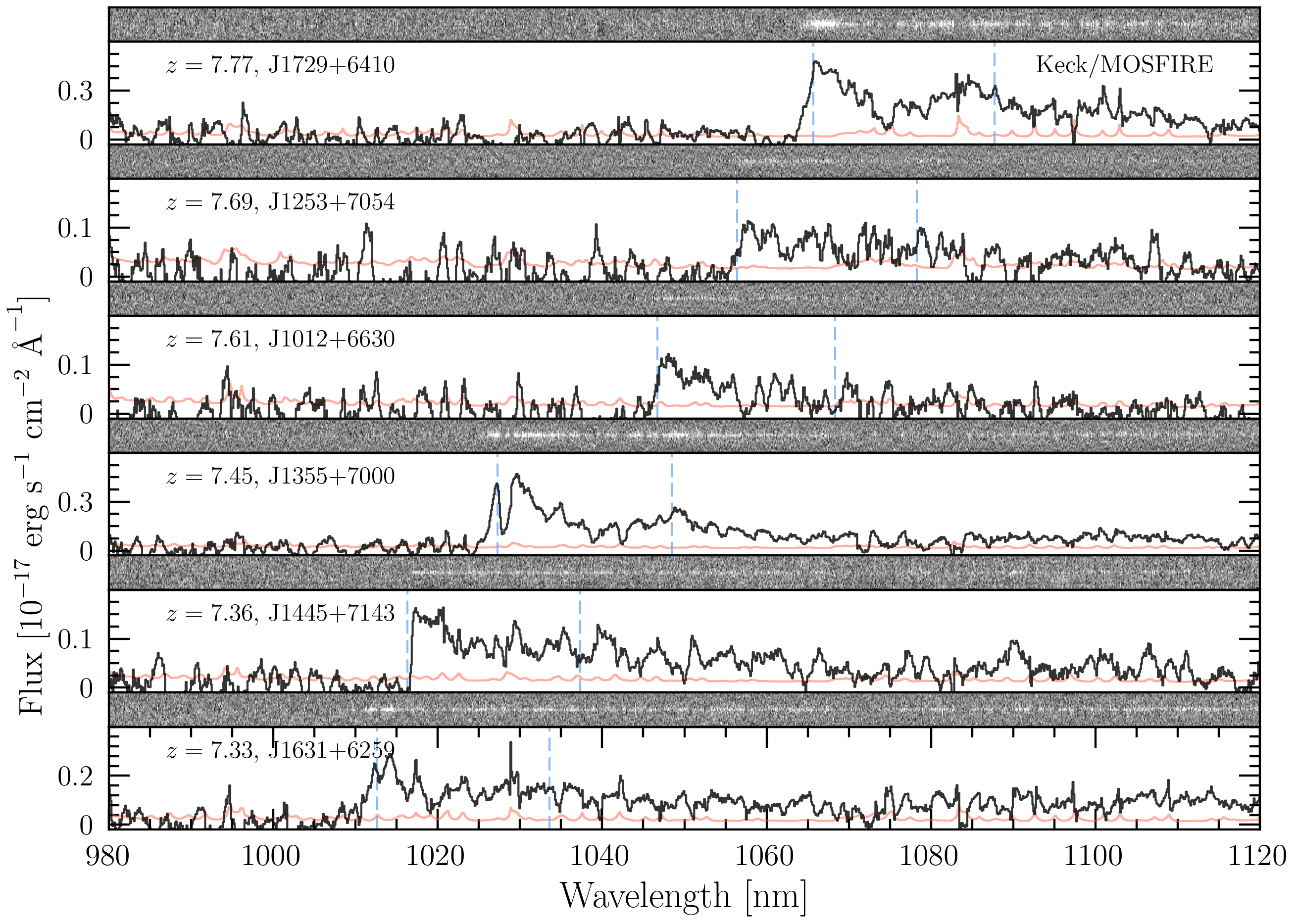}
   \end{center}
\caption{Reduced 2D/1D discovery spectra of Keck/MOSFIRE. The black curves are the fluxed spectra, smoothed with inverse-variance-weights with a boxcar of 9 pixel. The red curves are the noise vectors with the same smoothing. The light blue dashed lines mark the positions of the \lya and \nv lines at the tentative redshifts for each quasar. These redshifts are estimated from the \lya lines.}
\label{fig:spec2d_mosfire}
\end{figure*}

\begin{figure*}[htbp!]
  \begin{center}
    \includegraphics[angle=0,width=0.92\columnwidth]{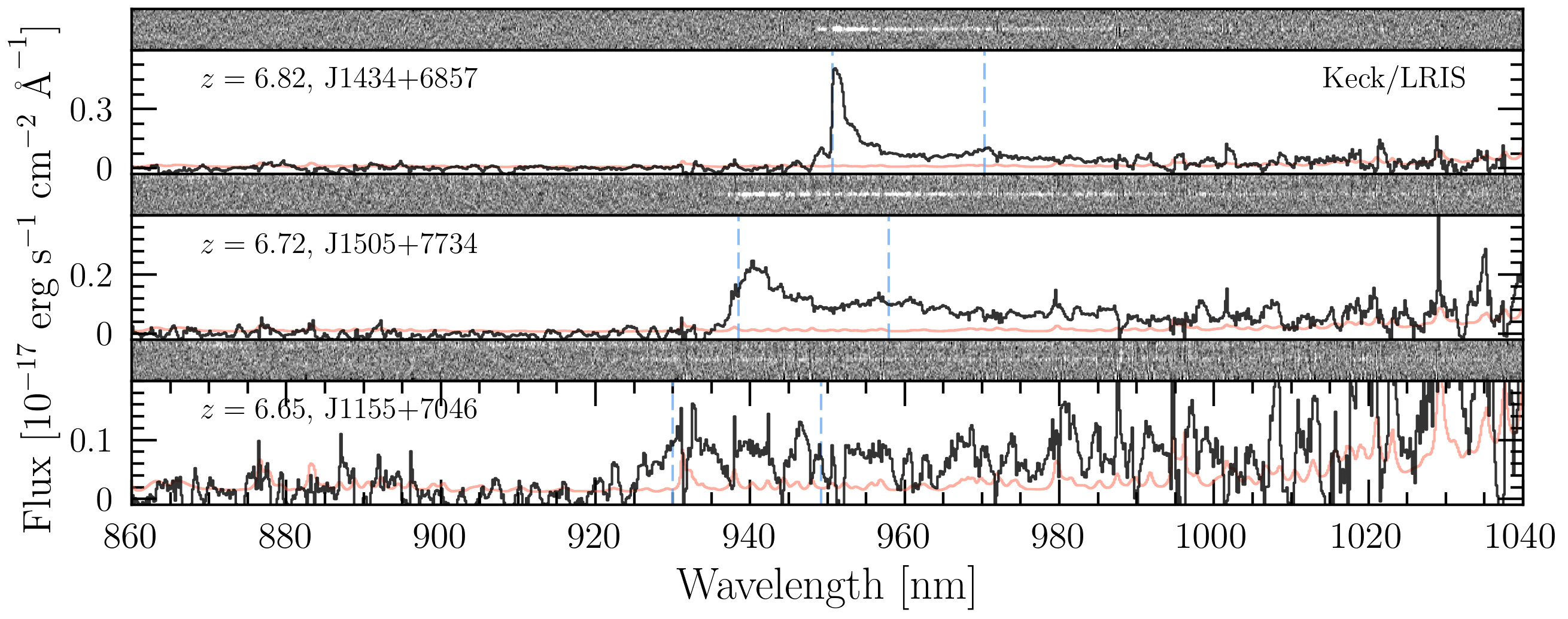}
   \end{center}
\caption{Reduced 2D/1D discovery spectra of Keck/LRIS. The spectra are inverse-variance-smoothed with a boxcar of 7 pixel.}
\label{fig:spec2d_lris}
\end{figure*}

\begin{figure*}[htbp!]
  \begin{center}
    \includegraphics[angle=0,width=0.92\columnwidth]{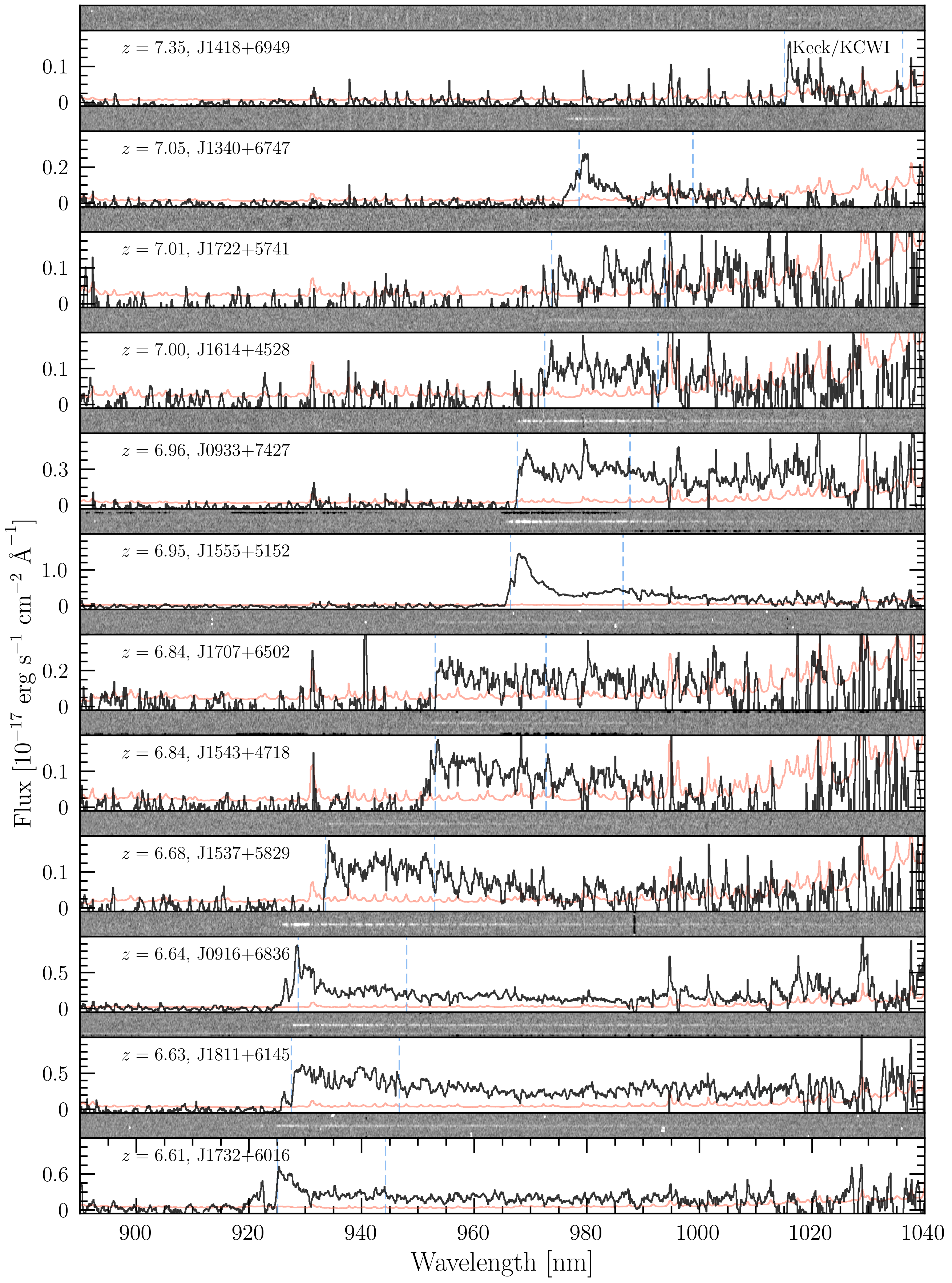}
   \end{center}
\caption{The reduced 2D and 1D discovery spectra of Keck/KCWI. The spectra are inverse-variance-smoothed with a boxcar of 7 pixel.}
\label{fig:spec2d_kcwi}
\end{figure*}

\begin{figure*}[htbp!]
  \begin{center}
    \includegraphics[angle=0,width=0.92\columnwidth]{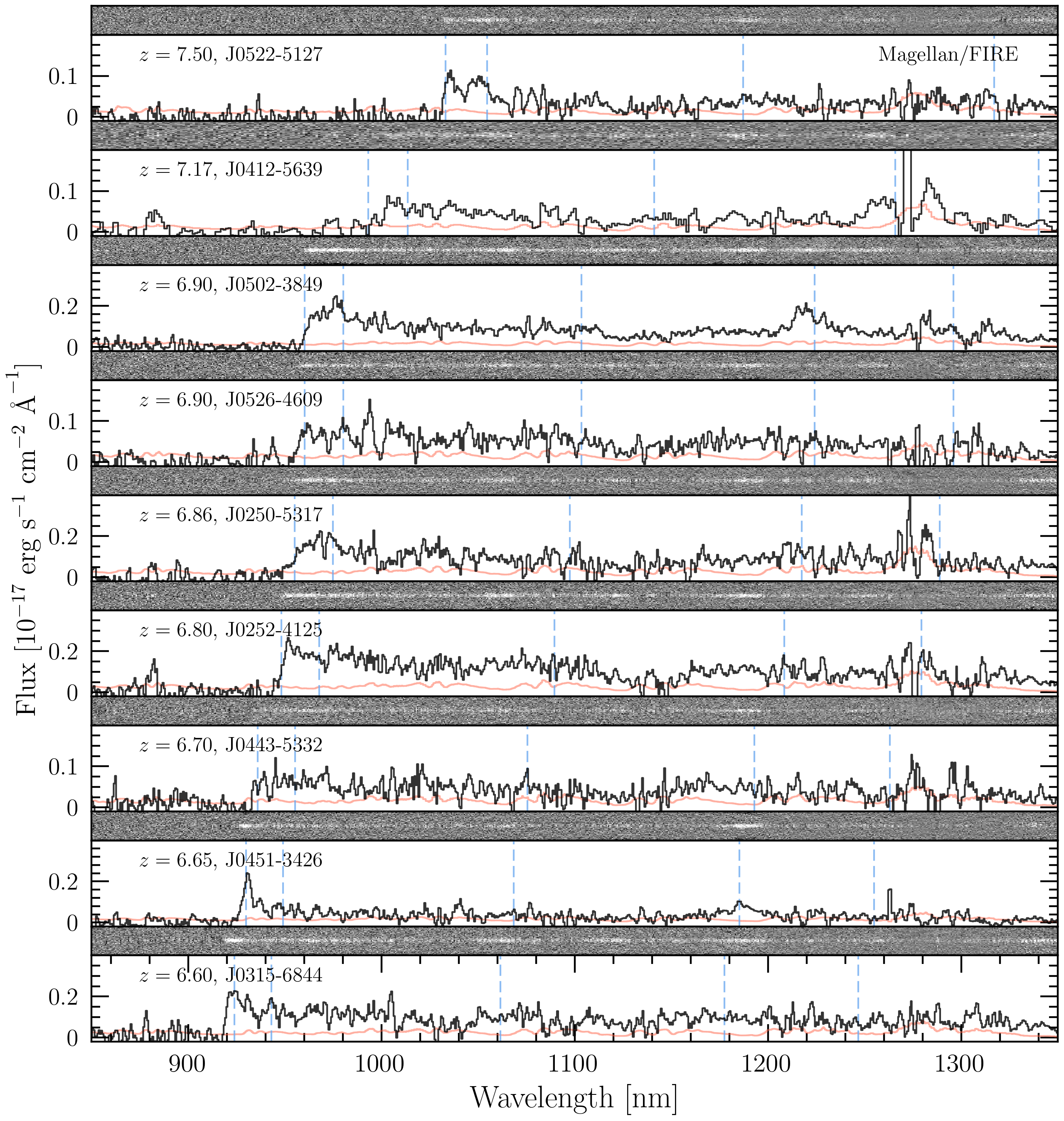}
   \end{center}
\caption{Reduced 2D/1D discovery spectra of Magellan/FIRE. The spectra are inverse-variance-smoothed with a boxcar of 3 pixel. The green dashed lines mark the positions of the \lya, \nv, \siiv, \civ, and \heii lines at the tentative redshifts for each quasar. These redshifts are estimated from the \lya lines.}

\label{fig:spec2d_fire}
\end{figure*}

\clearpage
\section{Reduced 2D/1D spectra of some examples of contaminants}\label{sec:append_contaminant}

\begin{figure*}[htbp!]
  \begin{center}
    \includegraphics[angle=0,width=0.95\columnwidth]{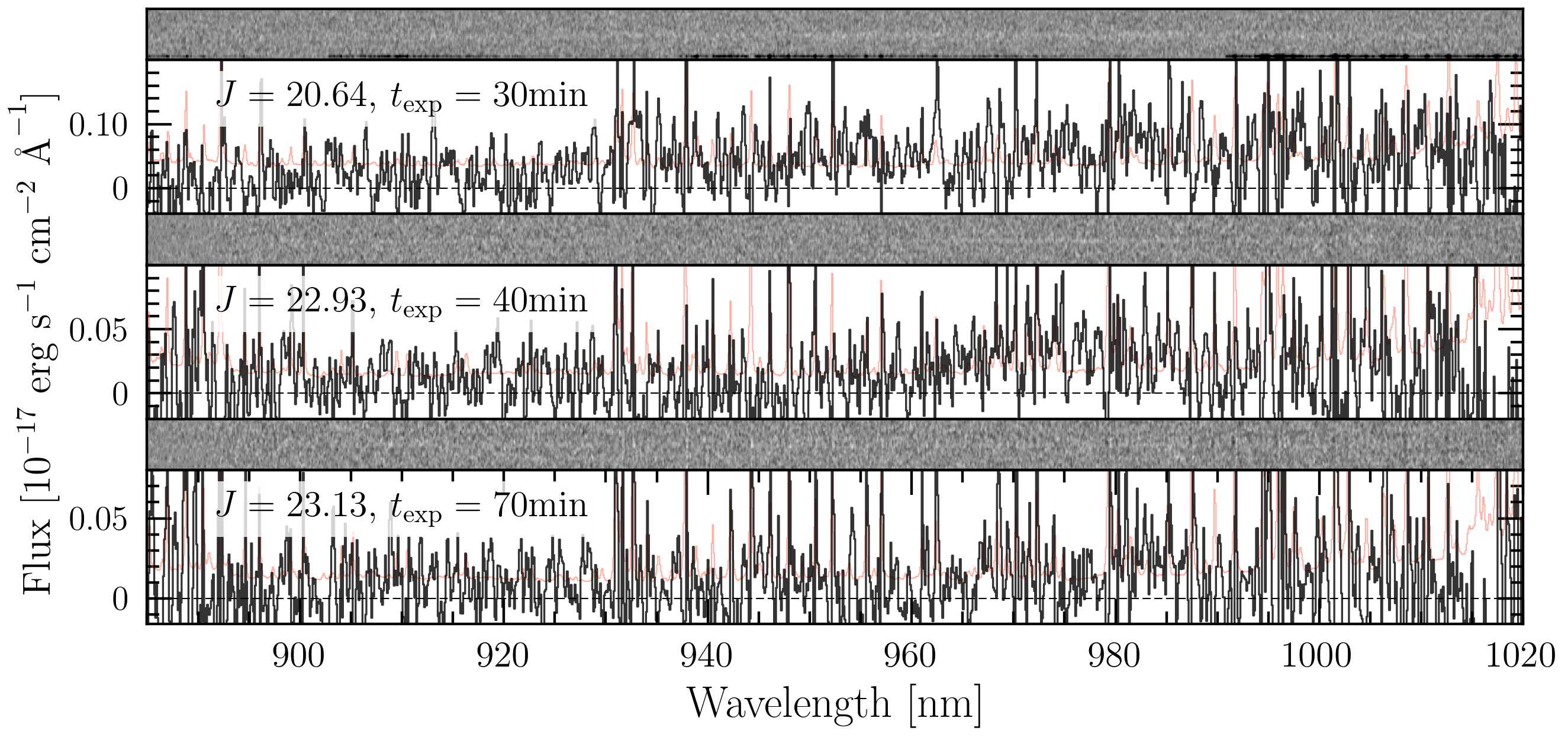}
   \end{center}
\caption{Reduced 2D/1D spectra of three contaminants observed with Keck/KCWI. The total exposure time and the \JE-band magnitude for each target are listed in each panel. The black curves are the fluxed spectra, smoothed with an inverse-variance-weighting and a boxcar of 3 pixel. The red curves are the noise vectors with the same smoothing.}
\label{fig:spec2d_kcwi_cont}
\end{figure*}

\begin{figure*}[htbp!]
  \begin{center}
    \includegraphics[angle=0,width=0.95\columnwidth]{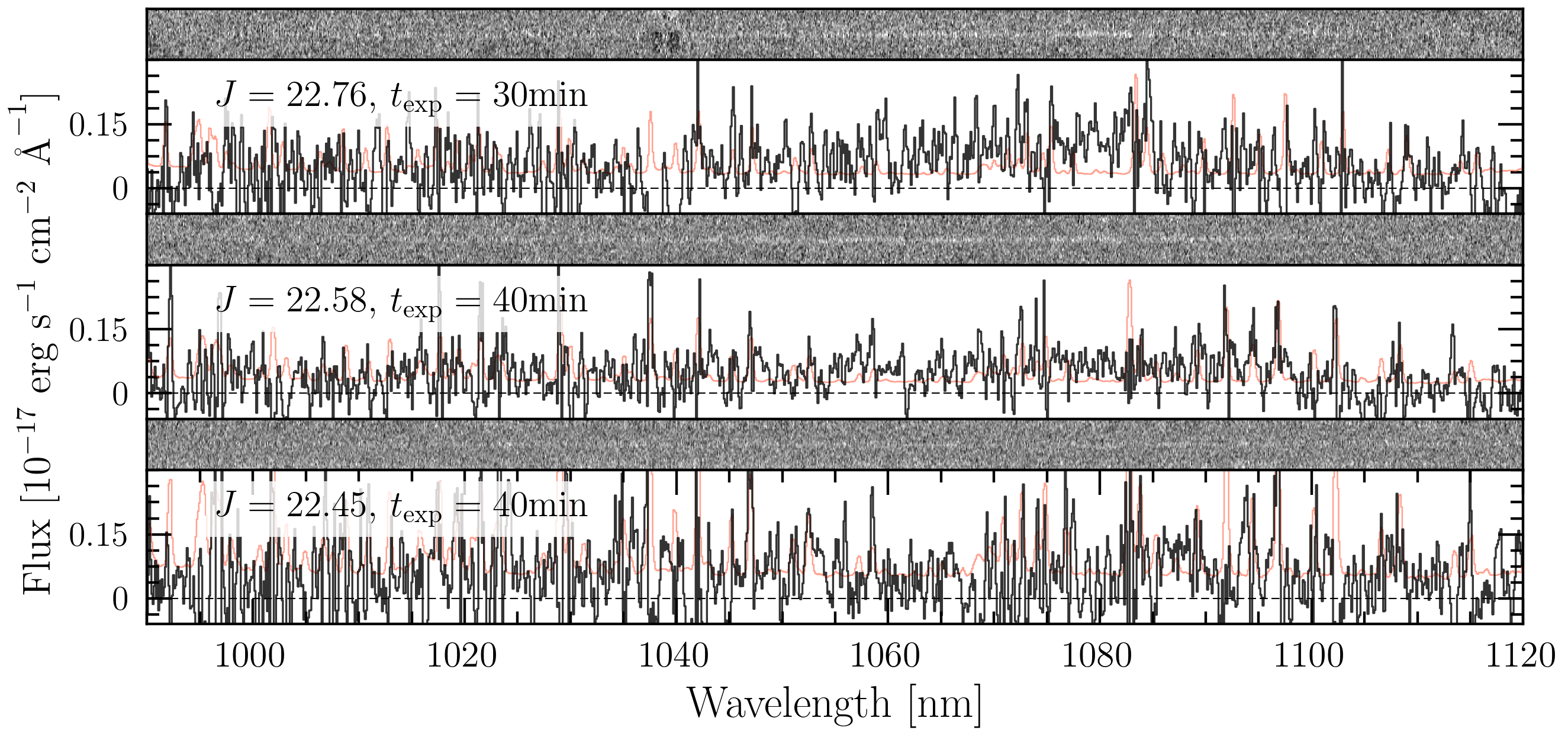}
   \end{center}
\caption{Same as Fig.~\ref{fig:spec2d_kcwi_cont}, but for Keck/MOSFIRE.}
\label{fig:spec2d_mosfire_cont}
\end{figure*}

\begin{figure*}[htbp!]
  \begin{center}
    \includegraphics[angle=0,width=0.95\columnwidth]{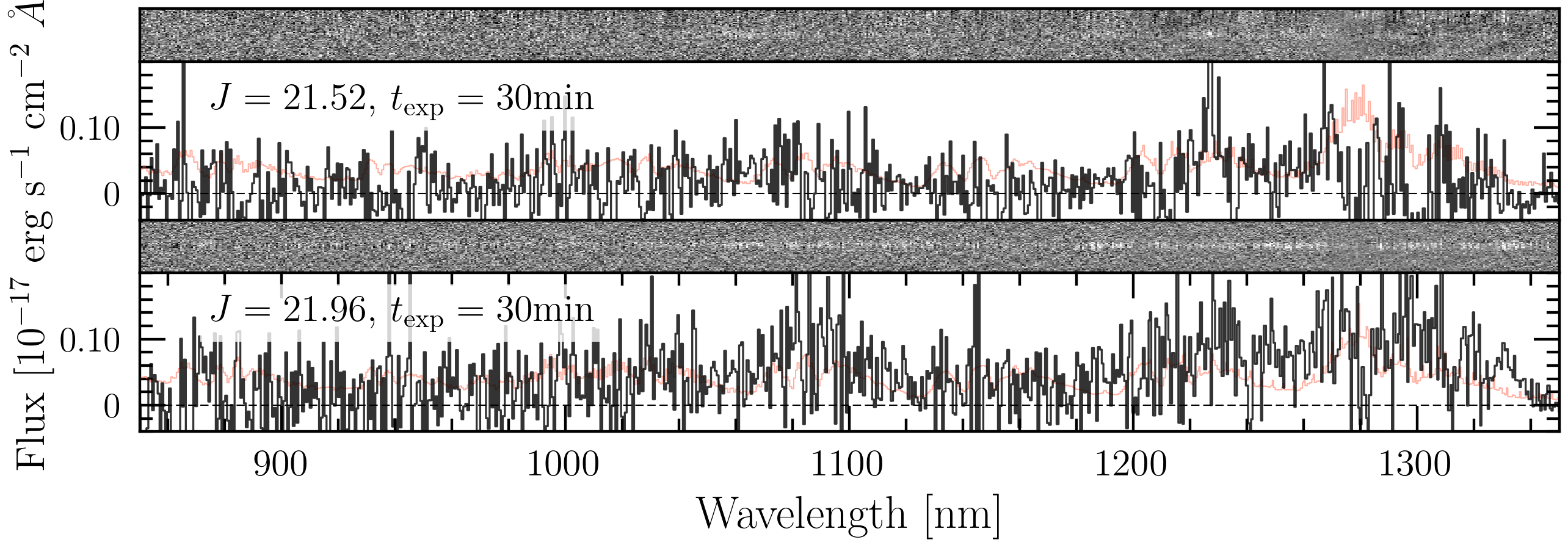}
   \end{center}
\caption{Same as Fig.~\ref{fig:spec2d_kcwi_cont}, but for Magellan/FIRE.}
\label{fig:spec2d_fire_cont}
\label{LastPage}
\end{figure*}

\begin{table*}
\centering
\caption{List of identified contaminants. Some of them were tentatively identified as brown dwarfs.}\label{table:contaminant_list}
\begin{threeparttable}[b]
{
\begin{tabular}{lcccr}
\toprule\toprule
Name & RA & Dec & Instruments & Notes \\
\midrule
EUCL\,J015220.18$-$632724.6 & 01:52:20.18 & $-$63:27:24.6 & FIRE & dwarf \\
EUCL\,J030218.49$-$535654.0 & 03:02:18.49 & $-$53:56:54.0 & FIRE & dwarf \\
EUCL\,J034256.06$-$541719.3 & 03:42:56.06 & $-$54:17:19.3 & FIRE & dwarf \\
EUCL\,J034303.88$-$552508.2 & 03:43:03.88 & $-$55:25:08.2 & FIRE & dwarf \\
EUCL\,J034343.36$-$544243.6 & 03:43:43.36 & $-$54:42:43.6 & FIRE & dwarf \\
EUCL\,J034510.85$-$582115.9 & 03:45:10.85 & $-$58:21:15.9 & FIRE & dwarf \\
EUCL\,J035110.64$-$483806.7 & 03:51:10.64 & $-$48:38:06.7 & FIRE & dwarf? \\
EUCL\,J052656.21$-$545056.4 & 05:26:56.21 & $-$54:50:56.4 & FIRE & dwarf \\
EUCL\,J052929.24$-$461251.6 & 05:29:29.24 & $-$46:12:51.6 & FIRE & dwarf \\
EUCL\,J053001.69$-$523428.6 & 05:30:01.69 & $-$52:34:28.6 & FIRE & dwarf \\
EUCL\,J061759.55$-$492927.2 & 06:17:59.55 & $-$49:29:27.2 & FIRE & dwarf \\
EUCL\,J084714.14$+$641938.7 & 08:47:14.14 & $+$64:19:38.7 & KCWI; MOSFIRE & dwarf? \\
EUCL\,J091849.67$+$660353.1 & 09:18:49.67 & $+$66:03:53.1 & MOSFIRE & dwarf \\
EUCL\,J092049.59$+$711423.7 & 09:20:49.59 & $+$71:14:23.7 & KCWI & contaminant \\
EUCL\,J092540.76$+$693549.5 & 09:25:40.76 & $+$69:35:49.5 & KCWI & contaminant \\
EUCL\,J093005.83$+$720814.5 & 09:30:05.83 & $+$72:08:14.5 & KCWI; MOSFIRE & dwarf? \\
EUCL\,J094603.30$+$630522.1 & 09:46:03.30 & $+$63:05:22.1 & KCWI & dwarf? \\
EUCL\,J094617.98$+$652346.6 & 09:46:17.98 & $+$65:23:46.6 & KCWI & dwarf? \\
EUCL\,J100835.25$+$691935.8 & 10:08:35.25 & $+$69:19:35.8 & KCWI & dwarf \\
EUCL\,J102324.58$+$703158.6 & 10:23:24.58 & $+$70:31:58.6 & KCWI & dwarf? \\
EUCL\,J114858.03$+$743225.2 & 11:48:58.03 & $+$74:32:25.2 & LRIS & contaminant \\
EUCL\,J115159.14$+$682156.9 & 11:51:59.14 & $+$68:21:56.9 & KCWI & dwarf \\
EUCL\,J115853.41$+$683101.2 & 11:58:53.41 & $+$68:31:01.2 & KCWI & contaminant \\
EUCL\,J122605.56$+$731439.4 & 12:26:05.56 & $+$73:14:39.4 & KCWI & contaminant \\
EUCL\,J124115.79$+$662306.7 & 12:41:15.79 & $+$66:23:06.7 & KCWI & contaminant \\
EUCL\,J131224.33$+$670729.5 & 13:12:24.33 & $+$67:07:29.5 & LRIS & dwarf \\
EUCL\,J131809.35$+$700558.3 & 13:18:09.35 & $+$70:05:58.3 & KCWI & dwarf \\
EUCL\,J133810.37$+$685317.0 & 13:38:10.37 & $+$68:53:17.0 & LRIS & contaminant \\
EUCL\,J142441.75$+$693037.4 & 14:24:41.75 & $+$69:30:37.4 & KCWI & contaminant \\
EUCL\,J143351.31$+$540844.7 & 14:33:51.31 & $+$54:08:44.7 & MOSFIRE & contaminant \\
EUCL\,J144051.62$+$684954.5 & 14:40:51.62 & $+$68:49:54.5 & KCWI & contaminant \\
EUCL\,J144852.41$+$732435.6 & 14:48:52.41 & $+$73:24:35.6 & LRIS & contaminant \\
EUCL\,J154127.20$+$714353.1 & 15:41:27.20 & $+$71:43:53.1 & LRIS & dwarf? \\
EUCL\,J162903.17$+$683127.6 & 16:29:03.17 & $+$68:31:27.6 & MOSFIRE & contaminant \\
EUCL\,J170812.71$+$563448.0 & 17:08:12.71 & $+$56:34:48.0 & MOSFIRE & contaminant \\
EUCL\,J175348.34$+$665640.3 & 17:53:48.34 & $+$66:56:40.3 & LRIS & dwarf \\
EUCL\,J180128.23$+$651012.8 & 18:01:28.23 & $+$65:10:12.8 & LRIS & dwarf \\
EUCL\,J183032.83$+$585324.0 & 18:30:32.83 & $+$58:53:24.0 & MOSFIRE & dwarf \\
EUCL\,J215435.66$+$180214.9 & 21:54:35.66 & $+$18:02:14.9 & LRIS & dwarf \\
\bottomrule
\end{tabular}
}
% \begin{tablenotes}
%     \item [a] No detection in this observation. 
% \end{tablenotes}
\end{threeparttable}    
\end{table*}

\end{appendix}

\end{document}